\documentclass[fleqn,usenatbib]{mnras}
\usepackage{newtxtext,newtxmath}
\usepackage[T1]{fontenc}

\DeclareRobustCommand{\VAN}[3]{#2}
\let\VANthebibliography\thebibliography
\def\thebibliography{\DeclareRobustCommand{\VAN}[3]{##3}\VANthebibliography}

\usepackage{graphicx}	
\usepackage{amsmath}	
\usepackage{verbatim}   
\usepackage{longtable}
\usepackage{multicol,lipsum}
\usepackage{subfigure}
\usepackage{adjustbox}
\usepackage{longtable}
\usepackage{caption}
\usepackage{float}
\usepackage{placeins}
\usepackage{filecontents}
\usepackage{changepage}   
\usepackage{flafter}
\title[X-ray timing and spectral analysis of reverberating active galactic nuclei]{X-ray timing and spectral analysis of reverberating active galactic nuclei}

\author[S. Hancock et al.]{
S. Hancock,$^{1}$\thanks{E-mail: steff.hancock@bristol.ac.uk}  
A. J. Young,$^{1}$ and P. Chainakun$^{2,3}$ \\
$^{1}$HH Wills Physics Laboratory, Tyndall Avenue, Bristol BS8 1TL, UK\\
$^2$School of Physics, Institute of Science, Suranaree University of Technology, Nakhon Ratchasima 30000, Thailand\\
$^3$Centre of Excellence in High Energy Physics and Astrophysics, Suranaree University of Technology, Nakhon Ratchasima 30000, Thailand\\
}

\date{Accepted XXX. Received YYY; in original form ZZZ}

\pubyear{2021}

\begin{document}
\label{firstpage}
\pagerange{\pageref{firstpage}--\pageref{lastpage}}
\maketitle

\begin{abstract}
We use the publicly available \textit{XMM-Newton} archive to select a sample of 20 active galactic nuclei (AGN) known to exhibit reverberation signatures caused by the reflection of X-rays from the corona off the accretion disc that feeds the central black hole engine. Inverse Compton scattering by energetic electrons, coupled with accretion disc fluctuations give rise to the highly variable observed X-ray spectrum, the behaviour of which is still not fully understood. We use 121 observations in 3 -- 4 distinct spectral states for each source and calculate the time lags as a function of frequency. We fit the relativistic reflection model \texttt{RELXILL} and explore parameter correlations. The known scaling relationship between the black hole mass and time lag is well recovered and the continuum flux is coupled strongly to the disc reflection flux. We also find that 1H~0707-495 and IRAS~13224-3809 are well described using reflection and absorption modelling in a variety of flux states. The reflection fraction is strongly coupled to the power law photon index and may be linked to dynamics of the emitting region. The data reveals hints of the power law evolutionary turnover when the 2 -- 10 keV Eddington fraction is $\sim0.02$, the origin of which is not fully understood. Finally, we report the covering fraction is inversely correlated with the flux and power law photon index in IRAS 13224-3809. These findings support recent studies of 1H 0707-495 where the covering fraction may contribute to the observed variability via flux modulations from non-uniform orbiting clouds. 
\end{abstract}
\begin{keywords}
X-rays: individual: 1H~0707-495; X-rays: individual: IRAS13224-3809; galaxies: active – galaxies:
\end{keywords}

\section{Introduction}

Active galactic nuclei (AGN) are thought to be powered by an optically thick, geometrically thin accretion disk \citep{1973A&A....24..337S} which releases gravitational energy as multi-temperature black body radiation, peaking in the optical and UV regions of the spectrum. The spectra of AGN are known to be composed of direct continuum emission, however, cold gas can can absorb and reflect some of the X-ray continuum. Evidence for this was first reported from Ginga observations of multiple Seyfert galaxies \citep{1990Natur.344..132P}. The cold, optically thick and high column density material seen through fluorescence and reflection confirms the presence of an accretion disk and this phenomenon can also explain the observed iron K line \citep{1991MNRAS.249..352G}.  The X-ray continuum is produced by inverse Compton (IC) scattering by a hot optically-thin electron population (i.e. the hot corona) which increases photon energies up to X-ray energy levels. These high energy X-rays are often reflected off the accretion disk with a modified spectrum which leads to the fluorescence of Fe K lines at 6.4 keV. 

The first relativistically broadened Fe K line was discovered in MCG-6-30-15 \citep{1995Natur.375..659T}, where the intrinsically narrow 6.4~keV Fe line was broadened and skewed to lower energies by Doppler and relativistic effects. Later, the first hint of reflection or reprocessed time-delay was seen in \emph{XMM-Newton} observations of Ark564 \citep{2007MNRAS.382..985M}. X-ray reverberation lags were robustly first discovered in 1H0707-495 \citep{2009Natur.459..540F} where the soft energy band (0.3 -- 1~keV) lagged behind the hard band (1 -- 4~keV) by 30s. Many reverberation lags have since been discovered \citep[e.g.][]{2011MNRAS.416L..94E,2013MNRAS.434.1129K,2013ApJ...767..121Z,2013MNRAS.431.2441D,2016MNRAS.462..511K}. 

The current understanding of the soft negative time lag is the signature of relativistic reflection that ‘reverberates’ or responds to continuum fluctuations and have been interpreted as the averaged light crossing time from the source to the reflecting region \citep{2013MNRAS.431.2441D}. The hard positive lags observed at lower frequencies have been explained by fluctuations of the accretion rate propagating from outer to inner radii, causing the hard X-rays produced at smaller radii to respond after soft X-rays produced at larger radii \cite[e.g.][]{2001MNRAS.327..799K,2006MNRAS.367..801A,2008MNRAS.388..211A,2017MNRAS.468.3663J}. These hard lags were well known to exist in X-ray binaries before they were discovered in AGN \cite[e.g.][]{Nowak1999}.

The X-ray spectrum of AGN contains several components which have been subject to many recent studies. The continuum radiation arising from IC scattering of lower energy photons emerges from a compact corona or extended X-ray region located on the rotational axis of the central black hole, which can be approximated by a power law continuum of $N(E) \propto E^{-\Gamma}$ where the photon index $\Gamma$ is determined by corona properties such as the optical depth and electron temperature \citep{Rybicki&Lightman2004}. The exact origin of this power law remains unclear, however \cite{1991ApJ...380L..51H} cite earlier models from the 1970s where the spectrum depends on temperature and optical depth thus the spectral shape could be reproduced by applying an `ad-hoc' choice of these parameters. They explored progression of this paradigm by suggesting a two phase `coupled' model where optically thick emission of the cool layers inject soft photons for Comptonisation and hard Comptonised photons contribute to the increase in temperature. 
Variations in luminosity are a result of fluctuations in the mass accretion rate. Propagation of perturbations in the accretion flow in the inward radial direction on viscous timescales cause the observed variation in X-ray flux \citep{2001MNRAS.327..799K}. AGN with a high accretion rate can produce soft X-rays that contribute to Compton cooling of the harder X-rays leading to a steeper spectrum at high energies and a larger value of~$\Gamma$, for example, see \cite{Pounds1995}. This relationship suggests that the accretion disc and X-ray corona are connected and there is clear evidence of a strong correlation between the UV/optical and X-ray luminosity also suggests the existence of a regulated mechanism between the disc and corona \citep[e.g.,][]{Liu2002,Grupe2010,2016MNRAS.459.3963C}. Furthermore, there are strong correlations between the hard X-ray photon index and the Eddington ratio ($L_{\text{Bol}} / L_{\text{Edd}}$) where $L_{\text{Edd}}$ is the Eddington luminosity and $L_{\text{Bol}}$ is the bolometric luminosity \citep[see e.g.,][]{2013MNRAS.433.2485B,2015MNRAS.447.1692Y}. 

Reverberation lags can be modelled using ray-tracing techniques which follow photon trajectories along Kerr geodesics to calculate the time of arrival of the direct continuum and the reflected photons \citep[e.g.][]{1992MNRAS.259..569K, 1999ApJ...514..164R, 2012MNRAS.420.1145C, 2016MNRAS.458..200W}. \cite{2013MNRAS.430..247W} developed general relativistic ray tracing simulations of the propagation of X-rays from both point-like and extended coronae, modelling the frequency dependence of the lag which provided the height of a point like corona above the accretion disc or its radial extent. The model included the travel time difference of the ray in general relativity to that in classical Euclidean space. 
\cite{2014MNRAS.438.2980C} built on this analysis by considering the frequency and energy dependence of the Fe K$\alpha$ reverberation lags from an X-ray point source in NGC 4151 and found that the energy dependence of the lags in the Fe K$\alpha$ line region depends on the height of the X-ray source and the black hole mass. Their findings indicated a source height of $7r_g$, where the gravitational radius $r_g = GM/c^2$ (where $G$ is the gravitational constant, $M$ is the black hole mass and $c$ is the speed of light) using the lamppost model. However the authors argued that this may be too simplistic in approach and that models considering radial and vertical components in an extended X-ray corona scenario would be more realistic. There have been many subsequent studies of the lamppost model  \citep[e.g.][]{2014MNRAS.439.3931E,2014MNRAS.445...56K,2015MNRAS.452..333C,2016A&A...594A..71E,2018MNRAS.480.2650C,2019MNRAS.488..324I,2020NatAs...4..597A} and the extended corona scenario \citep[e.g.][]{2016MNRAS.458..200W, chainakun_investigating_2017, 2019MNRAS.487..667C}. 

\cite{2021MNRAS.506.4960H} have studied the fundamental parameters of the X-ray corona using a multi-observatory approach, using \textit{XMM-Netwon}, \textit{NuSTAR} and the \textit{Swift-BAT} telescopes to cover the 0.3--195 keV energy ranges for a sample of 33 AGN that have simultaneous observations from each observatory. They investigated correlations between the corona parameters and the physical properties of AGN such as the black hole mass, and Eddington ratio. Despite recovering the relationship between the reflection coefficient $R$ and the photon index $\Gamma$ and the X-ray Baldwin-effect, they found no strong correlations between the compactness, black hole mass or coronal temperatures. Whilst admitting that the result was `puzzling', yet remained consistent with previous work and concluded that the fundamental physics of the parameters is not yet understood. 

Since the X-ray spectrum of AGN contains components of direct and reflected emission arising from the accretion disc and irradiation of a fraction of the primary X-rays on the accretion disc, the strength of the reflection can be established from the ratio of direct and reflected flux. This was developed in the relativistic reflection model \texttt{RELXILL} \citep{Garc_a_2014} which combines \texttt{XILLVER} reflection code and the relativistic line profiles code \texttt{RELLINE} \citep{Dauser2016}, where the reflection fraction, RF, is defined at the ratio of photons that hit the disk to those that reach infinity. The accretion disk ranges from the marginally stable radius $R_\text{in} = 1.24 r_g$ to $R_\text{out} = 400 r_g$ and relativistic light bending effects can lead to the appearance of a warped disk. \texttt{RELXILL} is the standard relativistic reflection model, modelling irradiation of accretion by a broken power law emissivity. Ionization of the accretion disk ranges from $\log \xi = 0$ (neutral) to $\log\xi = 4.7$ (heavily ionised) and the iron abundance $A_\text{Fe}$ of the material in the accretion disc is in units of solar abundance. 

\cite{2021MNRAS.508.1798P} have successfully described the variability of 1H 0707-495 using a series of publicly available phenomenological models to build on previous findings of the physical processes driving the observed variability from the accretion flow and from the line-of-sight absorption. Using the fractional variance and energy dependence of the variability in low and high frequency regimes, they found that the soft variance is driven by partial covering and the higher frequency changes are intrinsic to the source where the absorption and intrinsic variability are evident on longer timescales. 

This study utilises a sample selection informed by the work of \cite{2013MNRAS.431.2441D}, who first reported that the time lags scale with black hole mass and scale inversely with the frequency at which the lags occur. In addition, most of the sources have been previously studied by \cite{2016MNRAS.462..511K} who investigated the time lags in the Fe K band (between the 3 -- 4 and 5 -- 7 keV bands) and found that although the time lags were well correlated with the black hole mass, a stronger correlation was evident in the Fe K lag frequency. In this paper we use \textit{XMM-Newton} archival data, adding several more recent observations to our sample (e.g. for IRAS 13224-3809, NGC 4395 and NGC 5548) to explore the time lags as a function of frequency and investigate the disc reflection component using the \texttt{RELXILL} reflection model for 20 AGN known to exhibit reverberation signatures. We seek to confirm the basic properties such as their lag profiles and to look for correlations between the time lags, mass, accretion rate and other parameters such as the disc ionisation. We aim to contribute to the understanding of the observed variability. This work was carried out prior to an exploration of the extended corona scenario (paper in prep) by redeveloping the `two-blob' approach first presented by \cite{chainakun_investigating_2017}.

\section{Observations and data reduction}
The sample was selected from well studied AGN known to exhibit reverberation signatures with a preference for observations $> 40$~ks. These AGN and basic data are listed in the Appendix Table~\ref{lit_info} and the full sample is listed in Table~\ref{table:obs_log}. Each observation was downloaded from the \emph{XMM-Newton} archive and we select only the EPIC pn data due to the higher S/N in the 0.3--10 keV energy range and processing was conducted using \emph{XMM} Science Analysis System software version 18.0.0. Following the standard methods, the data were cleaned of background flaring activity which can lead to spurious results. This was conducted by creating the 10--12 keV light curve with \texttt{PATTERN==0} in 100\,s bins, and removing high particle flaring from the beginning and end of each observation which slightly reduced the effective duration. The light curves were then generated in 0.3--0.8 and 1--4 keV with \texttt{FLAG==0} and \texttt{PATTERN<=4}. These energy ranges were chosen because of the higher signal to noise ratio at lower energies and for comparison with the existing literature. We also note that moderate background flares do not significantly degrade the S/N ratio in the relatively soft bands we are considering here, so we have not removed intra-observation observation flares. Even if we were to completely disregard the few observations with some moderate intra-observation flaring, this would not change our conclusions. Each source was examined and source counts extracted using a radius of 35 arcsecs. The associated background was selected with the same radius far from the source, whilst remaining on the same CCD chip to create background subtracted soft and hard photon light curves. The photon Redistribution Matrix File (RMF) and Ancillary Response File (ARF) were calculated for all spectra using the \texttt{rmfgen} and \texttt{arfgen} tools and grouped to a minimum of 25 counts per spectral bin using the \texttt{GRPPHA} package.

\begin{table}
\centering
\caption[Pileup extraction]{Table of the observations requiring pileup removal, including the source name, observation ID and pileup core exclusion radius in arcsecs.}
\begin{tabular}{ccc}
\hline
{Source} & {Obs ID} & {Core exclusion} \\
\hline
1H 0707-495 & 0511580101 & 8"  \\   
& 0511580301 & 12" \\   
& 0653510401 & 13" \\      
& 0653510501 & 12" \\    
& 0653510601 & 12" \\ 
Ark 564 & 0006810101 & 5" \\
& 0206400101 & 5"  \\
& 0670130201 & 5"  \\ 
& 0670130501 & 6"  \\
& 0670130801 & 8.5" \\ 
& 0670130901 & 8.5" \\
PG 1211+143 & 0112610101 & 9"\\
& 0208020101 & 5"\\  
& 0502050101 & 10"\\  
& 0745110301 & 15"\\ 
& 0745110401 & 4"\\  
& 0745110501 & 5"\\
& 0745110601 & 10"\\
& 0745110701 & 5"\\
REJ 1034+396 & 0506440101 & 21"\\
\hline
\label{pileup_obs_log}
\end{tabular}
\end{table}

The soft and hard corrected light curves were then created using the task \texttt{epiclccorr} for the soft and hard energy bands. Background and source spectrum were then generated before running the \texttt{BACKSCALE} task. The SAS \texttt{epatplot} was employed to check each observation for photon pileup which can lead to spurious results. Several observations in the sample did show some mild-to-significant pileup and this was dealt with by removing the core at the centre of the source, increasing in size until the pileup fraction was negligible. Table~\ref{pileup_obs_log} lists all data requiring action for pileup issues and includes the radius of core removal in arc seconds. For the more interested reader, full details of this `reproducible' analysis is also available via online logbooks \citep{X-rayReverberationinAGN2019}.

\section{THE UNFOLDED SPECTRA OF AGN}

We begin by providing an overview of the variability of these AGN by inspecting the background subtracted unfolded energy spectrum, that is, photon counts cm$^{-2}$ s$^{-1}$ keV$^{-1}$ as a function of energy for the full 0.3 -- 10 keV energy range. The observations for all AGN were then grouped into similar spectral states as seen in Figure~\ref{fig:unfold-spec}. For example, the unfolded spectra of 1H 0707-495 revealed three distinct groupings of low, intermediate and high energy count rates. The high flux spectral group was taken from the tightly packed group at the upper region consisting of 11 observations where the soft energy photon count rate was $\gtrsim 2 \times 10^{-3}$ s$^{-1}$. Working down the photon count rate, the medium flux state was taken from the next two observations (seen in blue and black) below the high flux group and the low flux group was taken from the two lower observations (green and indigo). Final lags were calculated from the fully combined spectrum. This is a crude method of grouping similar flux states, however the main idea is to capture various similar spectral profile groups which will provide us multiple  `snapshots' of the observed variability for each source. 

\begin{figure*}
    \centering
    \includegraphics*[scale=0.55]{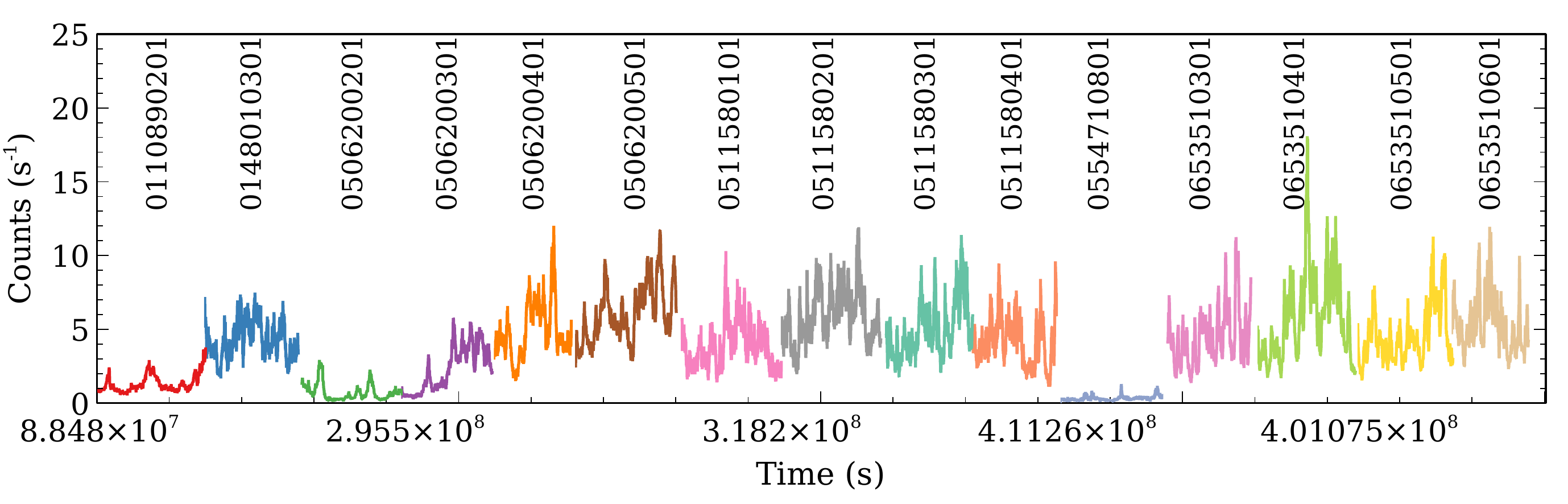}
    \includegraphics*[scale=0.55]{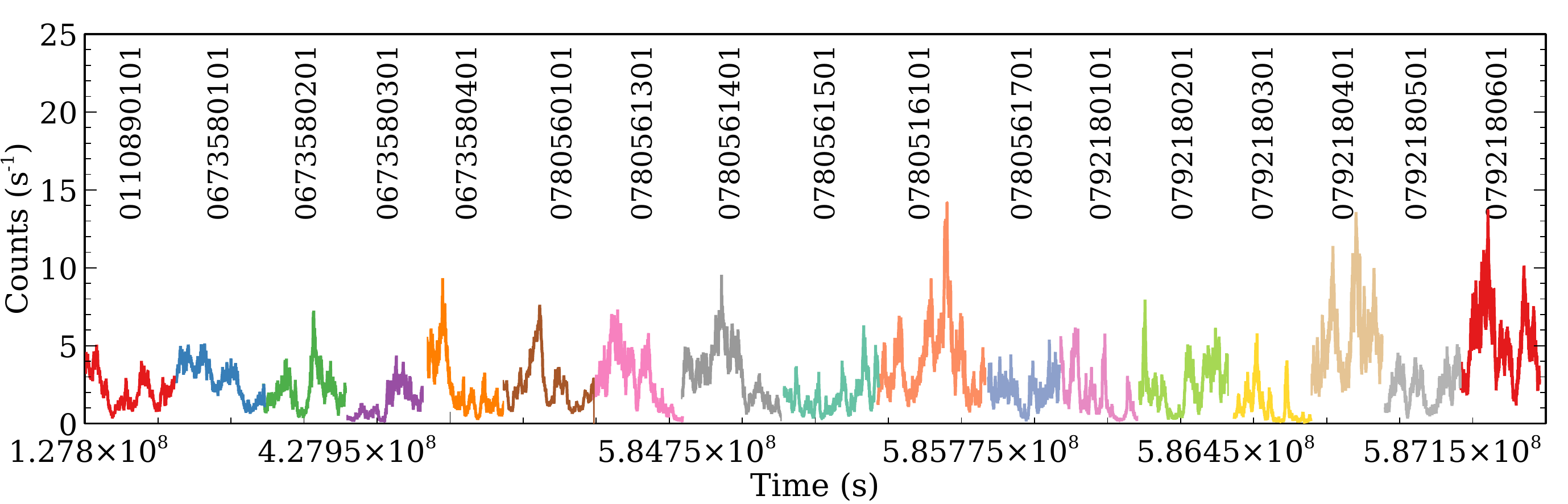}
    \caption{The background subtracted 0.3 -- 10 keV light curves for 1H~0707-495 \emph{(top panel)} and IRAS 13224-3809 \emph{(bottom panel)}. The observation ID numbers are shown above each data set. The x-axis is not to scale and some labels have been removed for clarity.}
    \label{fig:1h_lightcuvres2}
\end{figure*}

In order to capture lag profiles of the high and low photon count rate epochs, further examination was conducted by considering the background subtracted light curves in the full 0.3 -- 10 keV energy range as shown in Figure~\ref{fig:1h_lightcuvres2} for 1H 0707-495 and IRAS 13224-3809, demonstrating their strong variability. Subsequent inspection revealed that two of the observations in the lowest spectral group namely 0506200201 (green) and 0554710801 (indigo) contained some of the highest count rates not immediately obvious from the unfolded spectra, hence two further groups were created to encompass the observations containing a count rate above and below 5 cts s$^{-1}$. The same method was also applied to IRAS 13224-3809 as shown in Figure~\ref{fig:1h_lightcuvres2}, however for the remainder of AGN in the sample only the combined, low flux and high flux epochs were taken, with an additional medium flux state where deemed appropriate from the inspection of the unfolded flux (see e.g. Mrk 766 and NGC 3516). All AGN groups and the additional groups discussed above are listed against their specific observations in Table~\ref{table:obs_log}.

\begin{figure*}
    \begin{subfigure}{}
    \includegraphics[trim={3.5cm 1.5cm 6.5cm 3cm},clip,scale=0.31]{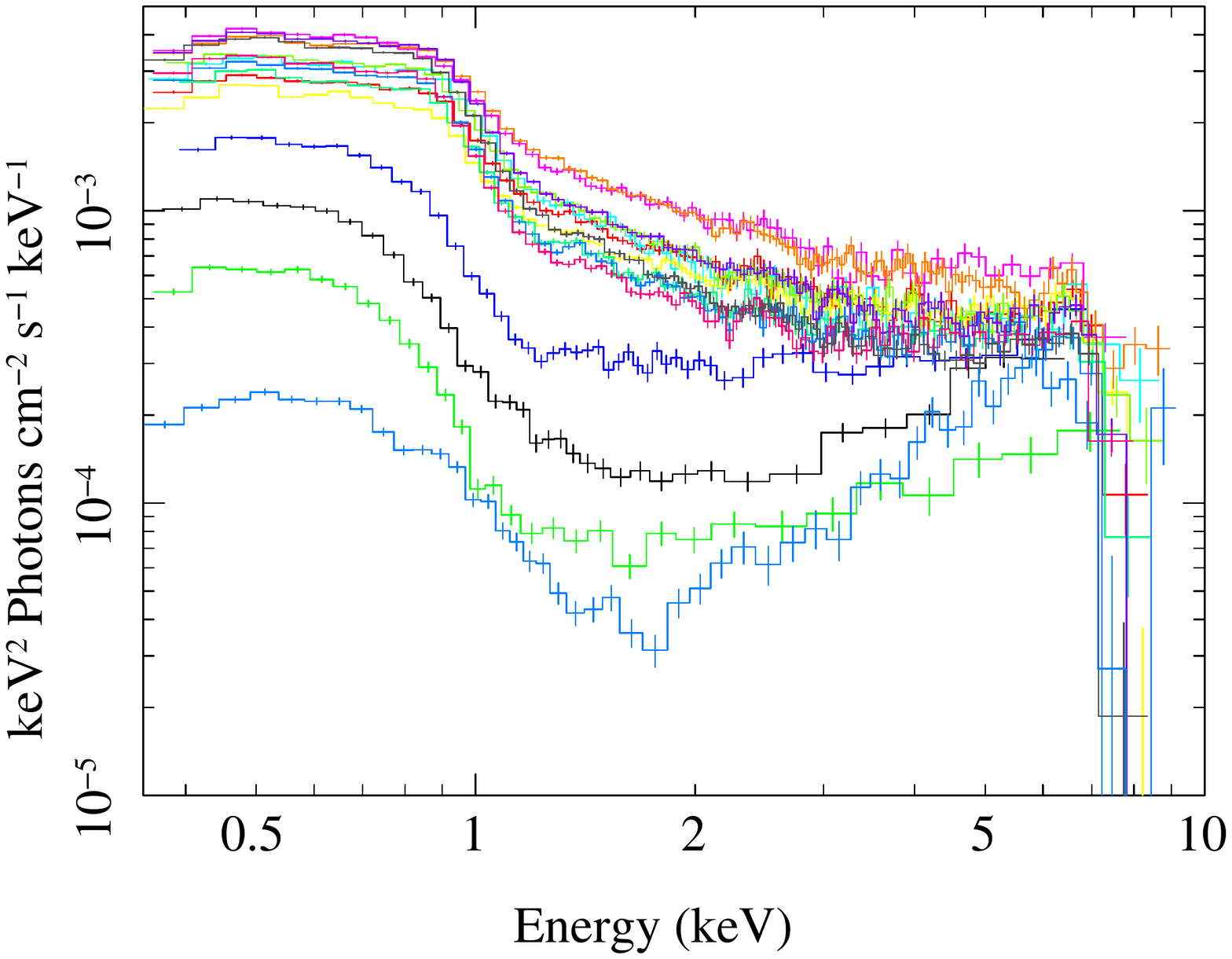}
    \put(-145,30){1H 0707-495}
    \includegraphics[trim={3.5cm 1.5cm 6.5cm 3cm},clip,scale=0.31]{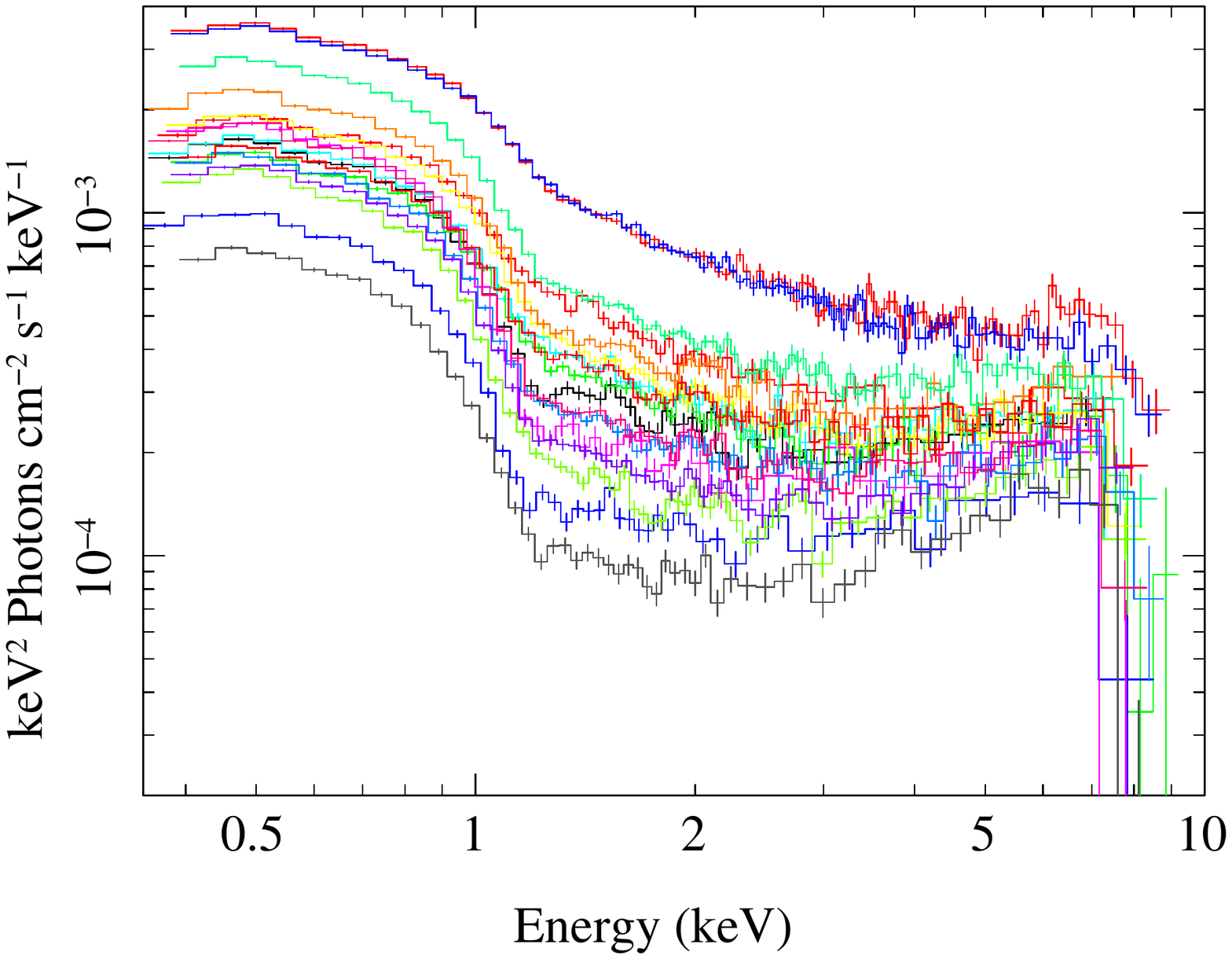}
    \put(-145,30){IRAS 13224-3809}
    \includegraphics[trim={3.5cm 1.5cm 6.5cm 3cm},clip,scale=0.31]{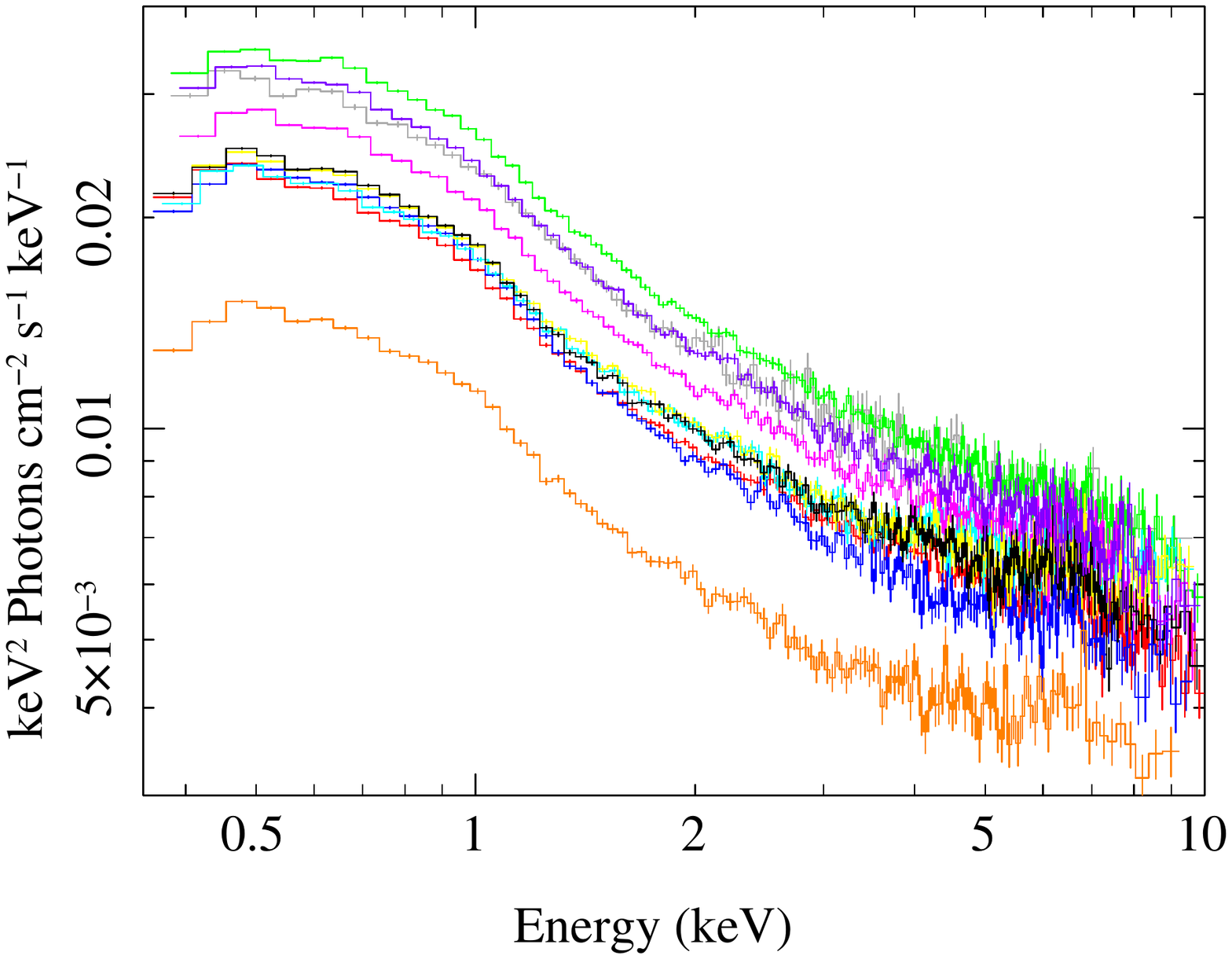}
    \put(-145,30){Ark 564}\\
    \includegraphics[trim={3.5cm 1.5cm 6.5cm 3cm},clip,scale=0.31]{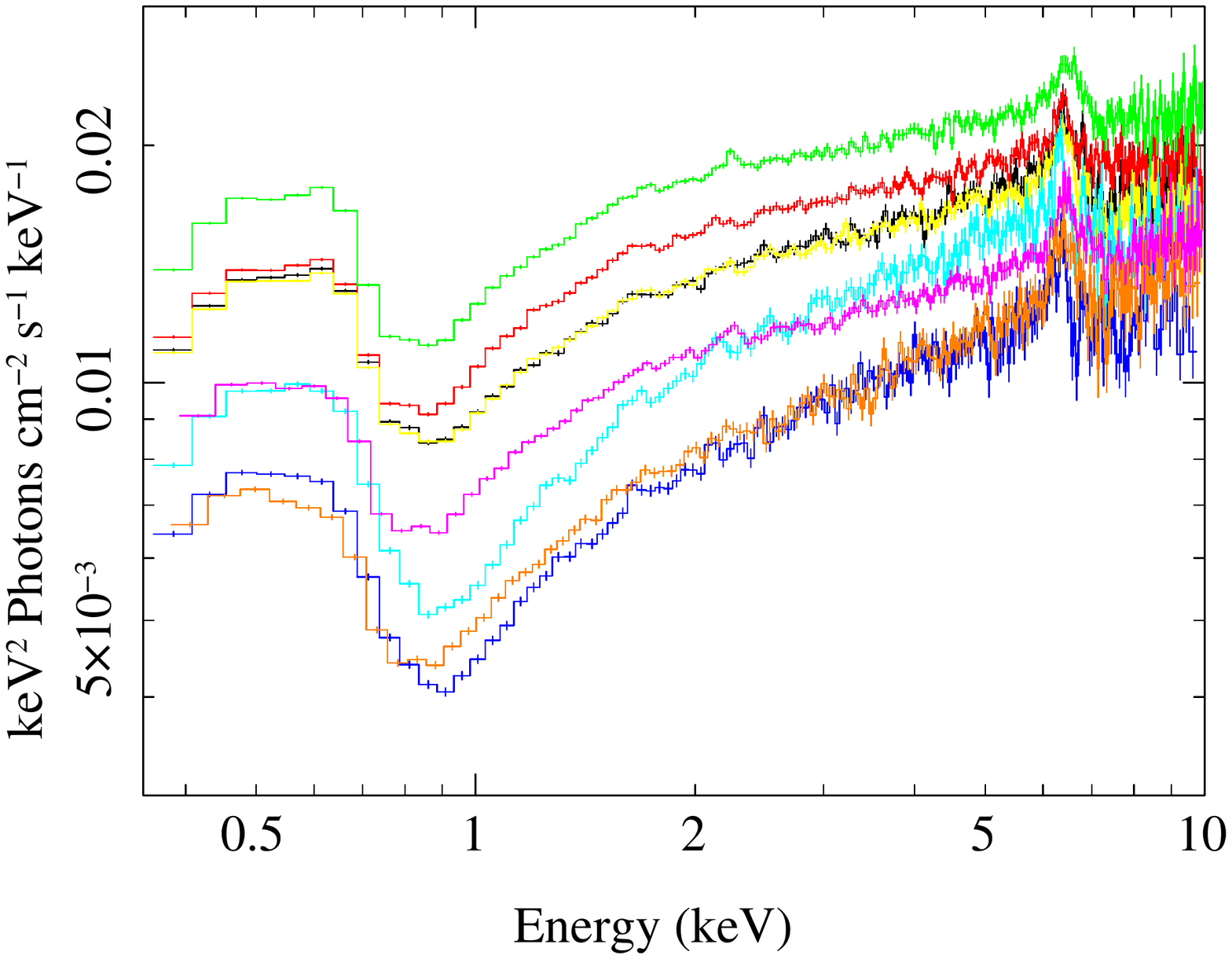}
    \put(-145,30){MCG-6-30-15}
    \includegraphics[trim={3.5cm 1.5cm 6.5cm 3cm},clip,scale=0.31]{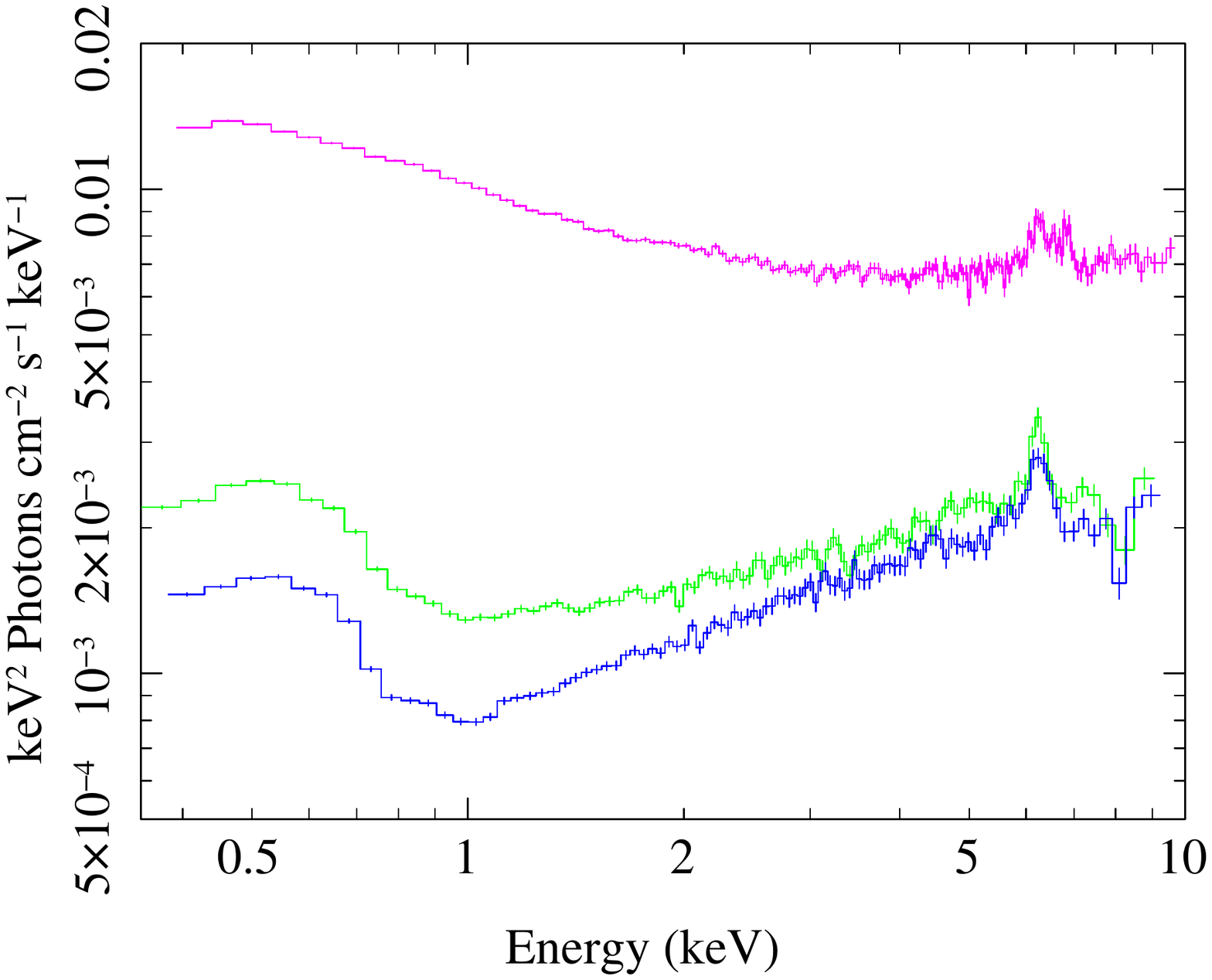}
    \put(-145,30){Mrk 335}
    \includegraphics[trim={3.5cm 1.5cm 6.5cm 3cm},clip,scale=0.31]{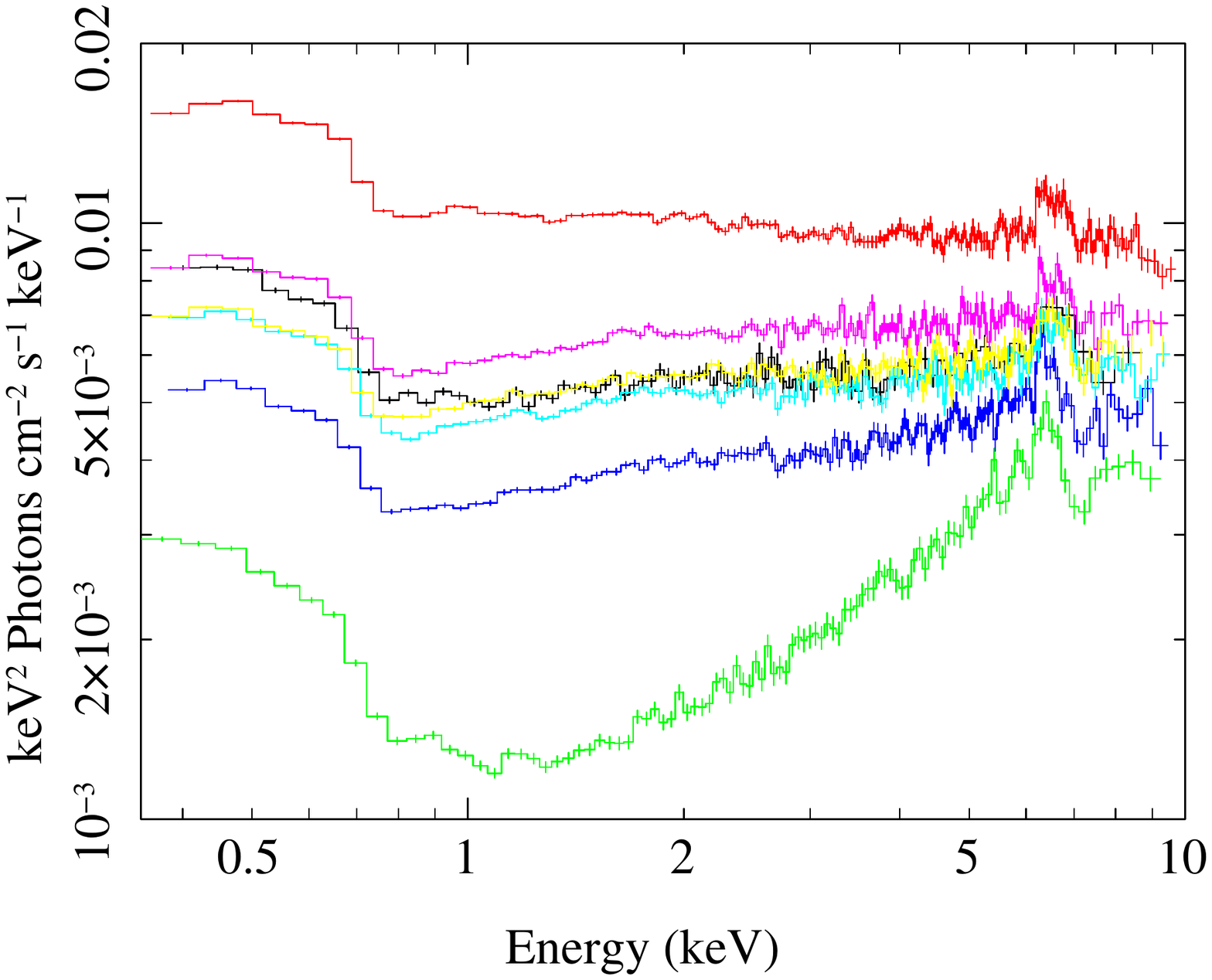}
    \put(-145,30){Mrk 766}\\
    \includegraphics[trim={3.5cm 1.5cm 6.5cm 3cm},clip,scale=0.31]{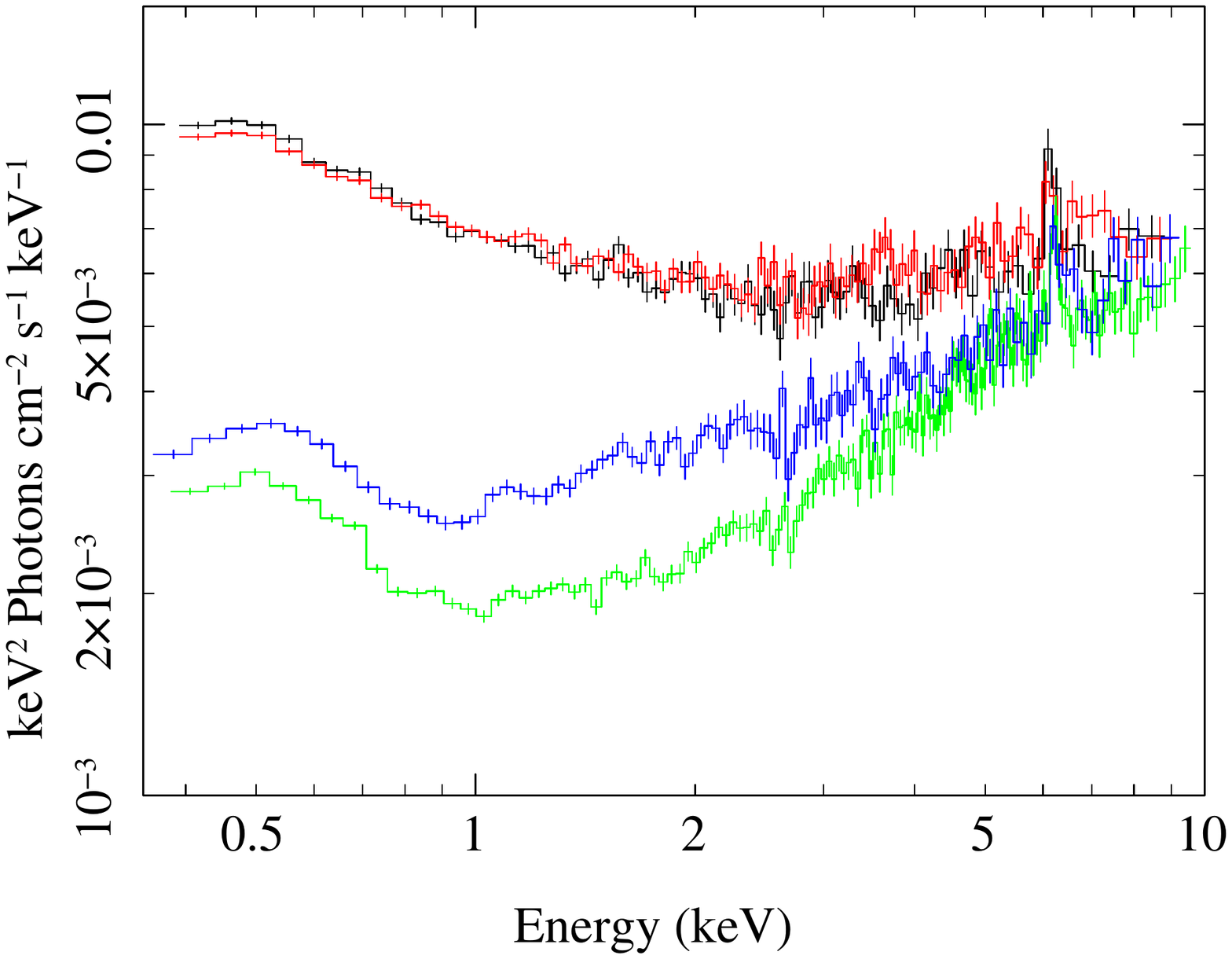}
    \put(-145,30){Mrk 841}
    \includegraphics[trim={3.5cm 1.5cm 6.5cm 3cm},clip,scale=0.31]{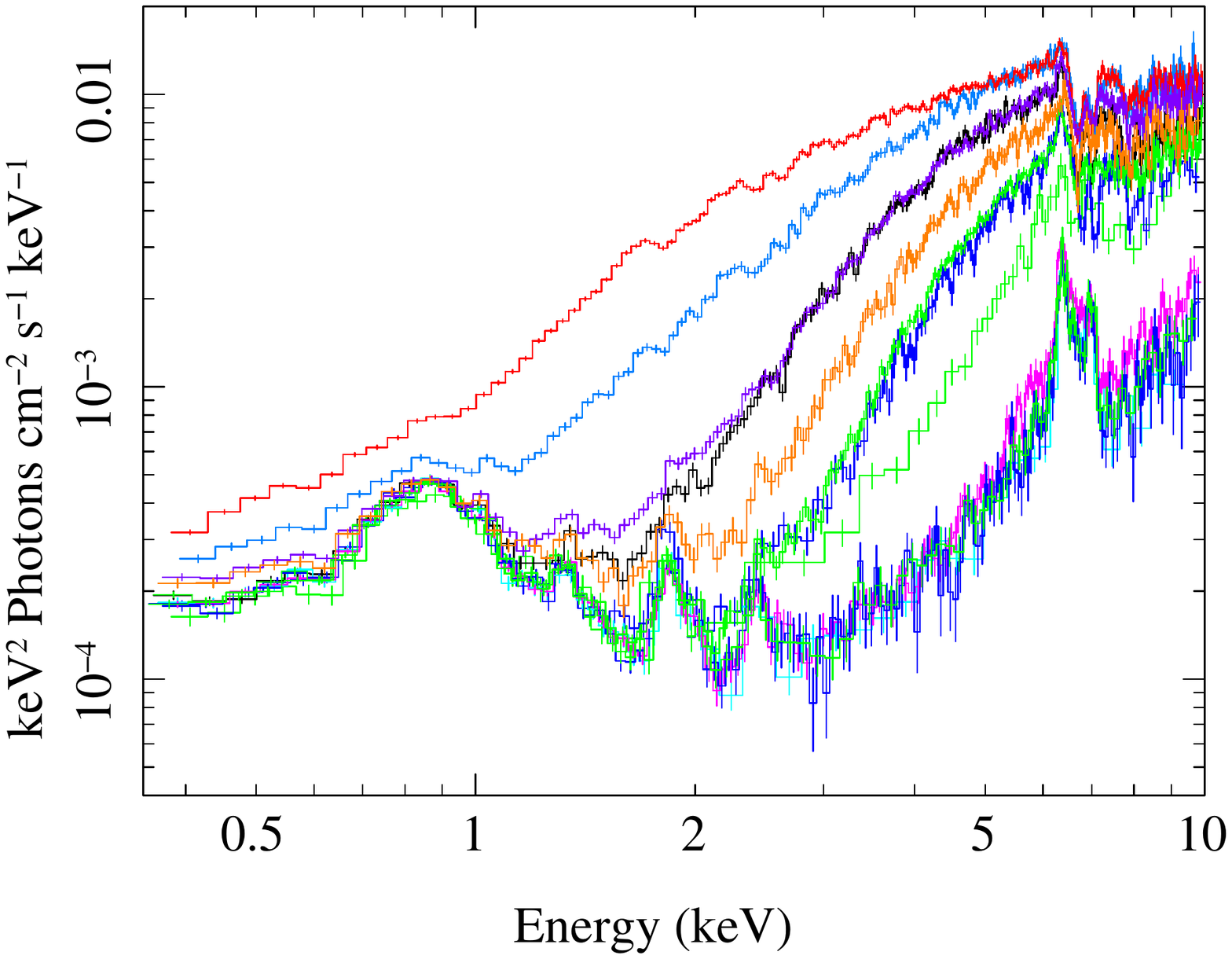}
    \put(-145,30){NGC 1365}
    \includegraphics[trim={3.5cm 1.5cm 6.5cm 3cm},clip,scale=0.31]{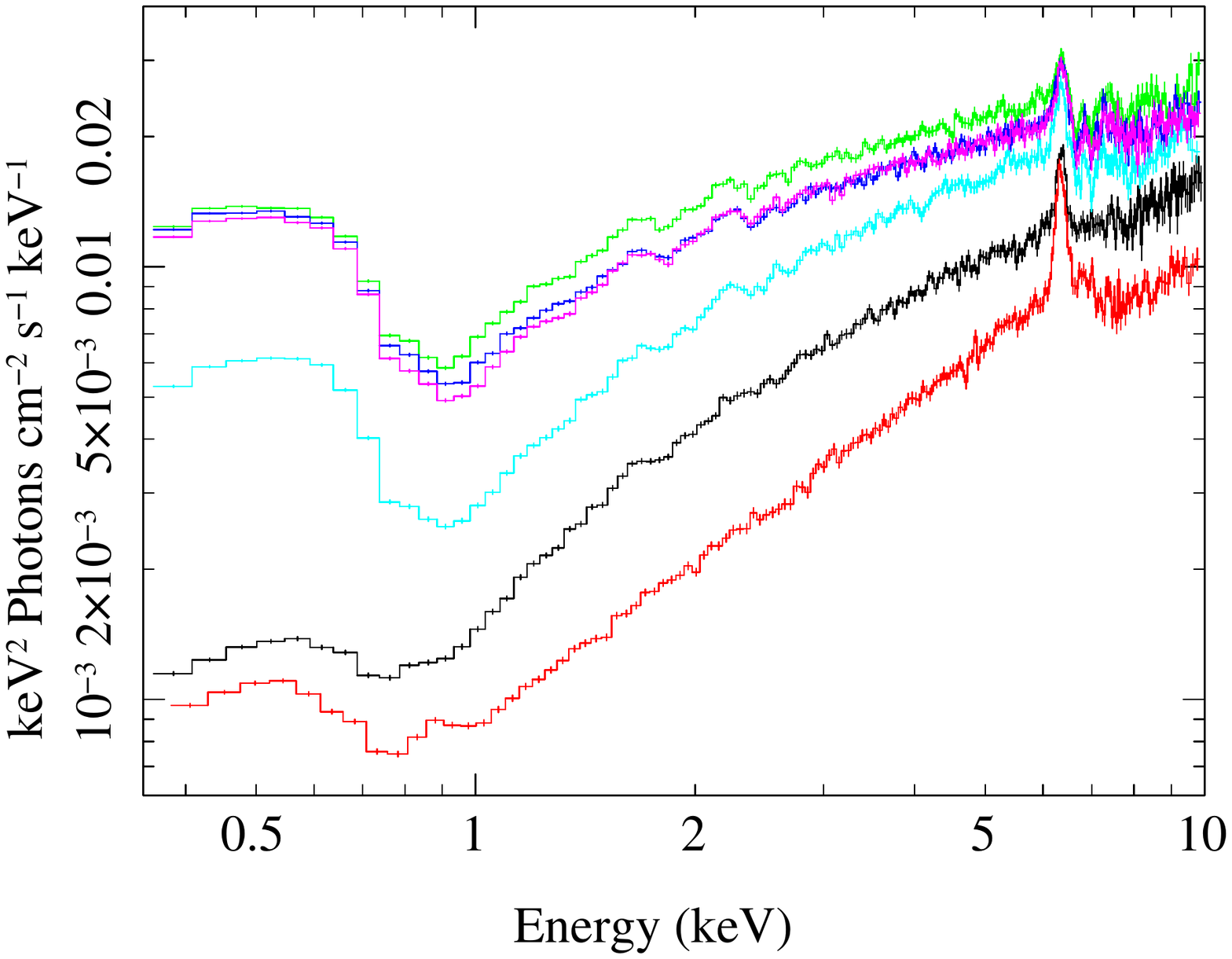}
    \put(-55,30){NGC 3516}\\
    \includegraphics[trim={3.5cm 1.5cm 6.5cm 3cm},clip,scale=0.31]{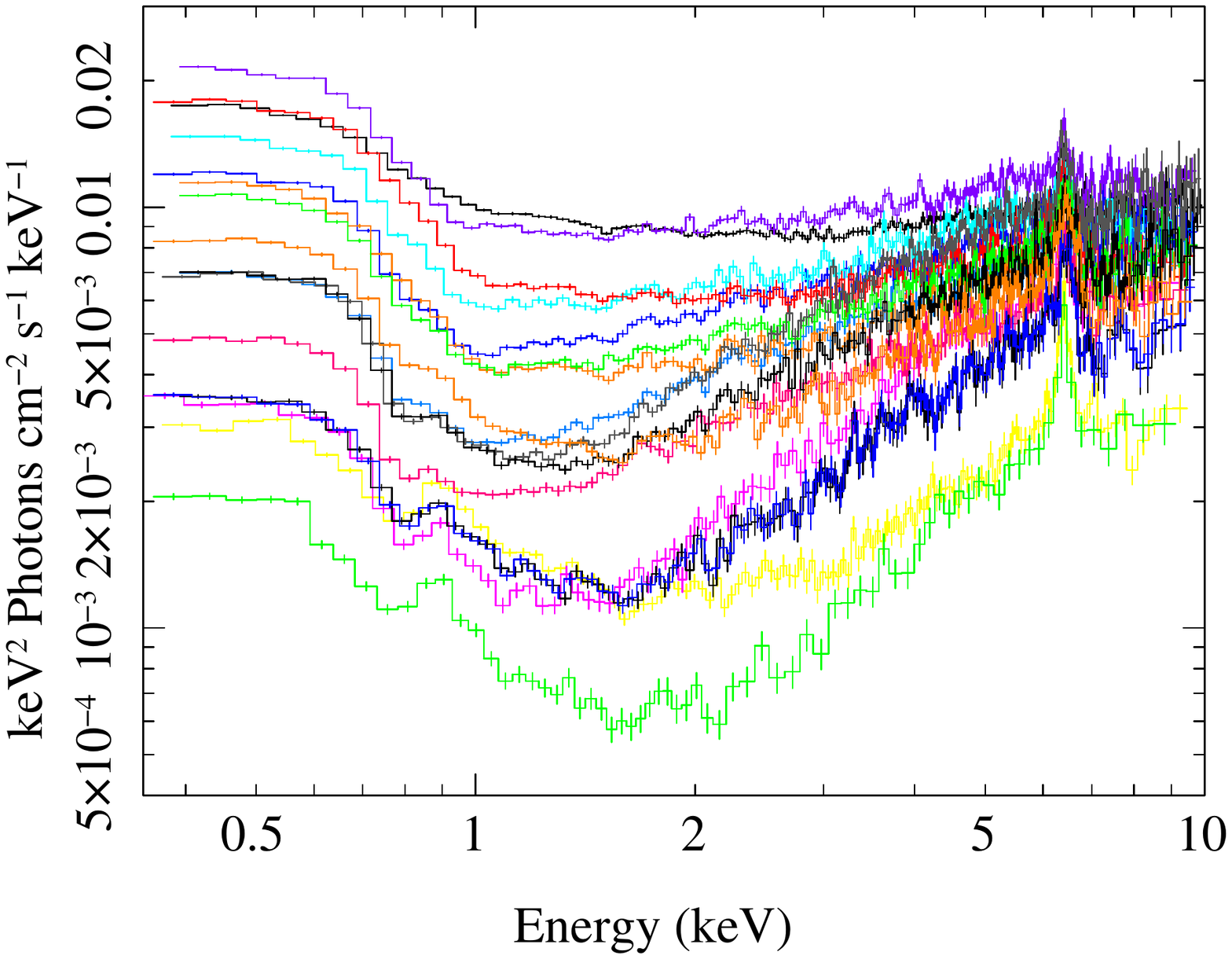}
    \put(-145,30){NGC 4051}
    \includegraphics[trim={3.5cm 1.5cm 6.5cm 3cm},clip,scale=0.31]{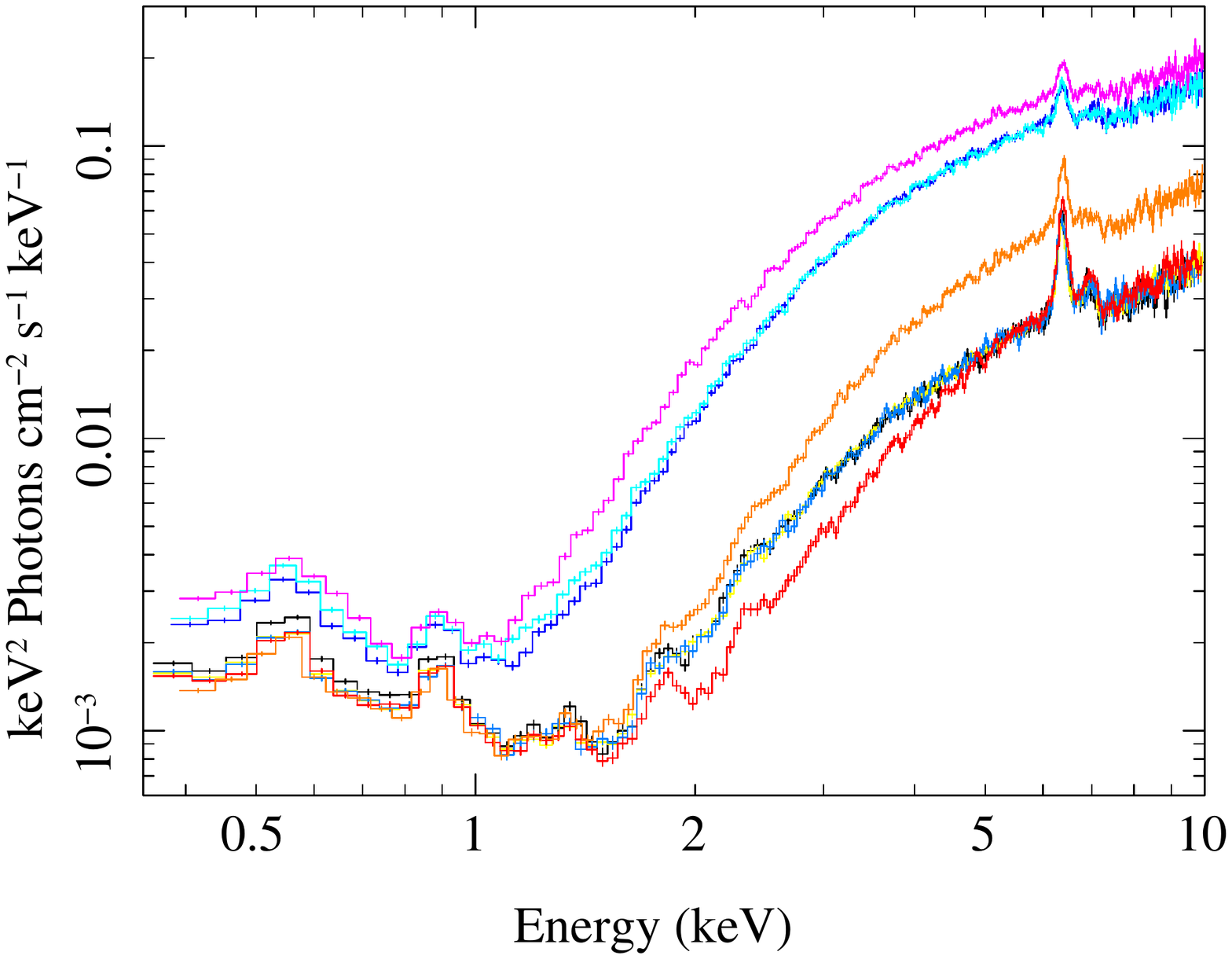}
    \put(-55,30){NGC 4151}
    \includegraphics[trim={3.5cm 1.5cm 6.5cm 3cm},clip,scale=0.31]{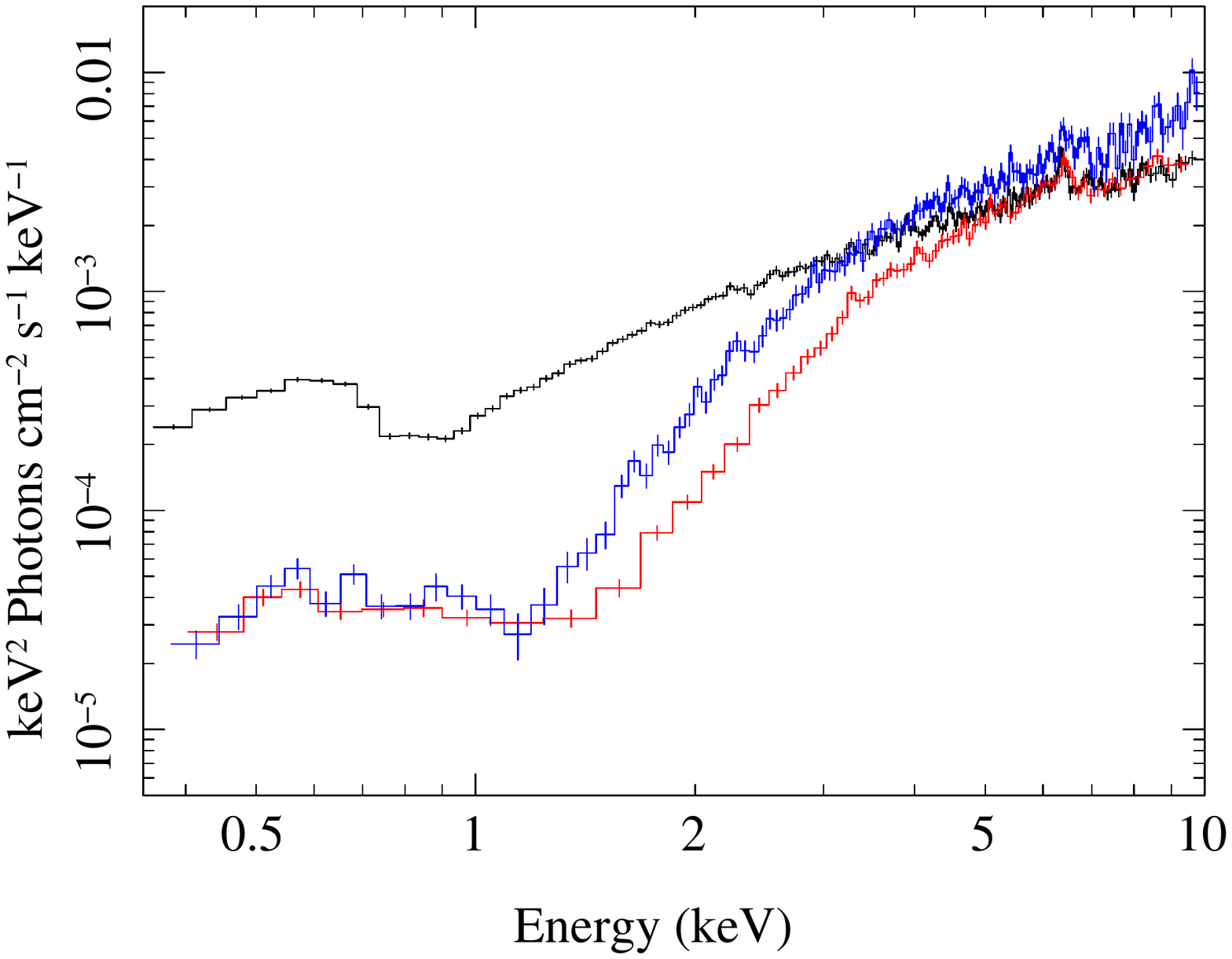}
    \put(-145,30){NGC 4395}   
    \end{subfigure}   
    \caption[The unfolded spectra of AGN]{The background-subtracted unfolded energy spectra of all AGN in the sample, as generated using the \texttt{ISIS} \texttt{plot\_unfold} package. The data have been binned to enhance clarity.}
\end{figure*}

\begin{figure*}\ContinuedFloat
    \centering
    \begin{subfigure}{}
    \includegraphics[trim={3.5cm 1.5cm 6.5cm 3cm},clip,scale=0.31]{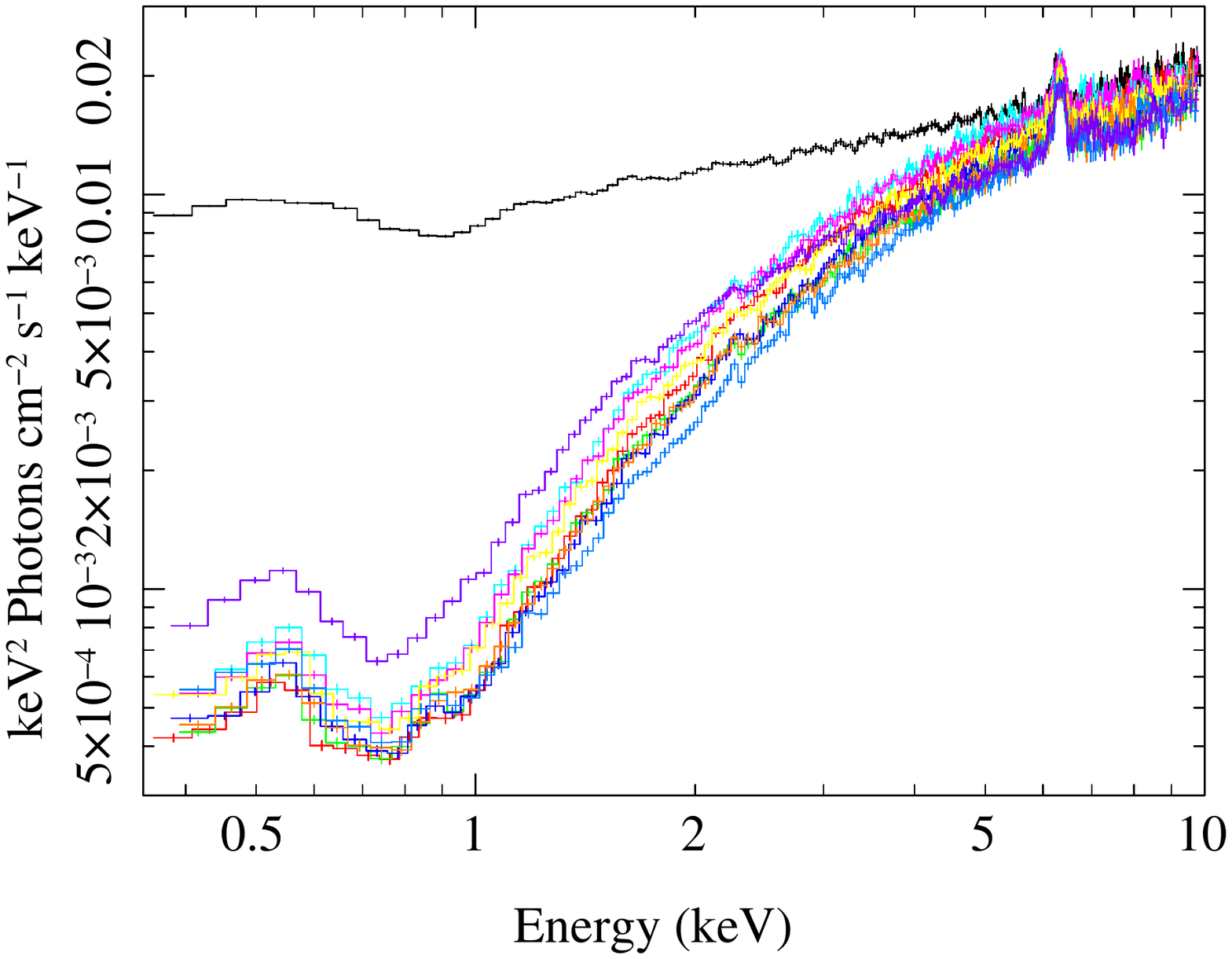}
    \put(-55,30){NGC 5548}
    \includegraphics[trim={3.5cm 1.5cm 6.5cm 3cm},clip,scale=0.31]{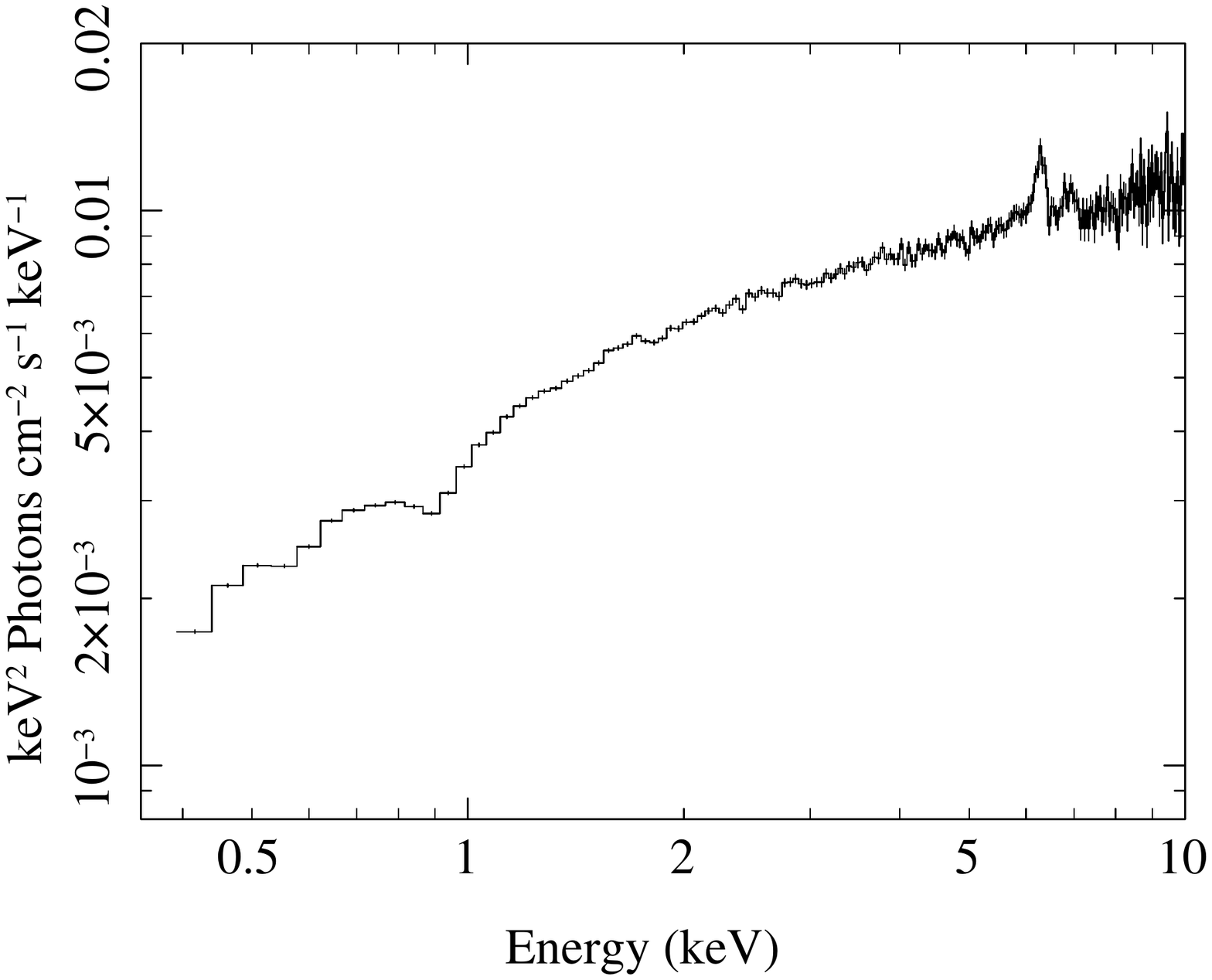}
    \put(-55,30){NGC 6860}
    \includegraphics[trim={3.5cm 1.5cm 6.5cm 3cm},clip,scale=0.31]{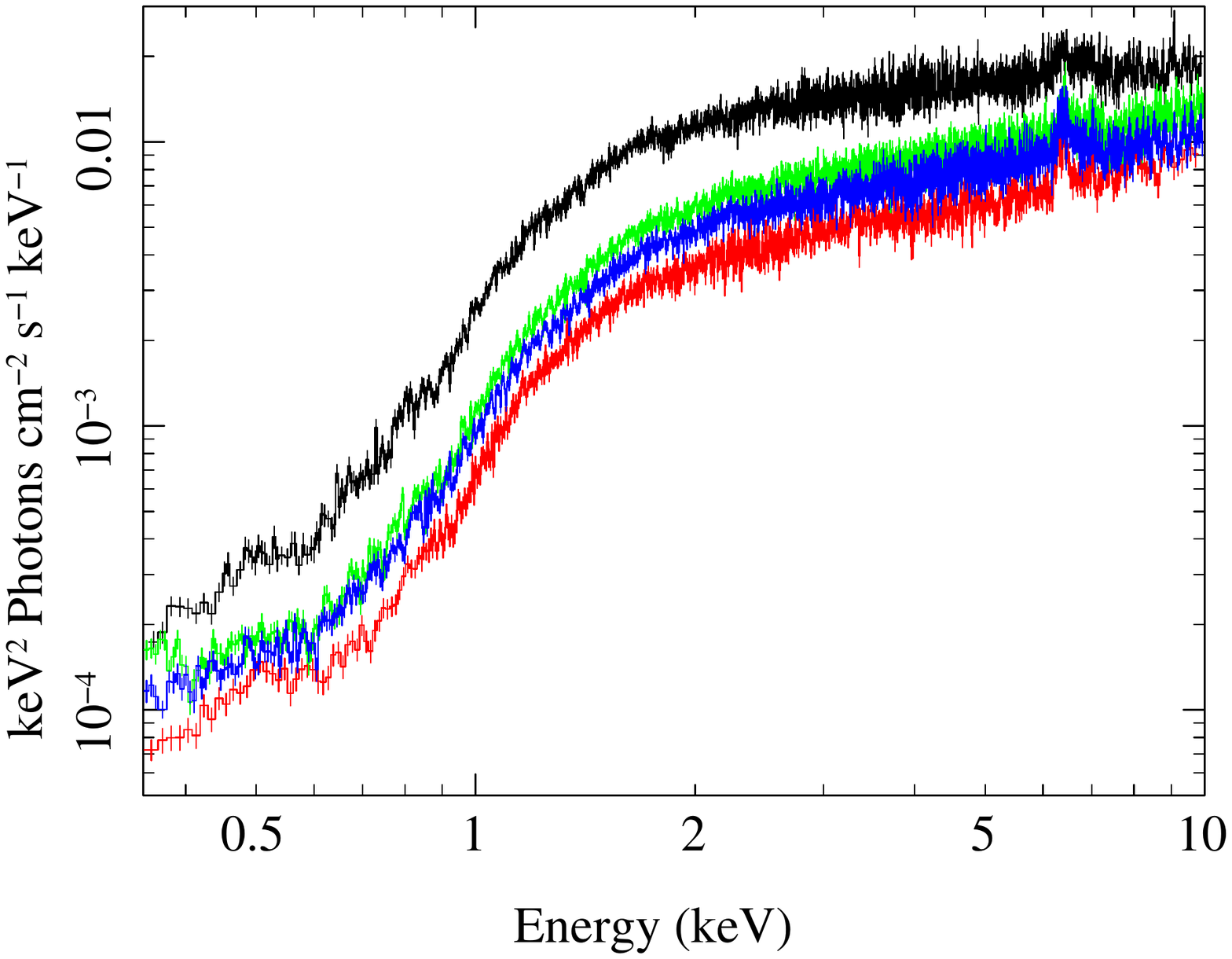}
    \put(-55,30){NGC 7314}\\
    \includegraphics[trim={3.5cm 1.5cm 6.5cm 3cm},clip,scale=0.31]{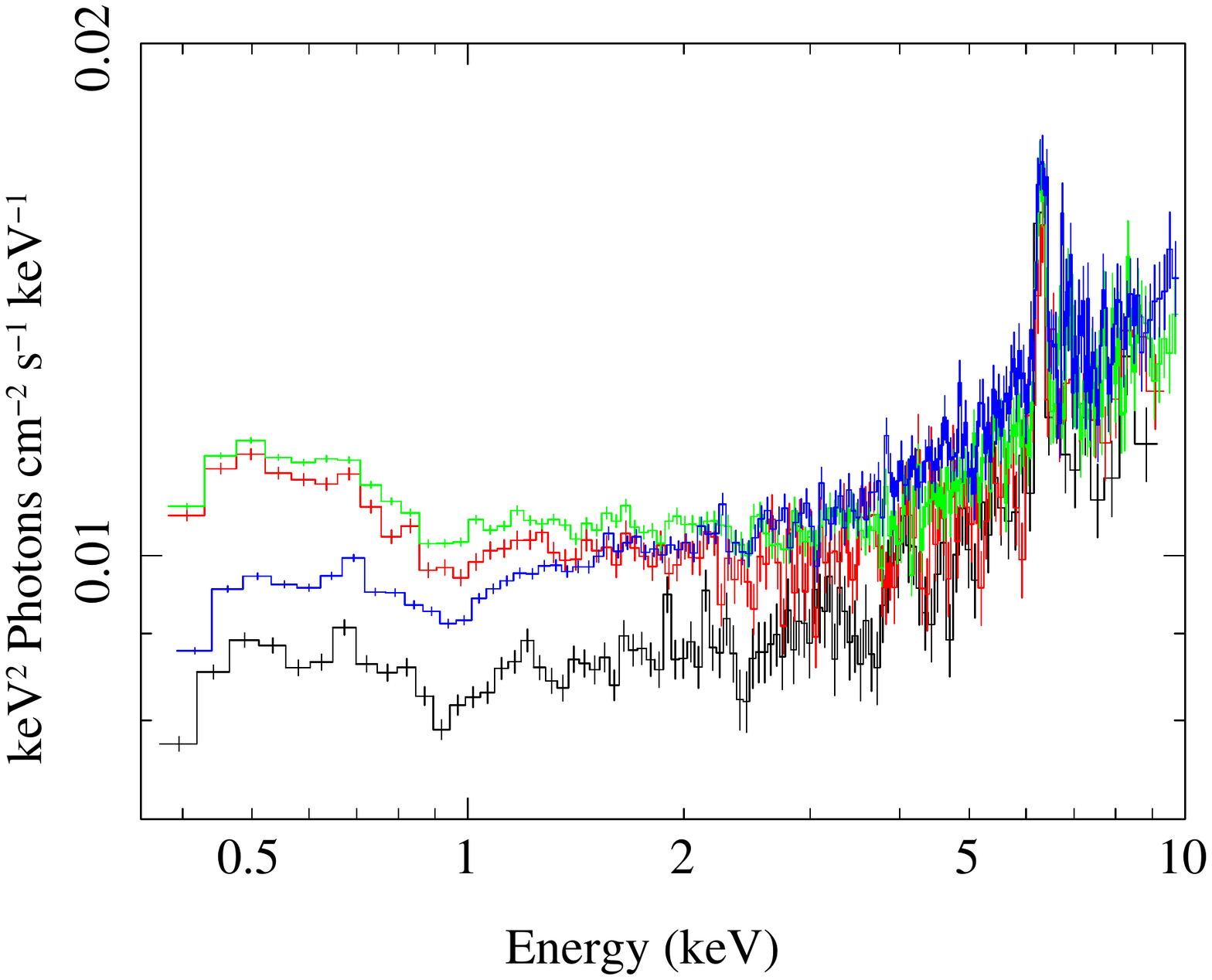}
    \put(-145,115){NGC 7469}
    \includegraphics[trim={3.5cm 1.5cm 6.5cm 3cm},clip,scale=0.31]{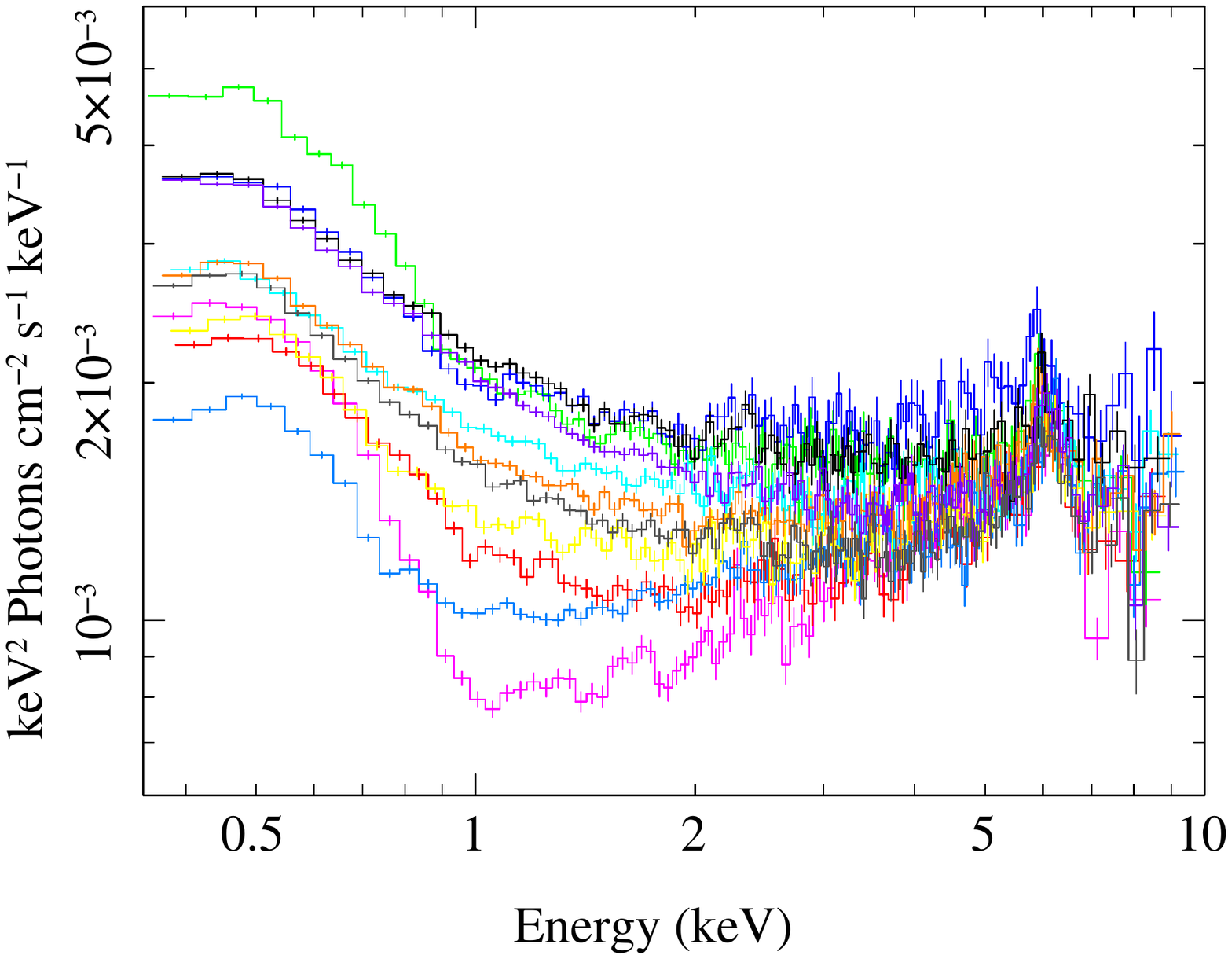}
    \put(-65,115){PG 1211+143}
    \includegraphics[trim={3.5cm 1.5cm 6.5cm 3cm},clip,scale=0.31]{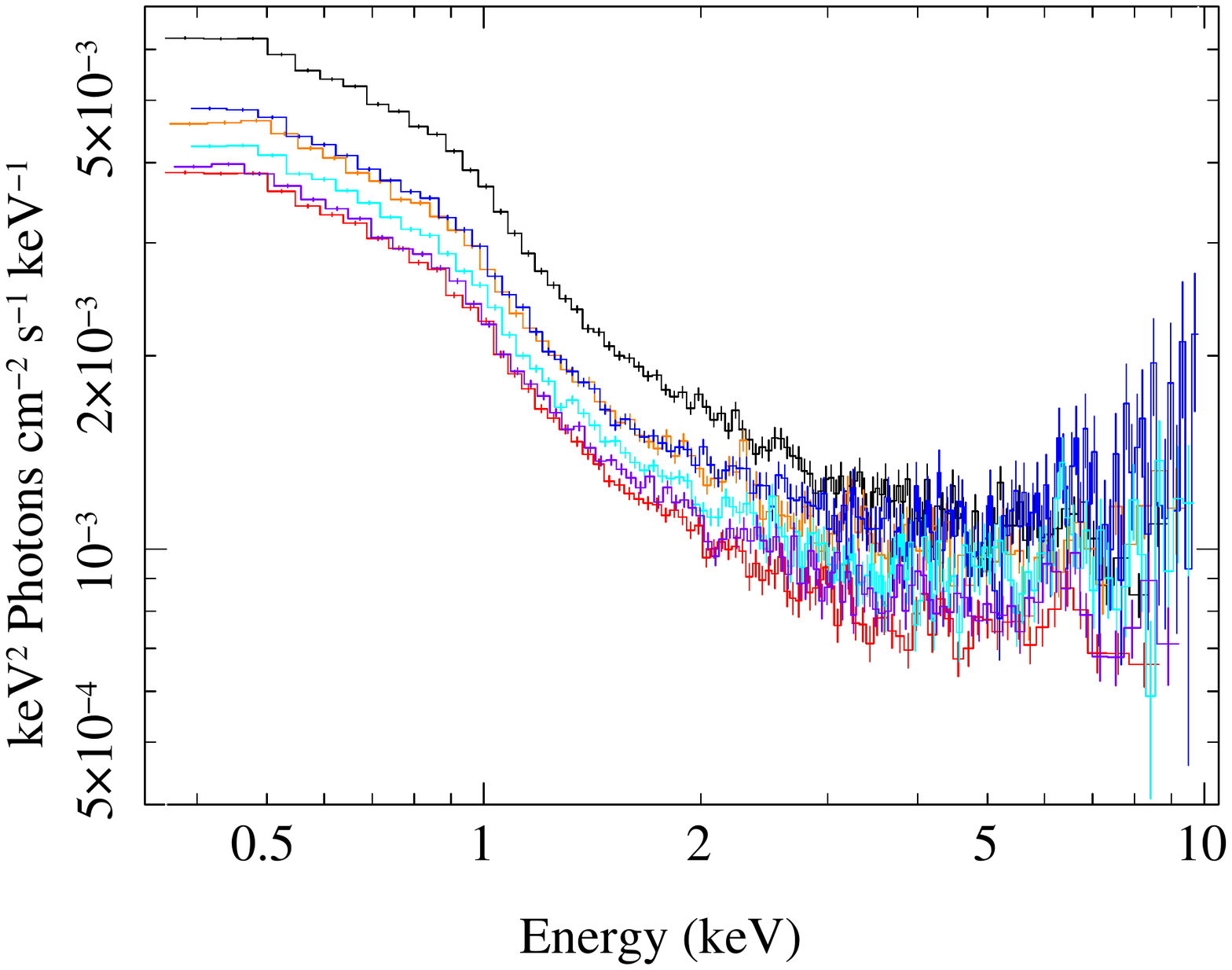}
    \put(-145,30){PG 1244+026}\\
    \includegraphics[trim={3.5cm 1.5cm 6.5cm 3cm},clip,scale=0.31]{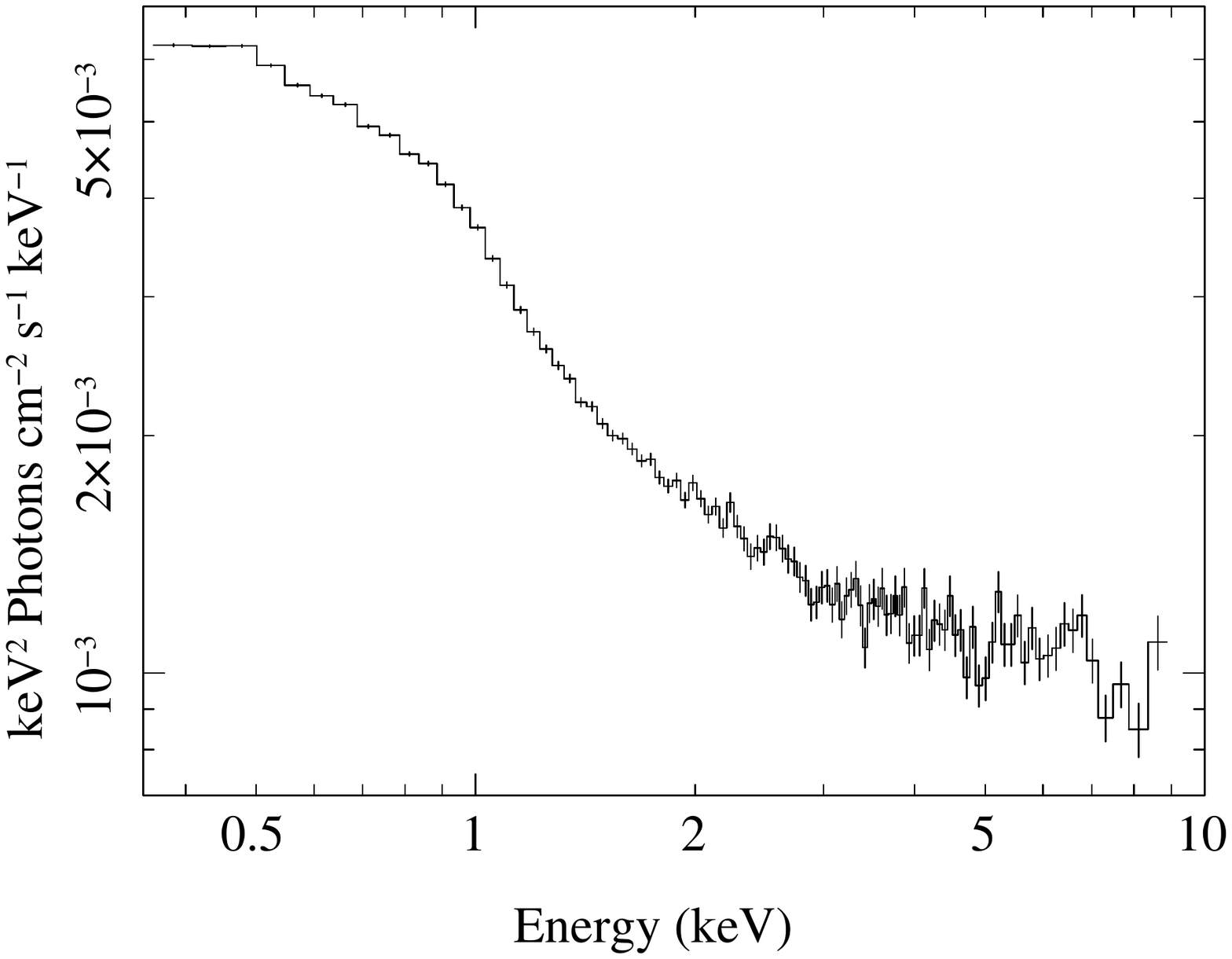}
    \put(-145,30){PG 1247+267}
    \includegraphics[trim={3.5cm 1.5cm 6.5cm 3cm},clip,scale=0.31]{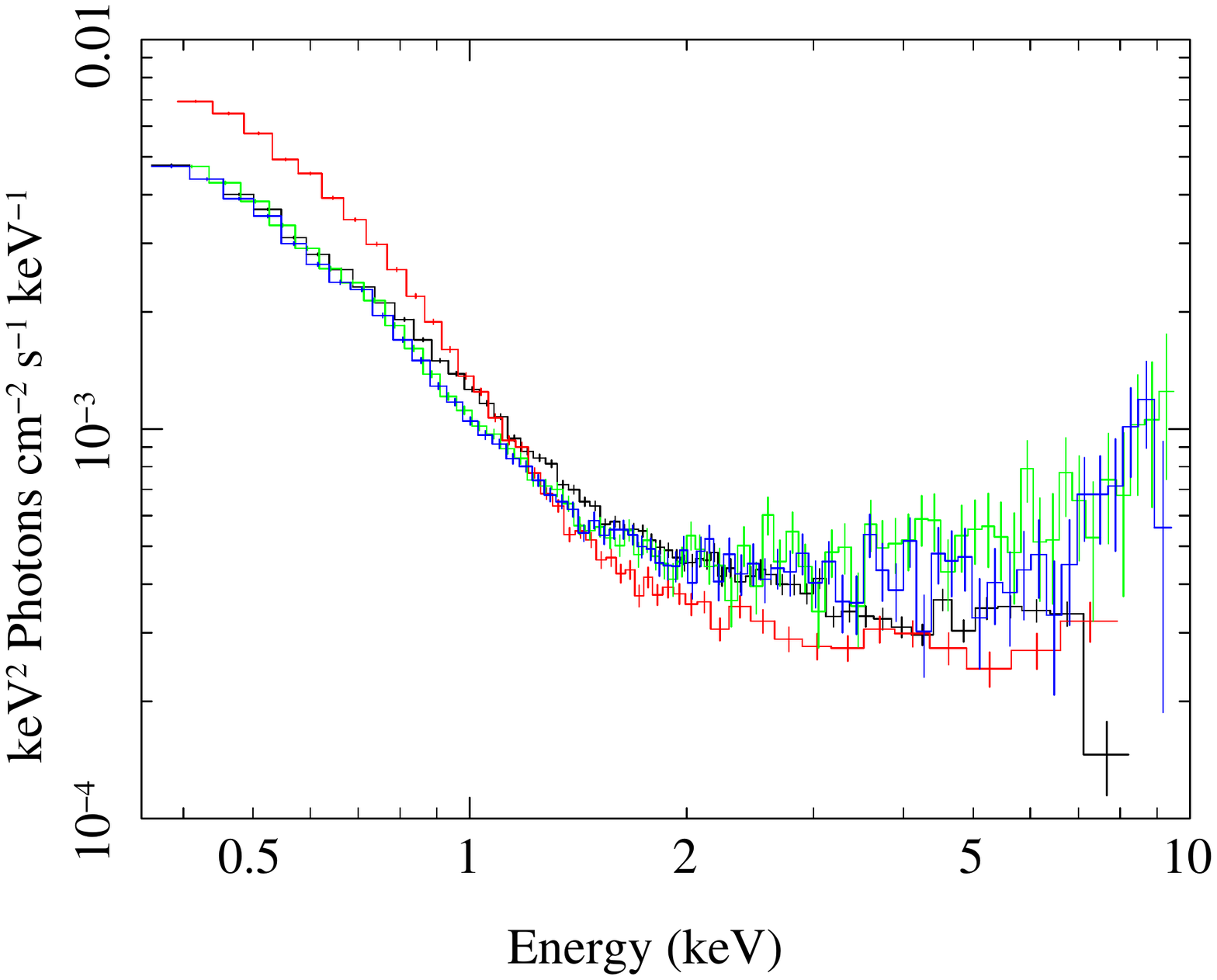}
    \put(-145,30){REJ 1034+396}
    \end{subfigure}
    \caption[The unfolded spectra of AGN]{The unfolded spectra of AGN (continued).}
\label{fig:unfold-spec}
\end{figure*}


\section{Estimating the time lags}
\label{Estimating_the_time_lags}
The time lags between the soft light curve $s(t)$ and the hard light curve $h(t)$ were computed using discrete Fourier transform (DFT) methods as outlined by \cite{Nowak1999} and \cite{Uttley2014}.
The DFT of a light curve $s(t)$ with $N$ contiguous time bins is given by
\begin{equation}
DFT_s(j)=\sum_{k=0}^{N-1}s(t_k) \exp\left(\frac{2\pi ijk}{N}\right)
\end{equation}
where $DFT_s (j)$ is the discrete Fourier transform at each Fourier frequency, $f_j=j/NT_s^{-1}$, for $n = 1, 2, 3, \ldots, N/2$, $T_s$ is the segment duration, and $t_k$ is the time corresponding to the $k^\text{th}$ bin of the light curve. The minimum frequency is the inverse of the time duration and the maximum frequency is the Nyquist frequency $f_\text{max}=1/(2\Delta t)$ where $\Delta t$ is the 10 s time bin size. The Fourier cross spectrum between two lightcurves $x(t)$ and $y(t)$ with DFTs $X_n$ and $Y_n$ was calculated using the real and imaginary components by $C_{XY,n}= X_n^*Y_n$. The use of the complex conjugate means that the phase of the cross spectrum gives the phase lag $\phi_n$ between the light curves and can be written $C_{XY,n}= A_{X,n} A_{Y,n} \exp (i\phi_n)$, where $A_{X,n}$ and $A_{Y,n}$ are the absolute amplitudes of the Fourier transform. Considering $\textit M$ segments and $\textit K$ frequencies per segment, the coherence may be used to calculate the errors on the phase lag $\Delta\phi$ by 
\begin{equation}
\Delta\phi(f_j) =  \sqrt{\frac{1-\gamma^2(f_j)}{2\gamma^2(f_j)KM}}.
\end{equation}
The time lag error $\Delta \tau$ can be calculated by $\Delta \tau=\Delta\phi/(2\pi f_j)$, thus the time lag $\tau$ was finally calculated by $\tau(f_j)=\phi(f_j)/(2\pi f_j)$.

We adhere to the most accurate method of lag estimation \citep{2016A&A...594A..71E, 2016A&A...591A.113E}, therefore all time lags were taken from pairs of light curves with the same duration using $M$ segments with duration $\gtrsim$ 20 ks, having a Gaussian distribution with known errors. They were calculated at frequencies where the coherence is equal to $1.2/(1 + 0.2M)$ which reduces the effects of the Poisson noise on the true value of the time lags. The results of the lag-frequency estimates are presented in Appendix Table~\ref{table:lag_results}, listing the maximum amplitudes of soft reverberation lags found and the frequencies where they occur. For swift reproduction, appropriate \texttt{FITS} files and code can be downloaded from the web pages\footnote{\url{http://www.star.bris.ac.uk/steff/downloads.html}}. In addition, all lags produced for this study are presented online in the timing analysis results\footnote{\url{http://www.star.bris.ac.uk/steff/timing_results.html}} and are also presented in the Supplementary material (online).

\subsection{Checking consistency of the time lags}

We checked that our lag estimates are consistent with those in the literature using robust lag production procedures. Specifically, our combined lag-frequency estimates for 1H 0707-495, MCG--6-30-15, Ark 564, Mrk 335, Mrk 766, Mrk 841, NGC 3516, NGC 4051, NGC 4395, NGC7469, PG1211+143 and REJ 1034+396 are consistent with the soft reverberation frequencies and amplitudes found by \citet[][DM13 hereafter]{2013MNRAS.431.2441D}. The lags found for NGC 5548 are much lower amplitude (but in the same frequency range) than those found by DM13 due to combining observations from 2013 and 2014 not available at the time of the DM13 study; the single observation (Obs ID: 0089960301) conducted in 2001 which is the sole higher spectral flux observation used here, hence the larger amplitude high-flux lags of $\sim 200$\,s at identical frequencies reported by DM13. The spectrum of this source changed dramatically between 2001 and 2014 and it should be noted that the low flux spectra group contains all observations except the 2001 observation and the lag frequency found was more comparable to the amplitude reported by DM13 although at a lower frequency. This apparent discrepancy is a consequence of averaging highly variable spectra and supports the grouping methods used throughout this paper. 
The single observation for NGC 6860 first produced a soft negative lag of $13.1 \pm 35.0$s at $9.11 \times 10^{-4}$ Hz and was generally inconsistent with the results reported by DM13, however a revised look at this data using their soft and hard energy bands (0.3-1 and 1-5 keV) reproduced a consistent result where the soft negative lag was found with amplitude $186.7 \pm 192.5$s at $1.94 \times 10^{-4}$ Hz. IRAS 13224-3809 was checked multiple times against some earlier literature and the lags found in the combined observations from 2002 -- 2011 were consistent with \cite{2013MNRAS.429.2917F} and all lags to include the observations taken in 2016 were strongly comparable to \cite{2020NatAs...4..597A}. For Ark 564 and Mrk 335, the lag-frequency was also found to be consistent with \cite{2013MNRAS.434.1129K} and the data for PG 1244+026 was checked against \cite{2014MNRAS.439L..26K} to find a comparable soft lag to that presented in the literature. 
The lags found in NGC 1365, NGC 4151 and NGC 7314 are comparable to those found by \cite{2016MNRAS.462..511K}. Finally, the lag-frequency for PG 1247+267 has no clear comparison in the literature and it should be noted that whilst the full sample contains AGN at low redshift $(z \ll1)$, PG 1247+267 is an ultra luminous source at $z = 2.043$ \citep{Bechtold_2002}.

\section{Spectral analysis}

Spectral analysis was conducted using the Interactive Spectral Interpretation System (ISIS) version 1.6.2-27 \citep{2000ASPC..216..591H}. When needed, multiple datasets were combined in ISIS using \texttt{combine\_datasets} which is conceptually similar to summing the datasets but with the models for the individual datasets being treated consistently and folded through the responses for each individual observation. The individual ARF and RMF files were not combined. The summed data are shown in the plots for clarity. The relativistic reflection model \texttt{RELXILL} was applied to all spectra and galactic absorption $n_H$ was obtained using the Colden Galactic Neutral Hydrogen Density Calculator \citep{TheChandraX-RayCenter2018} and applied via the interstellar medium absorption model for each source \citep{2000ApJ...542..914W}. 
Fitting was conducted systematically to each individual observation to include Galactic absorption and reflection components by adding the \texttt{RELXILL} model, initially in the general form (\texttt{tbabs*(powerlaw+relxill)}) in the 0.3--10 keV range. During the fitting procedure we explored the addition of a redshifted partial covering model which significantly improved the statistical $\chi^2$ results. The \texttt{RELXILL} cut off energy was fixed at 300 keV and the radius where emissivity changes from Index 1 to Index 2 was allowed to vary. The redshift was set to known values from literature for each AGN as outlined in Table~\ref{lit_info}. The black hole spin, $a$, can affect the accretion flow and spin-dependent locations such as the innermost stable circular orbit around black holes \citep[see detailed review by][]{2021ARA&A..59..117R} and a large number of accreting supermassive black holes have been found to be rapidly spinning. We find that 15/20 AGN in this sample have $a \gg 0.8$, therefore in order to reduce the number of free parameters we aligned this value for the subsequent reverberation model to be explored (Hancock et al. in prep) by fixing this parameter for a maximally spinning black hole. The inclination angle was also initially set to known values but was allowed to vary throughout the fitting procedure. Initial statistically poor fits were improved via the investigation of models similar to those used in literature for each AGN as detailed below. This was done to avoid an ad-hoc approach whilst using the wide range of model components available in \texttt{XSPEC}. This method was tested until good statistical descriptions were achieved, essentially building on known environmental parameters for each source.  

\subsection{Focusing on 1H 0707-495 and IRAS 13224-3809}

\begin{figure}
    \includegraphics[trim={0.5cm 0.0cm 1.8cm 2.5cm},clip,scale=0.31]{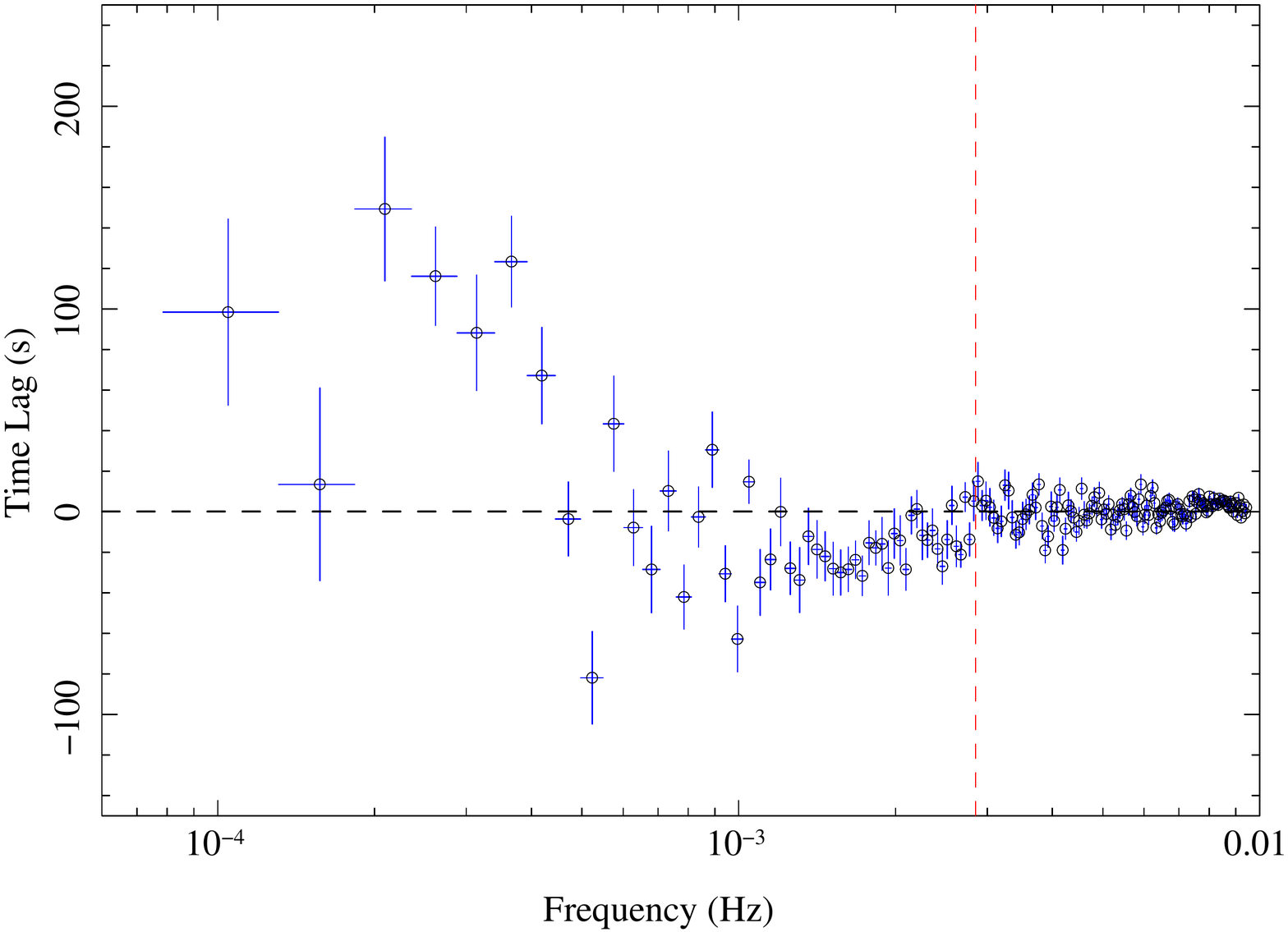} \\
    \includegraphics[trim={0.5cm 0.0cm 1.8cm 2.5cm},clip,scale=0.31]{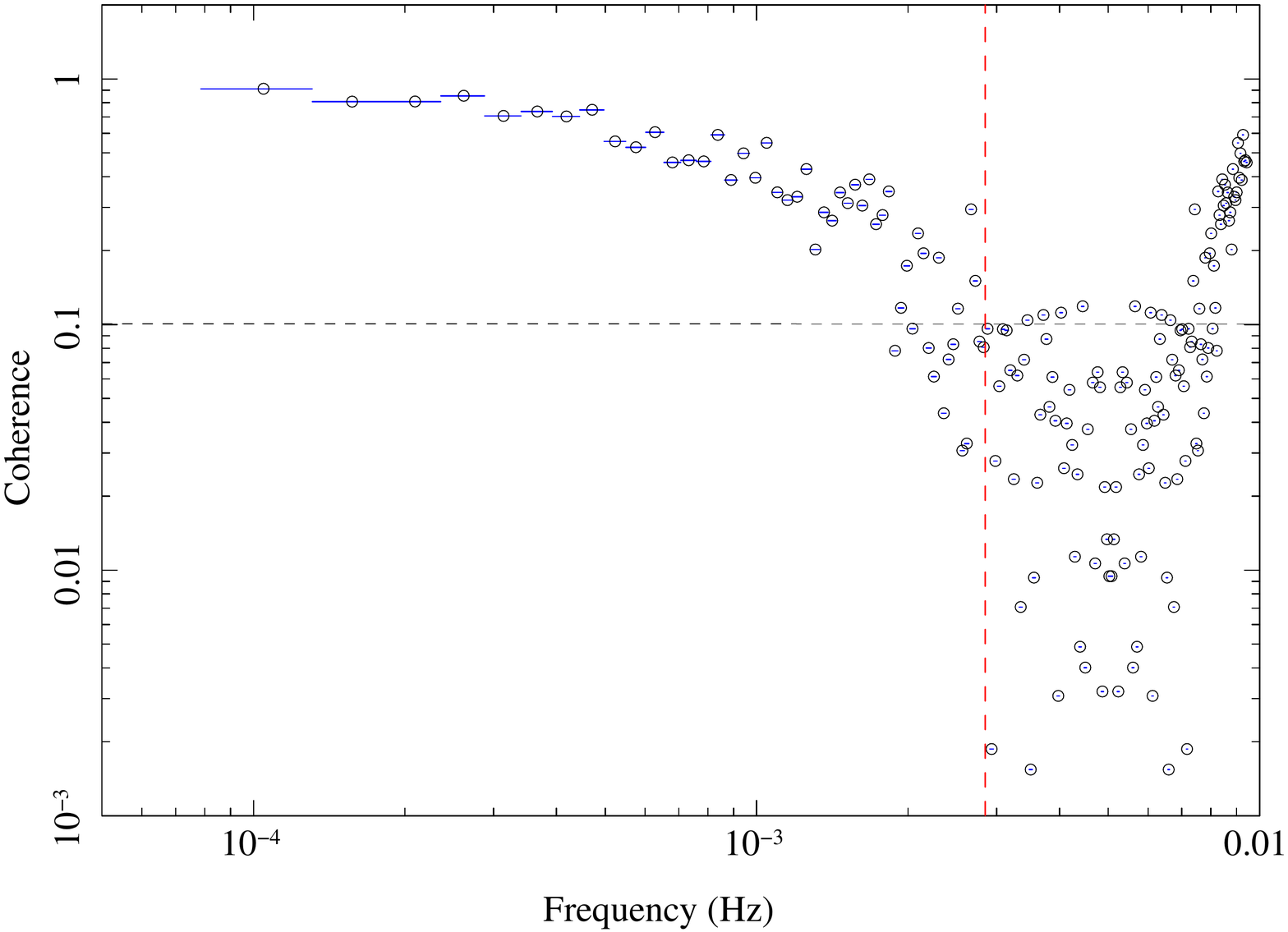}
    \caption{1H 0707-495 combined lag-frequency using 53 segments \emph{(top panel)} and the corresponding coherence \emph{(bottom panel)} limiting the maximum frequency $\nu_\text{max}$ shown by the red vertical dashed lined. The horizontal dashed line \emph{(bottom panel)} shows the coherence cut-off for these segments.}
    \label{1H0707-495_lag-coh}
\end{figure}

\begin{figure}
    \centering
    \includegraphics*[scale=0.48]{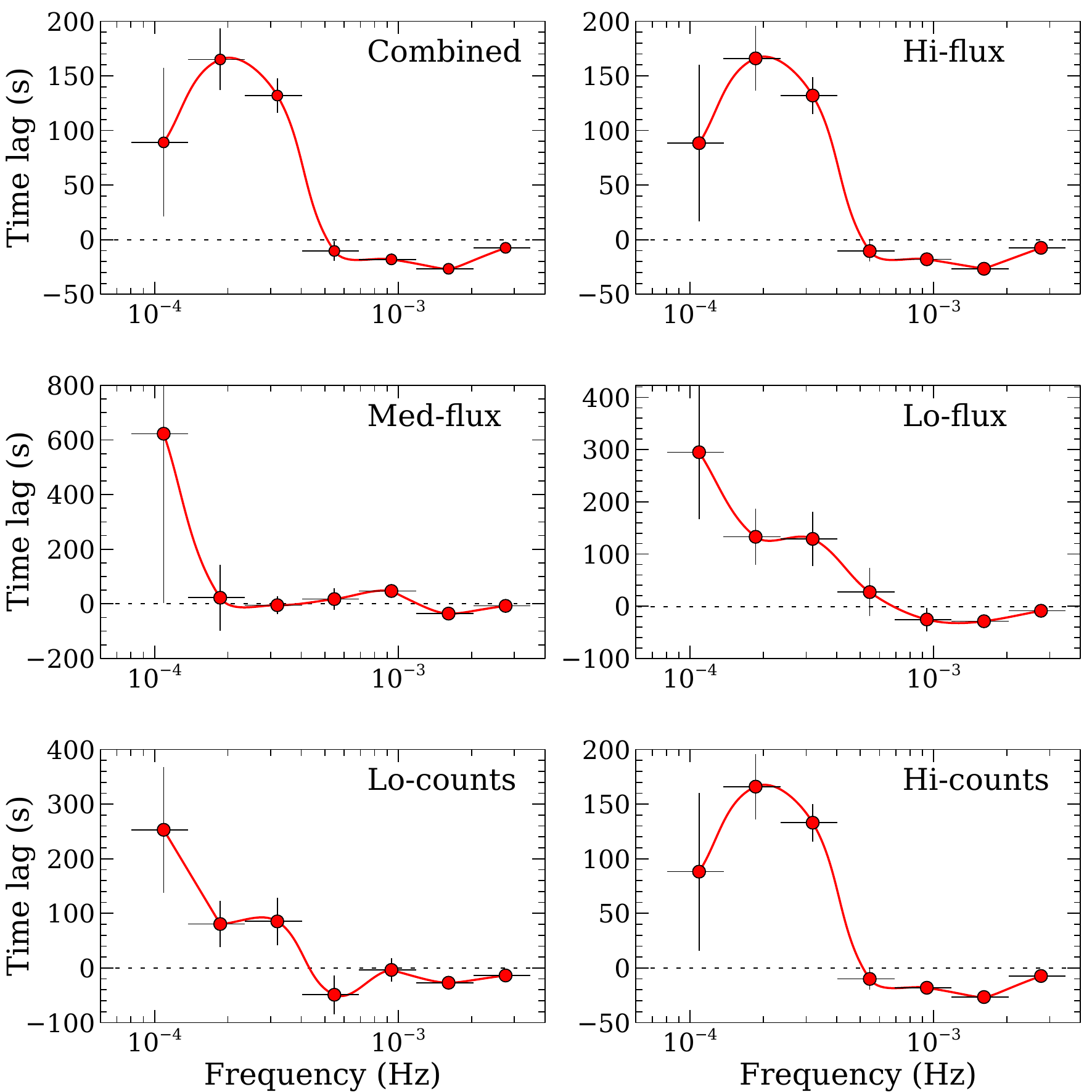}\\[4pt]
    \caption{The lag-frequency for 1H~0707-495 showing the calculated lags for each of the groups. Note that $x$-axis is identical for all groups.}
    \label{fig:1H_lo_hi_lags}
\end{figure}

1H 0707-495 and IRAS 13224-3809 are both highly variable AGN that have been well studied with an abundance of long observations available. Furthermore, they have remarkably similar spectra as we will see below from the spectral analysis therefore we examine these sources in greater detail before moving on to the remainder of the sample. Some spectra show evidence of absorption features around 8~keV. However, \citet{2004A&A...414..767K} pointed out the potential for false detection of an absorption feature at this energy as a result of spatial variation in the instrument background, and in particular the strong Cu K$\alpha$ emission lines around 8~keV.  We followed the approach of \citet{2017Natur.543...83P} to compare the spectra with and without the background. We found that the putative absorption features around 8~keV could be mostly explained by the instrument background. While weaker absorption features may remain after a careful analysis \citep[e.g.,][]{2012MNRAS.422.1914D, 2017Natur.543...83P}, since these lines are not the focus of this paper we have chosen to exclude the energy ranges 7.9--8.25~keV and 8.7--9.25~keV for 1H 0707-495 and 7.95--8.25~keV for IRAS 13224-3809. Excluding these energy ranges does not significantly change our best fit parameters. 
For 1H 0707-495 the fully combined lags were calculated up to a maximum frequency where the coherence value was 0.10, yielding $\nu_\text{max}$ = $2.8 \times 10^{-3}$ Hz and this method is represented in Figure~\ref{1H0707-495_lag-coh}. The same method was applied to IRAS 13224-3809 using 84 segments where the coherence value was 0.07 corresponding to $\nu_\text{max}$ = $2.9 \times 10^{-3}$ Hz. We begin by calculating the lags of the groups as outlined in Table~\ref{table:obs_log} and the lag-frequencies obtained are presented in Figures~\ref{fig:1H_lo_hi_lags} and \ref{fig:IRAS_lo_hi_lags} . The lag-frequencies for all other AGN are presented in the Supplementary material (online). 

\begin{figure}
    \centering
    \includegraphics*[scale=0.48]{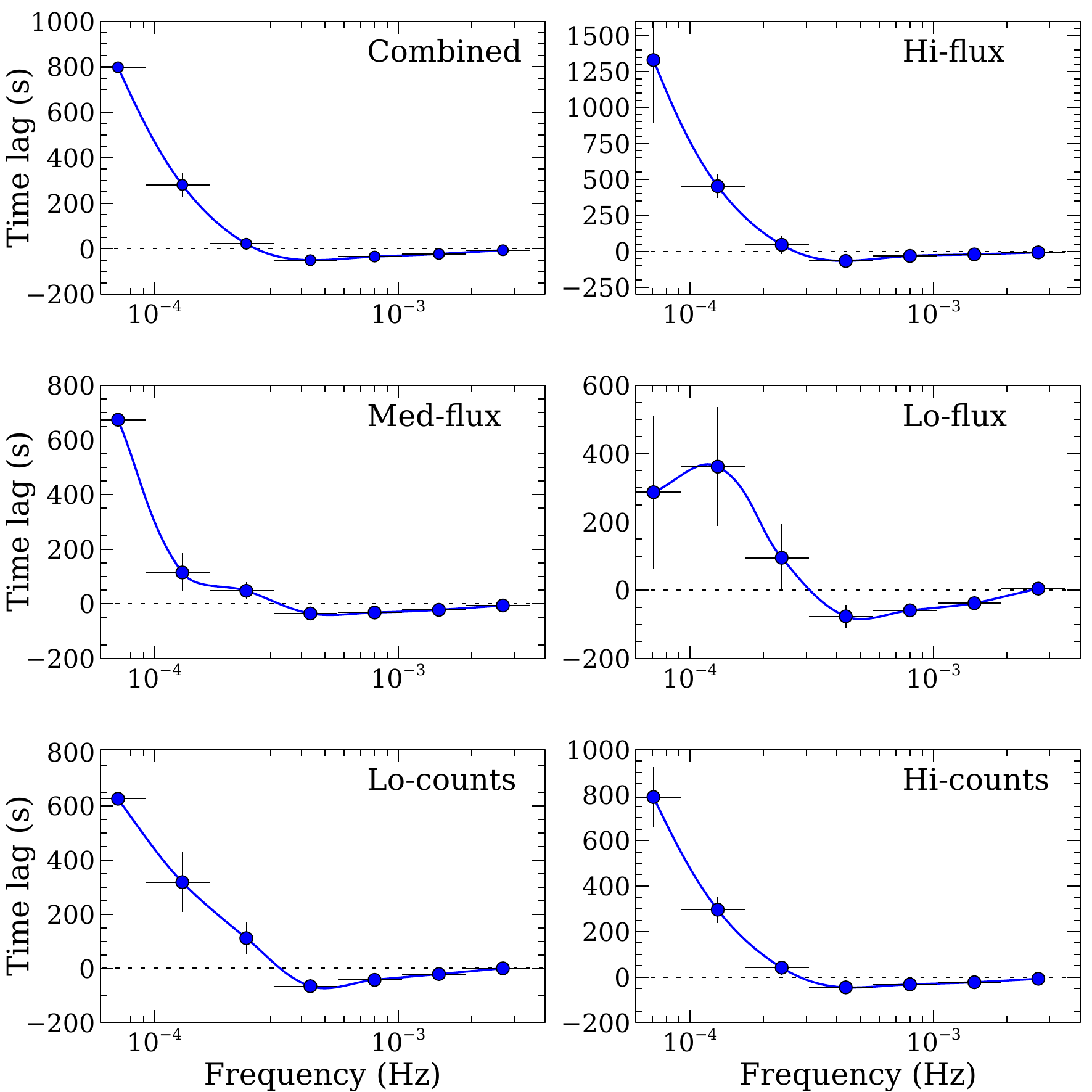}\\[4pt]
    \caption{The lag-frequency for IRAS~13224-3809 showing the calculated lags for each group. Note that $x$-axis is identical for all groups.}
    \label{fig:IRAS_lo_hi_lags}
\end{figure}

\subsection{1H 0707-495}

This AGN has been well documented to be dominated by relativistically blurred reflection at a moderate inclination angle \citep[see e.g.][]{2012MNRAS.422.1914D, 2009Natur.459..540F, 2013MNRAS.428.2795K, 2010MNRAS.401.2419Z}. The spectrum has also been suggested to be modified by absorption and contain a highly ionised outflow of up to $0.18 c$ \citep{2012MNRAS.422.1914D}. Alternative models have been explored and have similar qualitative results suggesting the spectral shape is absorption dominated \citep{2010MNRAS.408.1928M} and may also be subject to variable partial covering whereby the continuum spectrum is covered by two ionised absorption layers \citep{2014PASJ...66..122M}. 

We combine all 15 observations and add photo-electric absorption, power law, \texttt{relxill} and a disk-black body component in the 0.3 - 10 keV energy range. The statistical fits were poor. Soft excess dominates the region below $\sim$ 1 keV and above $\sim$ 7 keV the spectrum is dominated by a sharp absorption-like feature and there is insufficient data in this region to achieve a tightly constrained model fit. We use two absorption edges at $\sim$ 6.9 and 7.8 keV respectively to achieve at best $\chi^2 / \text{d.o.f.} = 10458/818$. The model was then fit between 1 -- 10 keV to exclude the soft excess band and added a red-shifted partial covering model and shifting the edge energies to 7.0 and 7.8 keV and reducing the absorption depth to $\sim$ 0.95 and 1.03 respectively and allowing them to vary. The redshift was tied to \texttt{zpcfabs} to achieve a best statistical fit where $\chi^2 / \text{d.o.f.} = 1460/683$. The model described the accretion disc environment to be mildly ionised with a high reflection fraction where the column density $n_H = 6.46^{+0.21}_{-0.19}\times10^{22}~\text{cm}^{-2}$ and the partial covering fraction is $0.64\pm0.02$ at an inclination angle of $\sim75^\circ$. This model, \texttt{zpcfabs*relxill*edge(1)*edge(2)}, was then used on all other spectral groups obtaining good fits and estimates of the 90\% confidence intervals for the fit parameters. It should be noted that the sharp absorbing regions above 7 keV have been well described by ultrafast outflow (UFO) modelling, \citep[see e.g.][]{2012MNRAS.422.1914D,2018MNRAS.481..947K}.
The model provided a reasonably good statistical fit. All groups showed steep photon indices from $\sim2.7 - 3.4$ from the lowest to highest counts rates respectively. We note that the inclination angle returned by the model is comparable with the findings of \cite{10.1111/j.1365-2966.2011.19676.x}. Full results are presented in Table~\ref{1HandIRAS_spec_results} and the combined fitted spectra are shown in Figure~\ref{fig:1H-IRAS-combspec}.

\subsection{IRAS 13224-3809}
\begin{figure*}
\centering
\includegraphics[trim={2.5cm 1.0cm 4.5cm 3.5cm},clip,scale=0.35]{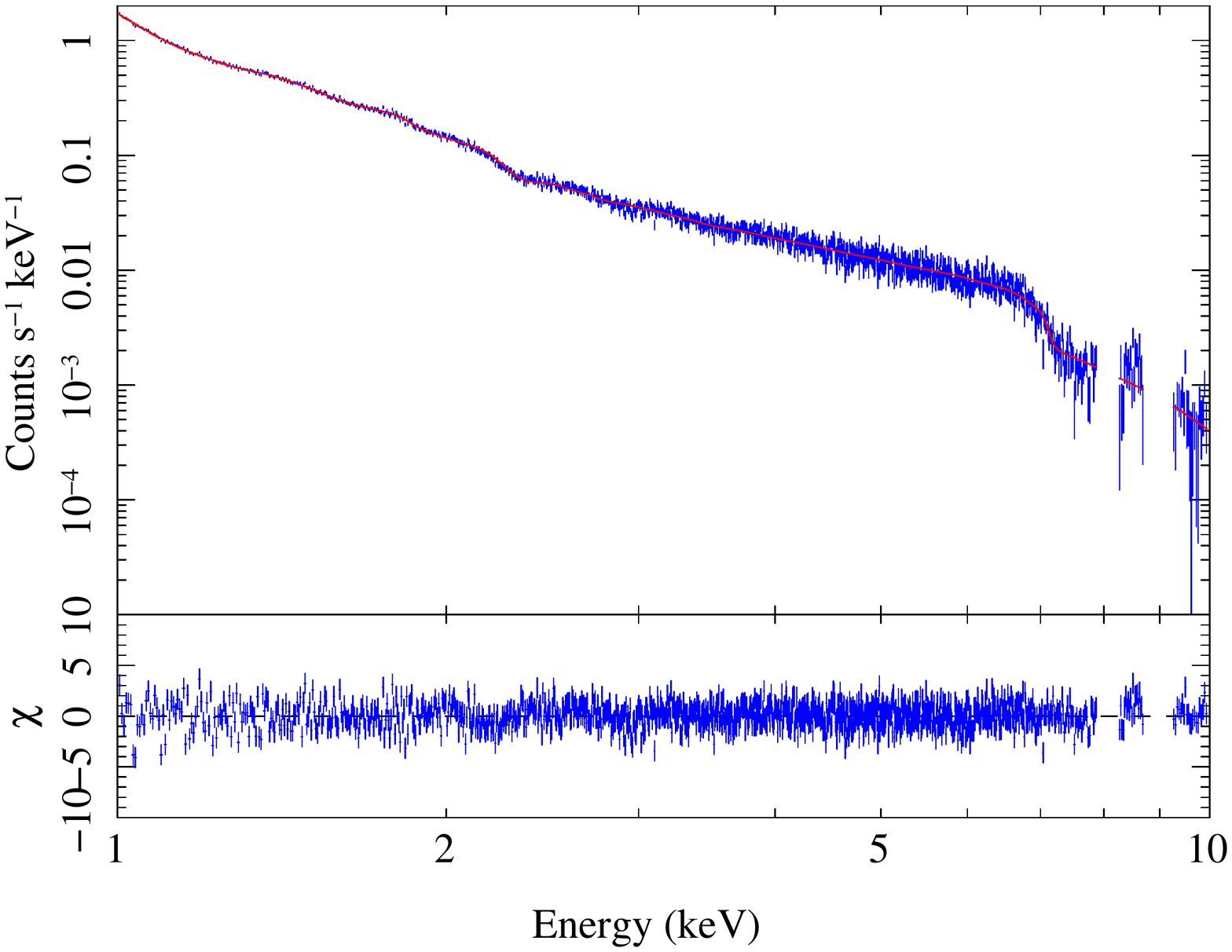}
\includegraphics[trim={2.5cm 1.0cm 4.5cm 3.5cm},clip,scale=0.35]{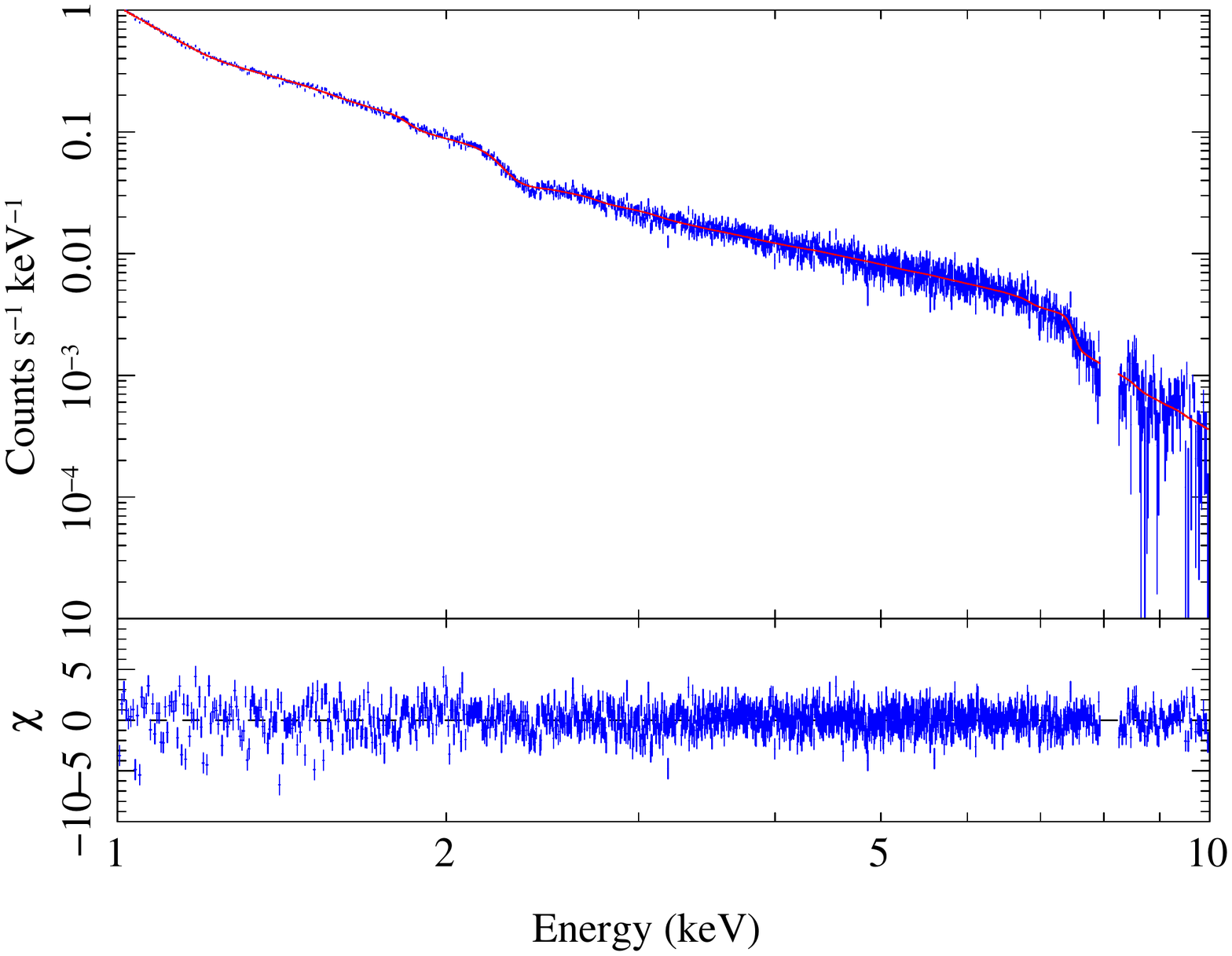}
\caption{The combined spectra of 1H~0707-495 \emph{(left panel)} and IRAS 13224-3809 \emph{(right panel)}, both AGN are fitted with \texttt{zpcfabs} and \texttt{RELXILL} model. Note we have excluded energies around 8~keV where there is a strong Cu~K$\alpha$ line in the background.}
\label{fig:1H-IRAS-combspec}
\end{figure*}

The unfolded spectra of the 17 observations as seen in Figure \ref{fig:unfold-spec} contain two well grouped spectra at the higher count rate in red and blue consisting of observations 0792180401 and 0792180601. The medium-counts group was taken from the top green down to bottom light green and consisted of 13 obs in total as outlined in Table~\ref{table:obs_log} and the two low-counts spectra at the bottom in blue and grey contain observations 0673580301 and 0792180301 was used for the low counts group. The associated lag-frequency was calculated for these groups and is presented in Figure \ref{fig:IRAS_lo_hi_lags}. The first striking feature is the large amplitude of the low frequency disc fluctuation lags extending $\sim300 - 1300$s in contrast to the moderate fluctuations of $\sim100 - 600$s as seen in 1H 0707-495 in  Figure~\ref{fig:1H_lo_hi_lags}. The soft reverberation lags however were comparable and both AGN exhibited larger reverberation lags when the photon count rate is lowest. The light curves for all IRAS 13224-3809 observations are shown in Figure~\ref{fig:1h_lightcuvres2} in the 0.3 -- 10 keV energy range. 
Thanks to the long observations conducted by \textit{XMM-Newton}, this represents an effective total observation time of over 1.9 Ms. Although the x-axis is not to scale, these light curves demonstrate the large amplitude of the variability of this AGN, often containing large photon count-rate spikes a few ks in duration. 
The spectral modelling for IRAS 13224-3809 was conducted using similar methods to those outlined above for 1H 0707-495. Once more we combine all spectra and apply a partial covering and \texttt{RELXILL} and also include absorption edges around the steep drop off at $\sim7.7$ keV. 
The best model fit was $\chi^2/\text{d.o.f.} = 2199/1281$ containing absorbed column density $n_H = 2.16^{+0.15}_{-0.17}\times10^{22}~\text{cm}^{-2}$ with a covering fraction of $61\pm0.02 \%$. Only one absorbing edge was required at $7.75\pm0.05$ keV. The reflection fraction tended to the model maximum and the disc inclination was $\sim78^\circ$. All other groups results are presented in Table~\ref{1HandIRAS_spec_results} and the combined fitted spectrum is shown in Figure~\ref{fig:1H-IRAS-combspec}.

\subsection{Other AGN in the sample}

All remaining AGN in the sample were inspected via the combined, high-flux and low-flux data groups except in cases where clear medium groupings were evident as seen in Mrk 766 and NGC 3516. Whilst most AGN were well fitted with the addition of the partial covering model, this was not required for Ark 564, Mrk~335, Mrk 841, MCG--6-30-15, PG 1211+143, PG 1244+026 and REJ 1034+396. Mrk 766 on the other hand required partial covering for the combined and low-flux groups only, fitting statistically better without this in the medium and high-flux groups. The 2 -- 10 keV luminosity $L$ was calculated in each case using the spectral model flux $F$, where $L = 4 \pi F_{(2-10\text{ keV})} D_L^2$ where $D_L$ is the luminosity distance obtained from \cite{NED2019} in cgs units with NED's default cosmology, $H_0 = 67.8 \text{ km s}^{-1}\text{ Mpc}^{-1}$, $\Omega_\text{matter} = 0.308$, $\Omega_\text{vacuum} = 0.692$. We find PG 1247+267 the most luminous of the sample where $L_{2-10\text{ keV}} = 1.27 \times 10^{46}$ erg s$^{-1}$. The strongest reflection was seen in Mrk 335, NGC 1365, NGC 4051 and NGC 4151 where the reflection fraction reached the maximum allowed by the model. Statistically good descriptions were obtained for all data groups and the reduced $\chi^2$ value was generally in the range 1.0 -- 2.5. All resultant parameters were initially plotted against $L_{2-10\text{ keV}}$ to explore the model behaviour. The full spectral modelling results for these AGN are presented in the Supplementary material (online). Several correlations were evident within the \texttt{RELXILL} components, power law flux, power law $\Gamma$, reflection fraction and covering fraction that may help to explain the variability observed in AGN. These correlations are explored in more detail and are discussed in turn below.  

\onecolumn

\begin{table*}
\centering
\caption[Relxill spectral fits]{Results for 1H 0707-495 and IRAS 13224-3809 groupings detailing the luminosity of the spectral model, the \texttt{RELXILL} model photon index $\Gamma$, ionisation parameter $\xi$, iron abundance $AFe$, reflection fraction RF, inclination angle $i$ in degrees and the covering fraction. The luminosity is corrected for Galactic absorption using the \texttt{tbabs} model. Note that error bar values of zero are the result of rounding to two decimal places. The full results for all AGN and individual observations are available in the Supplementary material (online).} 
\label{1HandIRAS_spec_results}
\begin{tabular}{cccccccccc}
\hline
Source & {Group} & $L_{2-10\text{ keV}}$ & {$\Gamma_\texttt{RELXILL}$}  & {$\log \xi$} & {$AF_e$} & {RF} & $i$ & {Cvr Frac} & {$\chi^2$ / d.o.f.}\\  \hline 
1H 0707-495 & Comb & $3.88 \times 10^{42}$ & $3.38^{+0.02}_{-0.01}$ & $2.37^{+0.05}_{-0.05}$ & $0.50^{+0.05}_{-0.00}$ & $2.14^{+0.15}_{-0.15}$ & $74.90^{+0.99}_{-1.52}$ & $0.65^{+0.02}_{-0.02}$ & 1533/683 \\ [3pt]
& Hi & $4.36 \times 10^{42}$ & $3.40^{+0.00}_{-0.02}$  &  $2.43^{+0.06}_{-0.05}$ & $0.50^{+0.04}_{-0.00}$ & $2.13^{+0.15}_{-0.14}$ & $76.68^{+0.69}_{-1.04}$ & $0.63^{+0.01}_{-0.01}$ &  1382/646\\ [3pt]
& Hi > 5 cts~s$^{-1}$ & $4.27 \times 10^{42}$ & $3.40^{+0.00}_{-0.02}$  &  $2.43^{+0.06}_{-0.05}$ & $0.50^{+0.01}_{-0.00}$ & $1.93^{+0.22}_{-0.18}$ & $76.93^{+0.83}_{-1.45}$ & $0.65^{+0.01}_{-0.03}$  & 995/690\\ [3pt]
& Med & $2.33 \times 10^{42}$ & $2.80^{+0.14}_{-0.34}$ &  $3.58^{+0.16}_{-0.32}$ & $10.00^{+0.00}_{-1.64}$ & $4.98^{+3.30}_{-2.70}$ & $76.10^{+1.09}_{-0.75}$ & $0.83^{+0.03}_{-0.16}$ & 446/411 \\ [3pt]
& Lo & $1.30 \times 10^{42}$ & $2.73^{+0.12}_{-0.16}$ &  $2.85^{+0.18}_{-0.15}$ & $9.59^{+0.41}_{-0.49}$ & $9.95^{+0.05}_{-4.98}$ & $80.00^{+0.00}_{-2.99}$ & $0.95^{+0.00}_{-0.01}$ & 539/437  \\ [3pt]
& Lo < 5 cts~s$^{-1}$ & $1.37 \times 10^{42}$ & $2.95^{+0.35}_{-0.29}$ &  $0.87^{+0.57}_{-0.29}$ & $10.00^{+0.00}_{-6.01}$ & $7.31^{+2.69}_{-3.85}$ & $80.00^{+0.00}_{-3.18}$ & $0.63^{+0.16}_{-0.02}$ & 1251/680\\ [3pt]  \hline 
IRAS 13224-3809 & Comb & $7.38 \times 10^{42}$ & $3.25^{+0.04}_{-0.02}$ & $2.13^{+0.06}_{-0.06}$ & $0.50^{+0.11}_{-0.00}$ & $3.20^{+0.36}_{-0.37}$ & $77.51^{+1.39}_{-1.18}$ & $0.60^{+0.02}_{-0.02}$ & 2199/1281 \\ [3pt]
& Hi & $1.40 \times 10^{43}$ & $3.17^{+0.09}_{-0.08}$ & $1.41^{+0.40}_{-0.10}$ & $2.79^{+1.40}_{-2.02}$ & $2.16^{+0.41}_{-0.24}$ & $77.82^{+2.18}_{-2.19}$ & $0.29^{+0.15}_{-0.09}$ & 1237/1000\\[3pt]
& Hi $>5$~cts~s$^{-1}$ & $7.42 \times 10^{42}$ & $3.13^{+0.09}_{-0.8}$ & $0.92^{+0.17}_{-0.04}$ & $0.50^{+0.24}_{-0.00}$ & $4.02^{+0.99}_{-1.02}$ & $77.29^{+1.48}_{-1.64}$ & $0.22^{+0.07}_{-0.05}$ & 1558/1353\\[3pt]
& Med & $6.08 \times 10^{42}$ & $3.27^{+0.00}_{-0.00}$ & $2.12^{+0.01}_{-0.00}$ & $0.86^{+0.02}_{-0.00}$ & $3.09^{+0.22}_{-0.21}$ & $72.30^{+0.34}_{-0.40}$ & $0.63^{+0.00}_{-0.00}$ &  1699/1455\\[3pt]
& Lo & $2.97 \times 10^{42}$ & $3.14^{+0.06}_{-0.21}$ & $1.88^{+0.17}_{-0.13}$ & $0.50^{+0.71}_{-0.00}$ & $9.24^{+0.76}_{-4.06}$ & $65.46^{+3.35}_{-2.32}$ & $0.74^{+0.03}_{-0.11}$ & 834/799 \\ [3pt]
& Lo $>5$~cts~s$^{-1}$ & $4.84 \times 10^{42}$ & $3.20^{+0.03}_{-0.04}$ & $1.83^{+0.06}_{-0.08}$ & $0.50^{+0.10}_{-0.00}$ & $2.92^{+0.37}_{-0.38}$ & $74.69^{+1.09}_{-1.41}$ & $0.64^{+0.00}_{-0.00}$ & 1579/1275 \\ \hline
\end{tabular}
\end{table*}

\begin{figure*}   
\centering
\includegraphics[trim={0.1cm 0.3cm 0cm 0.5cm},clip,scale=0.43]{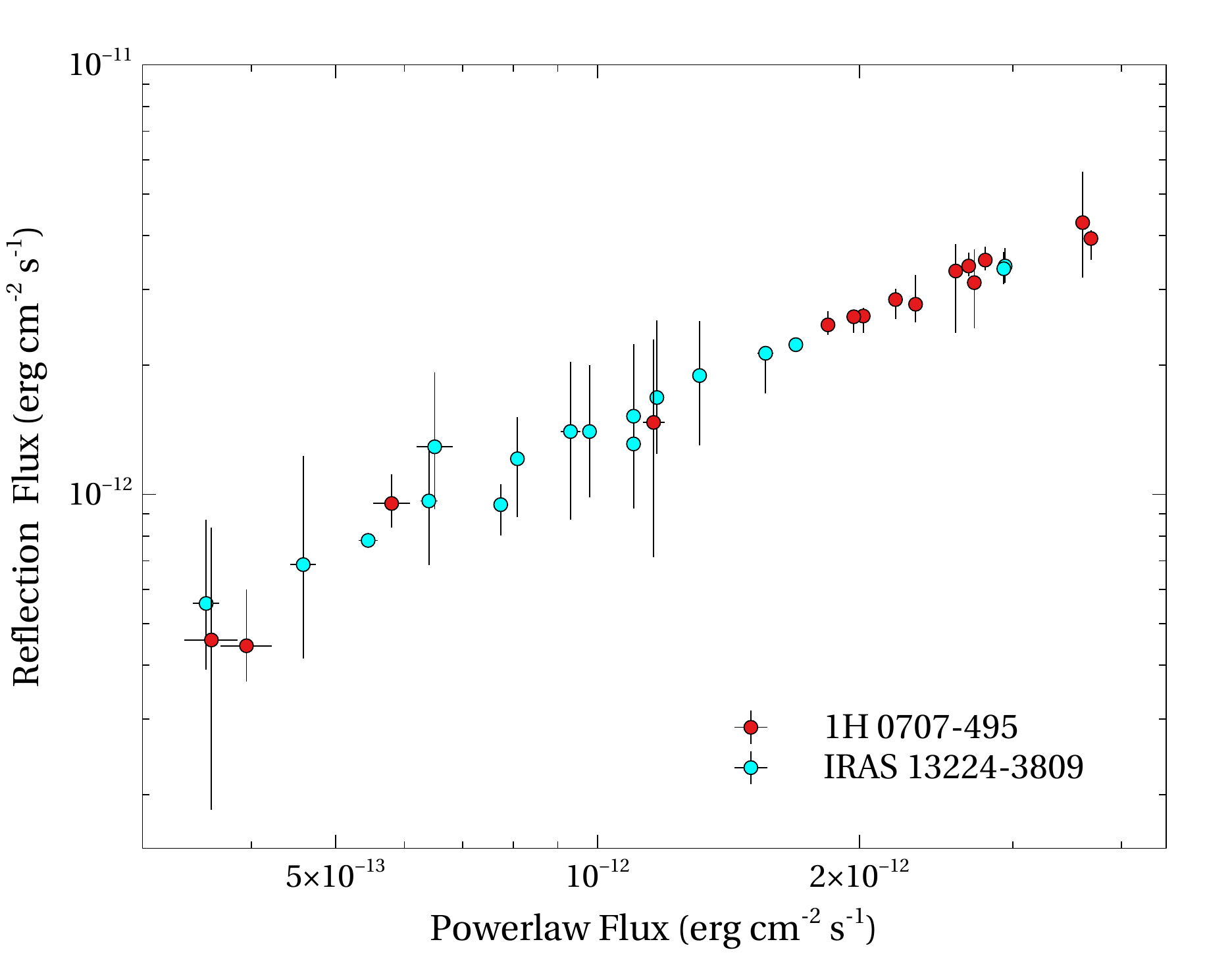}
    \put(-195,155){\large{$r_s = 0.99$}}
    \put(-195,145){\large{$p = 2.67 \times 10^{-26}$}} 
\includegraphics[trim={0.5cm 0cm 1.8cm 1.0cm},clip,scale=0.43]{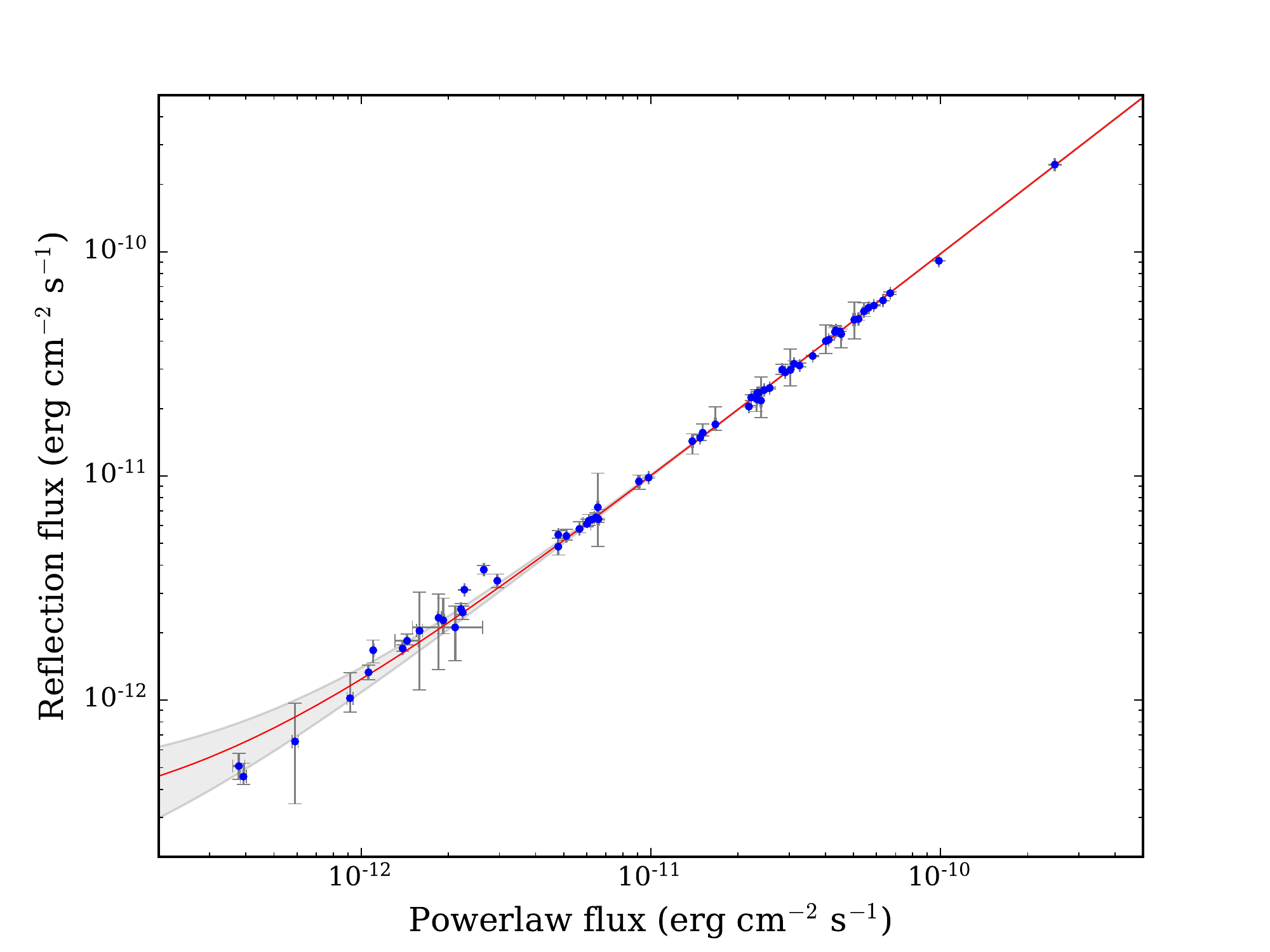}
    \put(-190,155){\large{$r_s = 0.99$}}
    \put(-190,145){\large{$p = 5.67 \times 10^{-76}$}} 
\caption{Continuum versus reflection components flux of all observations of 1H0707-495 (red) and IRAS 13224-3809 (cyan) showing the reflection flux correlated with the power-law flux in the 0.3-10 keV energy range \emph{(left panel)}. The same relationships is also shown for all 64 data groups in the sample \emph{(right panel)} showing the best line fit (red) and 1$\sigma$ region (grey).}
\label{1H0707-IRASflux}
\end{figure*}

\twocolumn

\section{Parameter correlations}

\subsection{Continuum and reflection flux} 

We inspect the dependence of the continuum flux and reflection flux by obtaining these components during this fitting phase to within the 90\% confidence limit in the 0.3--10 keV energy range. Focusing initially on 1H 0707-495 and IRAS 13224-3809 we find the Spearman's rank correlation $r_s = 0.99$ with a $p$-value of $2.67 \times 10^{-26}$. The slope and intercept was 1.11 and 2.72 $\times 10^{-13}$ respectively. The reflection scenario expects reflection component to correlate positively with the power-law component. These results are shown in Figure~\ref{1H0707-IRASflux} (left panel), accompanied by the same inspection for all other AGN groups (right panel) where the Spearman's rank correlation $r_s$ = 0.99 with a $p$-value of $5.67 \times 10^{-76}$. The slope and intercept was 0.98 and 2.63 $\times 10^{-13}$ respectively.

\subsection{Time lags and black hole mass}

\begin{figure*}
    \centering
    \includegraphics[trim={0cm 0.5cm 1.8cm 0.8cm},clip,scale=0.47]{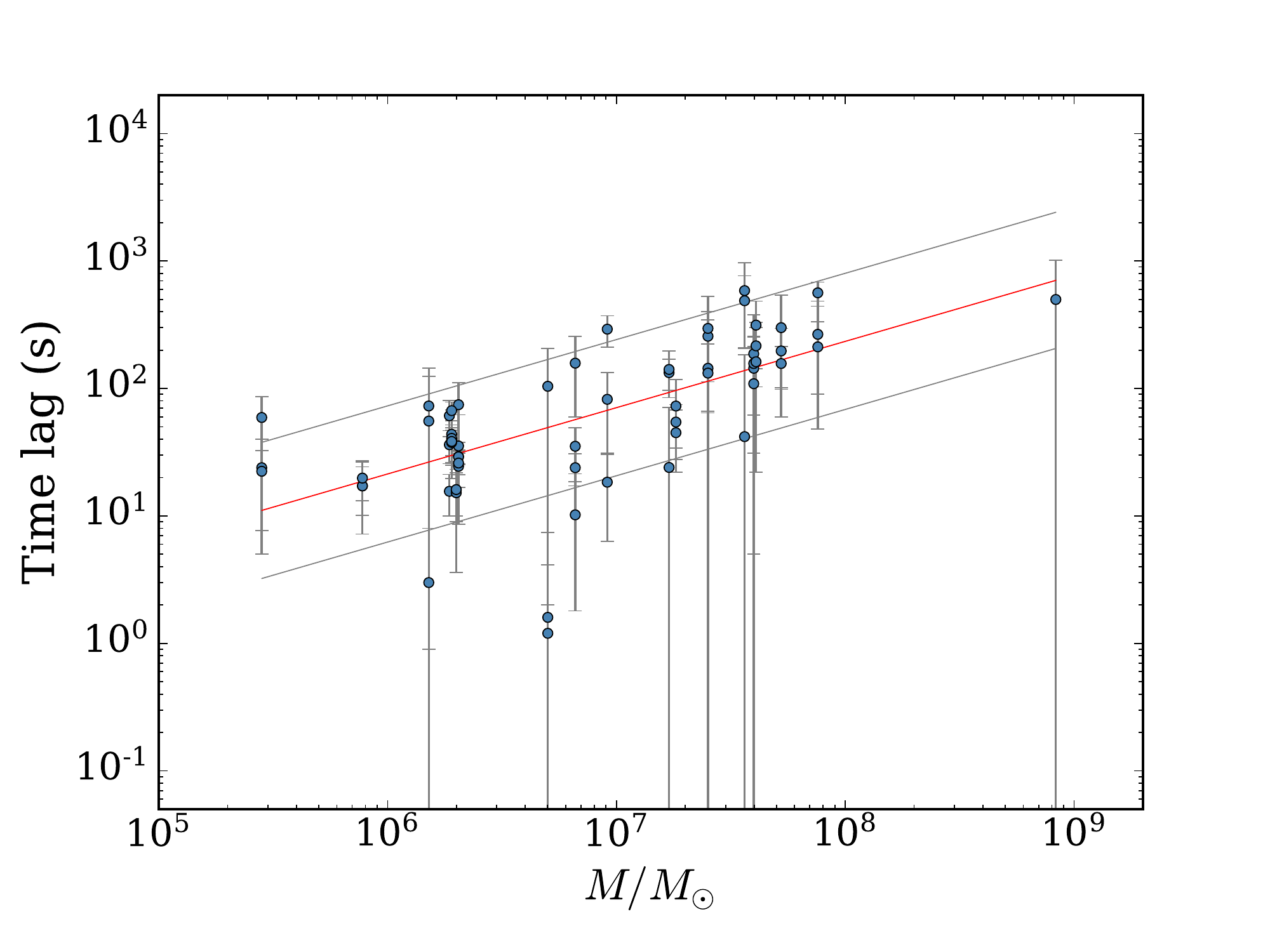}
    \put(-200,165){\large{$r_s = 0.72$}}
    \put(-200,155){\large{$p = 1.46 \times 10^{-11}$}} 
    \includegraphics[trim={0cm 0.5cm 1.8cm 0.8cm},clip,scale=0.47]{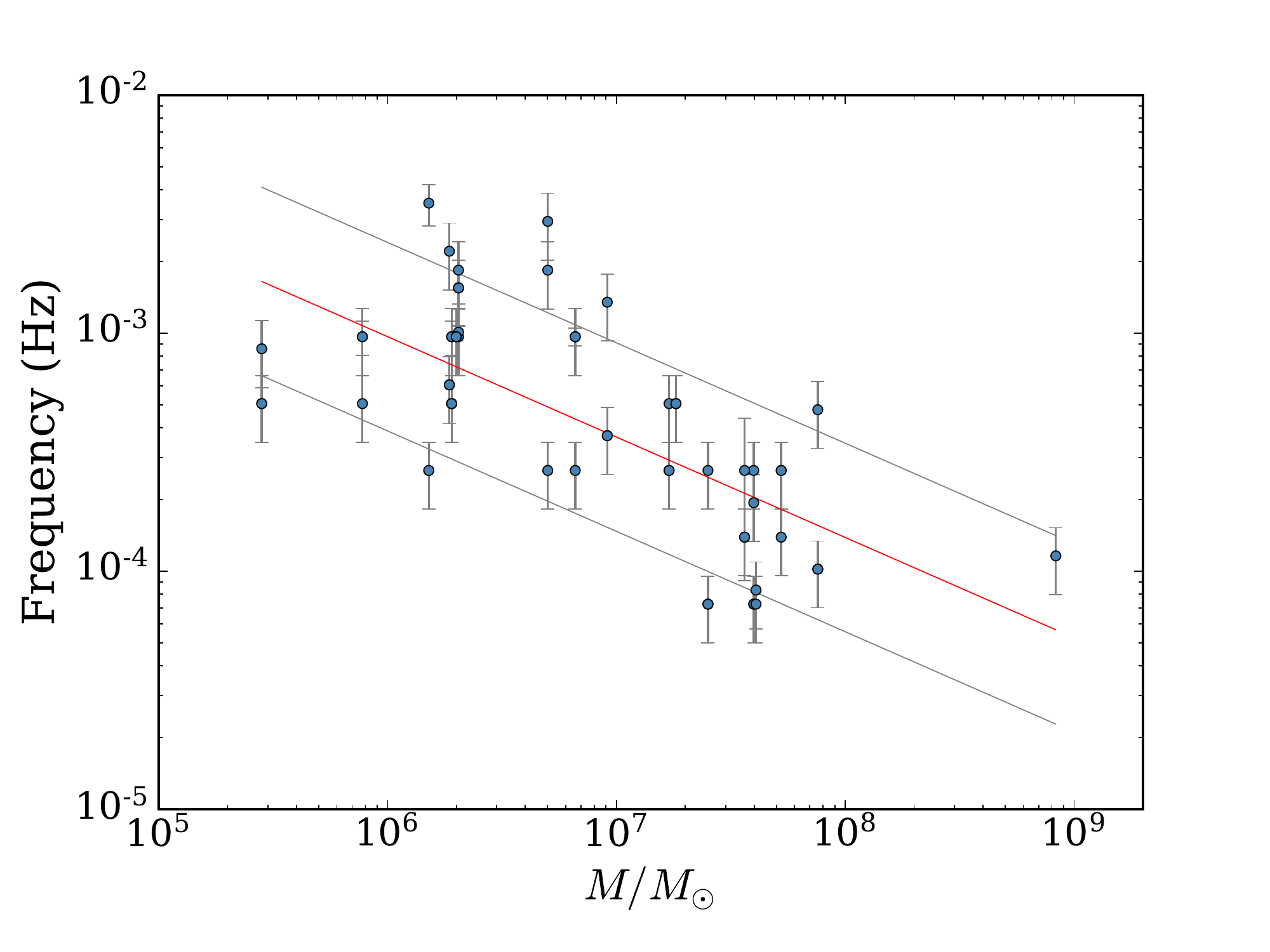}
    \put(-80,165){\large{$r_s = -0.66$}}
    \put(-80,155){\large{$p = 2.45 \times 10^{-9}$}}   
    \caption{The soft time lags \emph{(left panel)} and frequency at which the lags occur \emph{(right panel)} plotted against the black hole mass for all sample groups. The best line fit is shown in red and the 1$\sigma$ error bars are shown by the grey lines.}
    \label{fig:lag-mass-relation}
\end{figure*}

All groups and the associated lag-frequency results reported in Table~\ref{table:lag_results} were plotted against the black hole mass. The time lags correlated strongly with the mass where the Spearman's rank coefficient $r_s$ = 0.72 and $p = 1.46 \times 10^{-11}$. This correlation can be seen in the left panel in Figure~\ref{fig:lag-mass-relation}, shown with a best fit in red and the 1$\sigma$ region shown by the grey lines. 
The right panel of this figure presents the frequencies where these lags were found, which also correlates strongly with mass, having $r_s = -0.66$ and $p = 2.45 \times 10^{-9}$. These findings provide further support the mass and soft lag scaling relationship discovered by \cite{2013MNRAS.431.2441D}. We also take the lags as a function of the 2--10 keV luminosity and find that there is a moderate correlation with $r_s$ = 0.49 and $p = 4.61 \times 10^{-5}$. 

\subsection{Photon index $\Gamma$, luminosity and ionisation}
\begin{figure}
    \centering
    \includegraphics[trim={0cm 0.5cm 1.5cm 1.0cm},clip,scale=0.46]{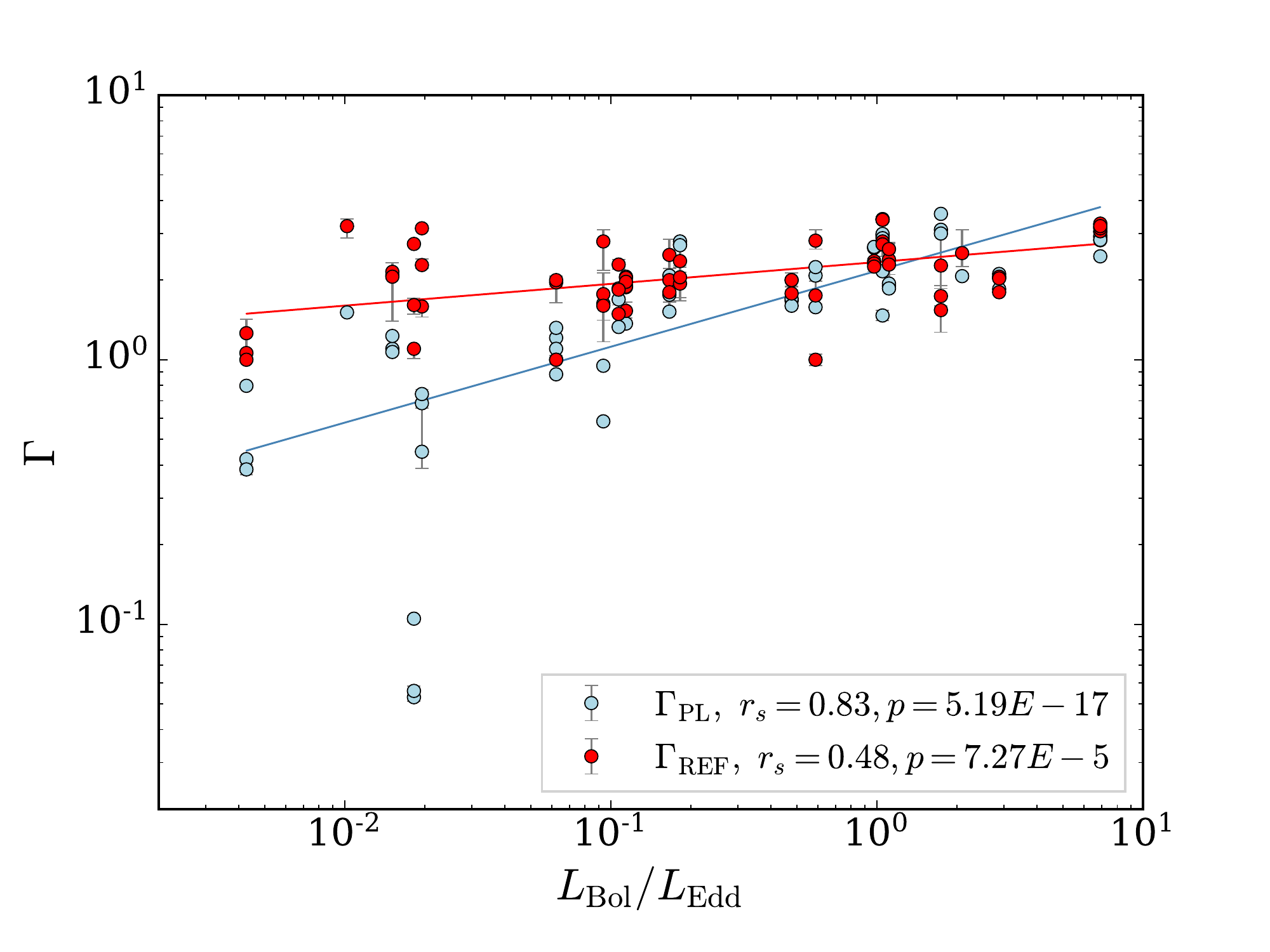}
    \caption{A plot of the $\Gamma_{\text{PL}}$ (light-blue) and $\Gamma_{\text{REF}}$ (red) against the Eddington ratio $\lambda$ with their respective best line fits for all sample groups.}
    \label{fig:gamma-lum-relation}
\end{figure}

We examine two regimes of photon index $\Gamma$. One from the continuum via a power law model photon index $\Gamma_{\text{PL}}$ and one from the independent \texttt{RELXILL} model fitting where the reflection fraction is set to -1 in order to return only the reflection component and associated power law photon index $\Gamma_{\text{REF}}$. Note that we do not require $\Gamma_\text{PL} \equiv \Gamma_\text{REF}$ which could represent an extended continuum source with a different continuum spectrum reaching infinity and being responsible for disc reflection \citep[see, e.g.,][]{chainakun_investigating_2017}.
Inspection of the power law photon index $\Gamma_{\text{PL}}$ and reflection photon index $\Gamma_{\text{REF}}$ yielded a correlation with the Eddington ratio $\lambda$ ($L_{\text{Bol}} / L _{\text{Edd}}$). The $\Gamma_{\text{PL}}$ correlated strongly with $\lambda$ where the Spearman's rank coefficient $r_s$ = 0.83 and $p = 5.19 \times 10^{-17}$. The mean $\Gamma_{\text{PL}}$ was 1.82 where $1\sigma = 0.88$. For $\Gamma_{\text{REF}}$ however, the correlation was much flatter and a moderate $r_s$ = 0.483 and $p = 7.27 \times 10^{-5}$ was returned along with mean value of 2.20 where $1\sigma = 0.67$. These relationships can be seen in Figure~\ref{fig:gamma-lum-relation}, showing both the $\Gamma_{\text{PL}}$ and  $\Gamma_{\text{REF}}$ fitted with a best line fit and the sigma regions are omitted for clarity.

We also inspected the behaviour of the ionisation parameter against the luminosity in the 0.3--10 keV energy range for all data and found no correlation. However, inspection of the combined data revealed a weak correlation between the reflection flux and ionisation where $r_s$ = 0.28 and $p = 0.23$ and between the power law flux and ionisation where $r_s$ = 0.30 and $p = 0.20$ and $1\sigma = 1.17$. 

\subsection{Photon index $\Gamma$ and covering fraction}

All data groups were inspected for behaviour of the $\Gamma_{\text{PL}}$ and $\Gamma_{\text{REF}}$ against the covering fraction and we find that the covering fraction is moderately inversely correlated with $\Gamma_{\text{PL}}$. The strongest relationship was found in the high-flux data where the Spearman's rank coefficient $r_s = -0.72$ and $p = 1.65 \times 10^{-2}$. The mean covering was 0.63 where $1\sigma = 0.22$. Moderate but weaker inverse correlations were found in the combined and low-flux data. For $\Gamma_{\text{REF}}$ however, no significant correlations were found. 

\begin{figure}
    \centering
    \includegraphics[trim={0.5cm 0.5cm 1.5cm 0.8cm},clip,scale=0.47]{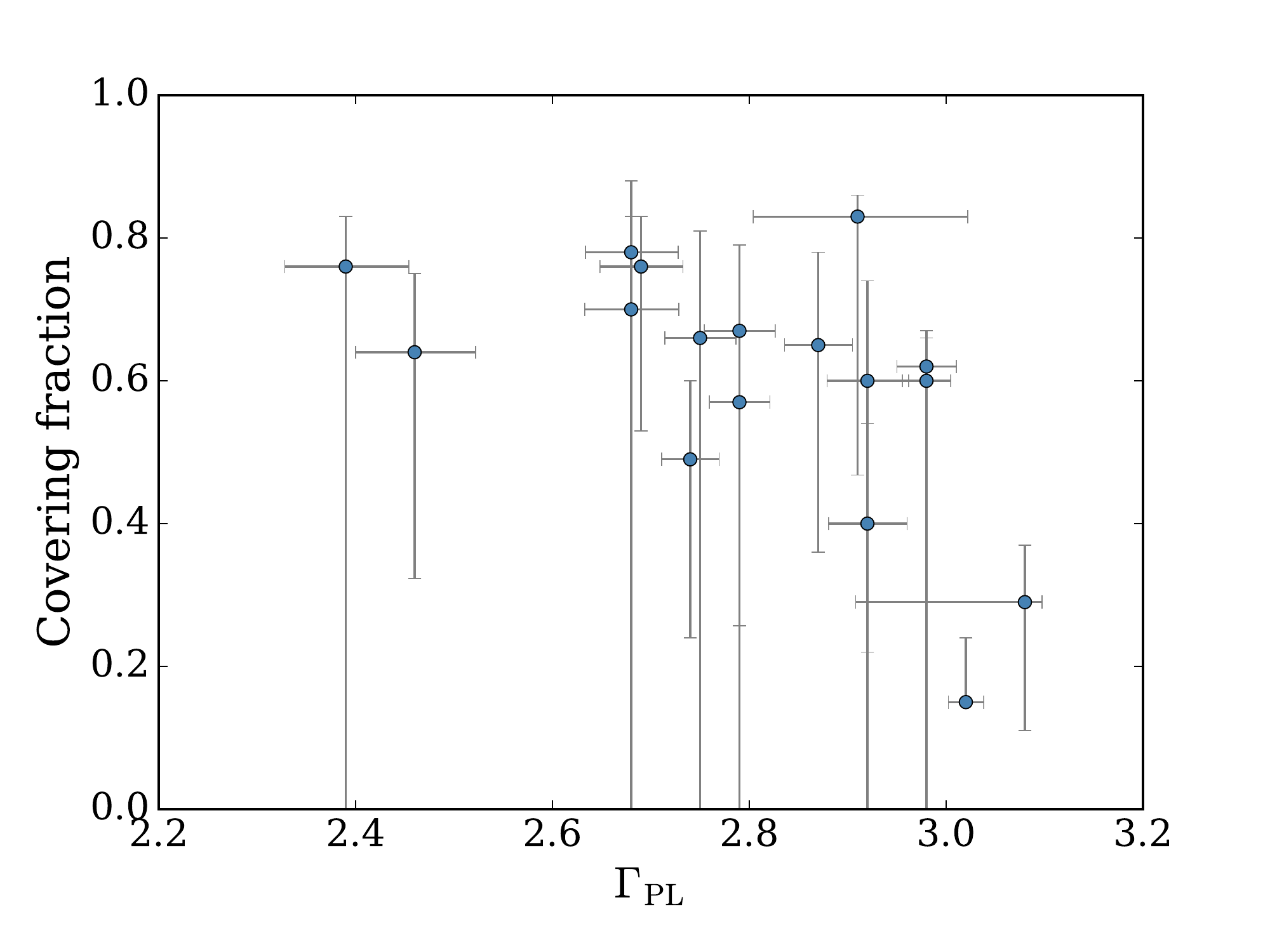}
    \put(-170,45){\large{$r_s = -0.63$}}
    \put(-170,35){\large{$p = 0.006$}}  \\   
    \includegraphics[trim={0.5cm 0cm 1.5cm 0.5cm},clip,scale=0.47]{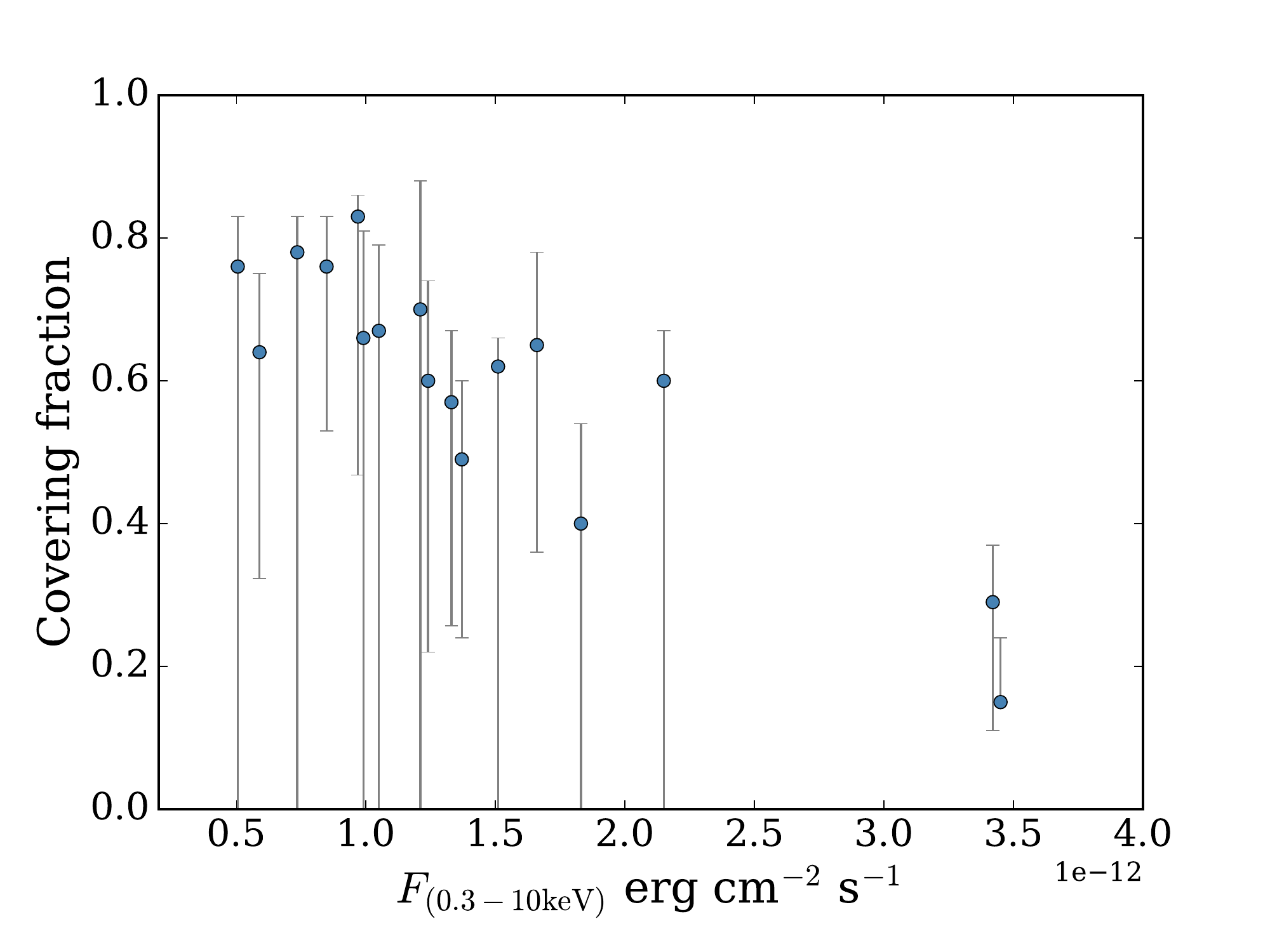}
    \put(-80,165){\large{$r_s = -0.74$}}
    \put(-80,155){\large{$p = 6.00 \times 10^{-4}$}} 
    \caption{A plot of the IRAS 13224-3809 covering fraction as a function of the $\Gamma_{\text{PL}}$ \emph{(top panel)} and the 0.3 -- 10 keV flux \emph{(bottom panel)} for the individual observations.}
    \label{fig:IRAS-cvr-relation}
\end{figure}

To investigate this further we explore all individual observations for 1H 0707-495 and IRAS 13224-3809 where the covering fraction was applied during the fitting procedure. For 1H 0707-495 we found weak anti-correlations between the covering fraction and $\Gamma_{\text{PL}}$. The 0.3--10 keV flux and 2--10 keV luminosity correlated weakly with the covering fraction where $r_s$ was around $-0.3$ and the $p$ value was $\sim0.25$ in both cases. However, the inspection of IRAS 13224-3809 revealed a strong correlation between the covering fraction and $\Gamma_{\text{PL}}$ where $r_s = -0.63$ and the $p$-value was $0.006$. We note that the mean covering fraction here was 0.60 where 1$\sigma$ was 0.17 and the errors are large as seen in top panel of Figure~\ref{fig:IRAS-cvr-relation}. If the covering fraction contributes to the variability of the spectrum then we also expect it to correlate with the flux and luminosity. Further exploration revealed the 0.3 -- 10 keV flux correlated strongly with the covering fraction where $r_s = -0.82$ and the $p$-value was $6.50 \times 10^{-5}$. Again we note the large errors on the covering fraction as presented in the bottom panel of Figure~\ref{fig:IRAS-cvr-relation}. The reflection flux also correlated strongly with the covering fraction where $r_s = -0.74$ and the $p$-value was $6.00 \times 10^{-4}$.

\subsection{The reflection fraction (RF)}

We examined the reflection fraction (RF) in all data groups and the individual observation spectra and find a significant correlation with the power law flux where $r_s = -0.70$ and $p$-value was $2.00 \times 10^{-3}$ for the IRAS 13224-3809 data. RF also correlated strongly with the reflection flux where $r_s = -0.61$ and the $p$-value was $9.00 \times 10^{-3}$. For 1H 0707-495, the $\Gamma_{\text{PL}}$ revealed a similar correlation between RF and the covering fraction of $r_s = -0.71$ and $p$-value of $3.00 \times 10^{-3}$. These correlations are discussed below and presented in Figure~\ref{fig:1H_IRAS_RF_plawGamma}. The mean RF for 1H 0707-495 was 2.71 and $\sigma = 0.56$ and IRAS 13224-3809 had mean RF of 2.80 and $\sigma = 0.18$. Examination of both AGN also revealed a moderate positive correlation between RF and the covering fraction where $r_s = 0.42$ and the $p$-value was $0.02$. These correlations do not hold when investigated using the full grouped sample results, but rather they are intrinsic to each source.

\subsection{Global correlations across the AGN sample}

Further analysis was conducted to check if the correlations described above hold across the AGN sample. This final stage was completed by independently running a similar phenomenological model to include Galactic absorption in the form `\texttt{tbabs*powerlaw*zpcfabs*relxill}'. The model was systematically fitted to all data groups and identical correlations were obtained. We find that all correlations hold with very similar Spearman rank values, hence the correlations reported here for all data groups appear to be globally obtained throughout the AGN sample. Fitting this model without the \texttt{zpcfabs} components did not significantly change the luminosity. 

\begin{figure}
    \centering
    \includegraphics[trim={0.1cm 0cm 0 0},clip,scale=0.55]{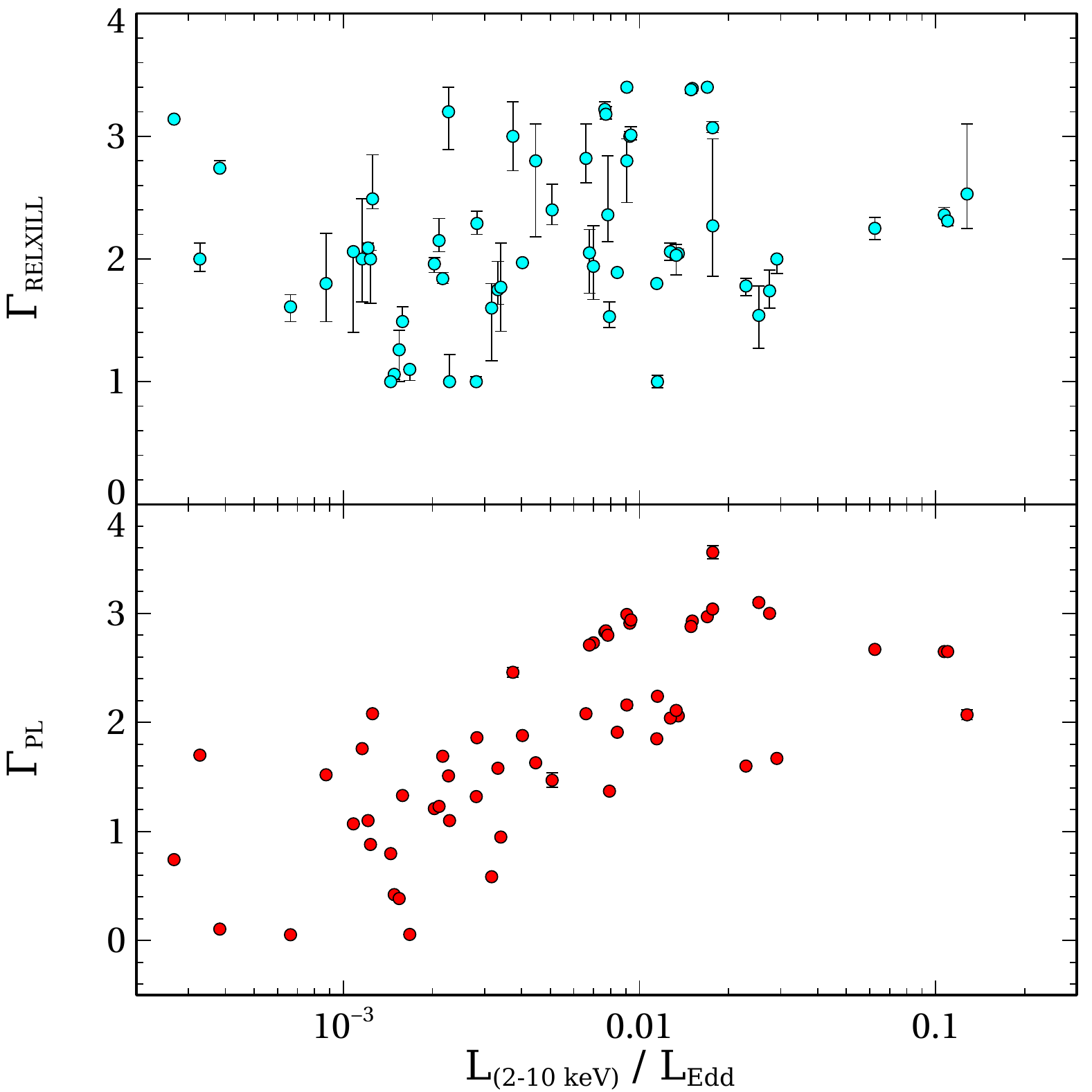}
    \caption{A plot of $\Gamma_{\text{PL}}$ and $\Gamma_{\text{REF}}$ as a function of the Eddington fraction for the grouped data for the second model (with Galactic absorption).}
    \label{fig:gamma-L-relation}
\end{figure}

We present one final interesting correlation between the Eddington fraction and the two photon indices $\Gamma_{\text{PL}}$ and $\Gamma_{\text{REF}}$. Both spectral models returned consistent correlation values and we consider the model relationships shown in Figure~\ref{fig:gamma-L-relation}, where $\Gamma_{\text{PL}}$ correlates strongly with the Eddington fraction where $r_s = 0.75$ and $p$-value $1.59 \times 10^{-12}$ and $\Gamma_{\text{REF}}$ correlates moderately with $r_s = 0.48$ and $p$-value $6.61 \times 10^{-5}$. There tentatively appears to be a difference between the power law photon indices when the 2 -- 10 keV Eddington fraction is $\lesssim 0.02$, in this case corresponding to a luminosity of $\lesssim 10^{43}$ erg s$^{-1}$.

\section{Discussion}

The soft reverberation lags for around 50\% of the AGN sample varied by a few tens of seconds when calculated between the combined, high and low flux states. The largest changes between these groups of $\sim 100 - 300$~s were found in Mrk 335, Mrk 766, Mrk 841, NGC 4151, NGC 5548, NGC 7314, 7469 and PG 1211+143. The largest time lags $\sim 500 - 600$~s were found in Mrk 841, NGC 4151 and PG 1247+267 and we have seen that these lags generally increase with the black hole mass and AGN luminosity. The correlations found between the mass and time lag and the frequency supports the findings of DM13 when using a different sample set where the black hole mass estimates were sourced from reverberation mapping techniques whenever possible \citep{2015PASP..127...67B}. Furthermore, most of these sources and similar mass estimates to this work have been previously reported by \cite{2016MNRAS.462..511K}. Nevertheless, \cite{2016MNRAS.462..511K} reported that the correlation of the frequency--mass relation is stronger than that of the lag–mass relation using the Fe K band lags between the 3 -- 4 keV and 5 -- 7 keV bands. Therefore they suggested using the frequency--mass relation to determine the black hole mass of AGN from the Fe K band lags. Interestingly, by grouping the data into similarly observed spectral states, we find that the lag--mass relation has a stronger correlation coefficient than the frequency--mass relation. The difference of these results arise due to the different averaging schemes used in our work and by DM13 and \cite{2016MNRAS.462..511K}. However, \cite{2022MNRAS.513..648C} suggested that the lag–mass correlation coefficient can decrease with the increasing number of newly-discovered reverberating AGNs. This could happen, for example, under the extended corona framework if the amplitude of the lags are more strongly affected by the coronal properties (e.g. being diluted due to the effects of coronal temperature and compactness). If this is the case, it might be still better to use the frequency--mass relation to determine the mass, keeping in mind that the dilution factor does not significantly change the frequency where the negative lags appear.

While the reverberation lag is thought to be caused by the light travel distance between the corona and the disc, the flux in the reflection-dominated band also includes a component of continuum emission and vice-versa and these are subject to dilution effects that will reduce the time lags \cite[e.g.][]{2013MNRAS.430..247W,2015MNRAS.452..333C,chainakun_investigating_2017}. Furthermore, \cite{2020MNRAS.498.3302W} investigated reverberation lags taking into account the effects of returning radiation and found that the soft X-ray lag could increase by $\sim 25$ percent. The source height implied by the standard lamppost model would then be overestimated if the returning radiation plays an important role. Due to the dilution and returning radiation effects, it is not appropriate to assume the lag amplitude directly corresponds to the distance of the X-ray source or corona above the disc. In fact, we will investigate the relations of the lags and the source geometry in our subsequent study of the extended corona scenario (in prep).

From the dependency of the lag-frequency seen in Figures~\ref{fig:1H_lo_hi_lags} and \ref{fig:IRAS_lo_hi_lags}, the soft lag amplitudes of 1H 0707-495 are generally small at $\sim30$s at frequencies $\sim1.5 \times 10^{-3}$ Hz, implying the corona is reasonably compact in agreement current literature \cite[e.g.][]{2020A&A...641A..89S}. In the case of IRAS 13224-3809, the result is comparable with \cite{2020MNRAS.498.3184C} where they analysed the combined spectral-timing data and found the trend of increasing source height with increasing luminosity. However, the clear trend of changing both amplitude and frequency where the maximal soft lag is found with changing flux states that relate to the dynamic of the corona \citep{2019Natur.565..198K} is not observed. In fact, by considering only the distinct high and low flux states in Table \ref{table:lag_results}, we find $\sim40\%$ of the sources whose maximum lag amplitude decreases when the source changes from high to low flux states, while the frequency of the maximum lag amplitude increases in $\sim50\%$ of the sources. There was no clear correlation between the time lags and the spectral model parameters.

The disc fluctuation lags varied highly throughout the sample with the most extreme variations seen in IRAS 13224-3809 and REJ 1034+396 where the positive fluctuations lags extend up to $\sim 1300$~s. A similar lag profile is seen in PG 1244+026 between $\sim 500 - 800$~s although not so highly variable between groups. The lag-frequency profile for these three AGN are very similar and their spectra contain heavy absorption in the 7 -- 10 keV energy range. Whilst this higher energy absorption feature is also evident in 1H 0707-495 and PG 1211+143 the lag-frequency does do not follow the same profile. This has been discussed by \cite{2016MNRAS.462..511K} where the absence of hard lags in NGC1365, Mrk 841 and NGC 3516 is also found here in this study. We find a similar profile in PG 1247+267, however the relatively short duration prevents lag estimation down to lower frequencies. This supports the need for more long observations ($\gtrsim 100$ ks).

We find that spectral fitting can be well constrained using both reflection and absorption modelling. 
The spectral analysis has shown that the reflection flux is strongly connected to the power law flux as expected in the reflection scenario. This is well constrained for all data in the sample and a focus into this relationship for 1H 0707-495 and IRAS 13224-3809 reveals a hint the reflection flux flattens out when the power law flux is higher (see Figure~\ref{1H0707-IRASflux}). Although this was not tested for all AGN due to the insufficient data, it has been suggested to be a consequence of light bending effects closer to the black hole \citep{2010MNRAS.401.2419Z}.  

The $\Gamma_{\text{PL}}$ also correlated strongly with the Eddington ratio $\lambda$ in line with previous studies such as \cite{2015MNRAS.448.1541S} and references therein. The $\Gamma_{\text{REF}}$ correlation was much flatter for the $\lambda$ range 0.003 -- 8.0 as seen in red in Figure~\ref{fig:gamma-lum-relation}. We note 3 outliers where $\Gamma_{\text{PL}}$ is very low (0.11, 0.56 and 0.53 respectively) as found in NGC 4151 that may be a hint of the $\Gamma_{\text{PL}}$ - $\lambda$ anti-correlation found in lower luminosity AGN \cite[e.g.][]{2009MNRAS.399..349G} yet this was still a well constrained moderate correlation. One would expect these correlations to be parallel to each other given the acute relationship between the different power law and reflection flux, however there appears to be no clear relationship between the photon index and luminosity for different sources, possibly due to changes in the accretion rate and/or disc geometry. The $\Gamma_{\text{PL}}$ has been found to significantly correlate with the Eddington fraction whilst testing the claim that $\Gamma$ can serve as an accretion rate indicator \citep{Shemmer2008} concluding that that the accretion rate largely determines the hard X-ray spectral slope across 4 orders of magnitude in AGN luminosity. Whilst we find much wider values of $\Gamma_{\text{PL}}$ than reported by \cite{Shemmer2008}, we also find that the $\Gamma_{\text{PL}}$ is much stronger correlated than $\Gamma_{\text{REF}}$ against the Eddington fraction and the same is seen against the Eddington ratio. This suggests the accretion rate is largely determined by the spectral slope of the continuum and the spectrum softens with increasing luminosity due to cooling via reprocessed emission of the hot plasma. The reflection component however, has an overall softer and flatter spectral profile and may be more determined by physical properties of the corona and a harder X-ray component that remains roughly constant during the observed timescales \citep[see][]{MCHARDY1999509}. 

The addition of a partial covering absorption improved the fitting procedures for 13/20 of AGN in the sample and was not required for Ark 564, MCG-6-30-15, Mrk335, Mrk 841, PG 1211+143, PG 1244+026 and REJ 1034+396. There is evidence that the covering fraction increases with a decrease in the $\Gamma_{\text{PL}}$ in the high flux data and this is also moderately evident in the combined and low flux data. Despite the large error bars, this suggests that the covering fraction may also be inversely connected to the power law and hence the reflection flux. Partial covering has been previously explored to help explain the large flux changes of AGN without the large changes in luminosity and inner disc temperature. In particular the flux changes seen in 1H 0707-495 have been explained by a changing covering fraction due to the motion of non-uniform orbiting clouds where the local minimum of the covering corresponds to flares or flux peaks. This explanation could modulate the observed flux without changing the spectral profile  \citep{2004PASJ...56L...9T} and is clearly evident in most of the AGN explored in this study and has been supported by subsequent studies \cite[e.g.][]{2014PASJ...66..122M,2009MNRAS.393L...1R}. Further investigation revealed that whilst this is not significant in a sample of AGN, strong correlations were found in the IRAS 13224-3809 covering fraction as a function of the $\Gamma_{\text{PL}}$ and the spectral model flux (Figure~\ref{fig:IRAS-cvr-relation}).  Findings in this study supports the suggestion that the variability seen in IRAS 13224-3809 is also due to non-uniform orbiting absorbing structures \citep{1997MNRAS.289..393B}. Therefore, both variability that is intrinsic to the source and that caused by absorption should play an important role in IRAS 13224-3809. Perhaps, they dominate on different timescales and vary among different flux states, such as what was seen in 1H 0707-495 \citep{2021MNRAS.508.1798P}.  On the other hand, it does not explain the variability seen in 7/20 AGN where the covering fraction was negligible and deemed unnecessary, hence alternative mechanisms must be in play. 

\begin{figure}
    \centering
    \includegraphics[trim={0.2cm 0.5cm 1.5cm 0.5cm},clip,scale=0.47]{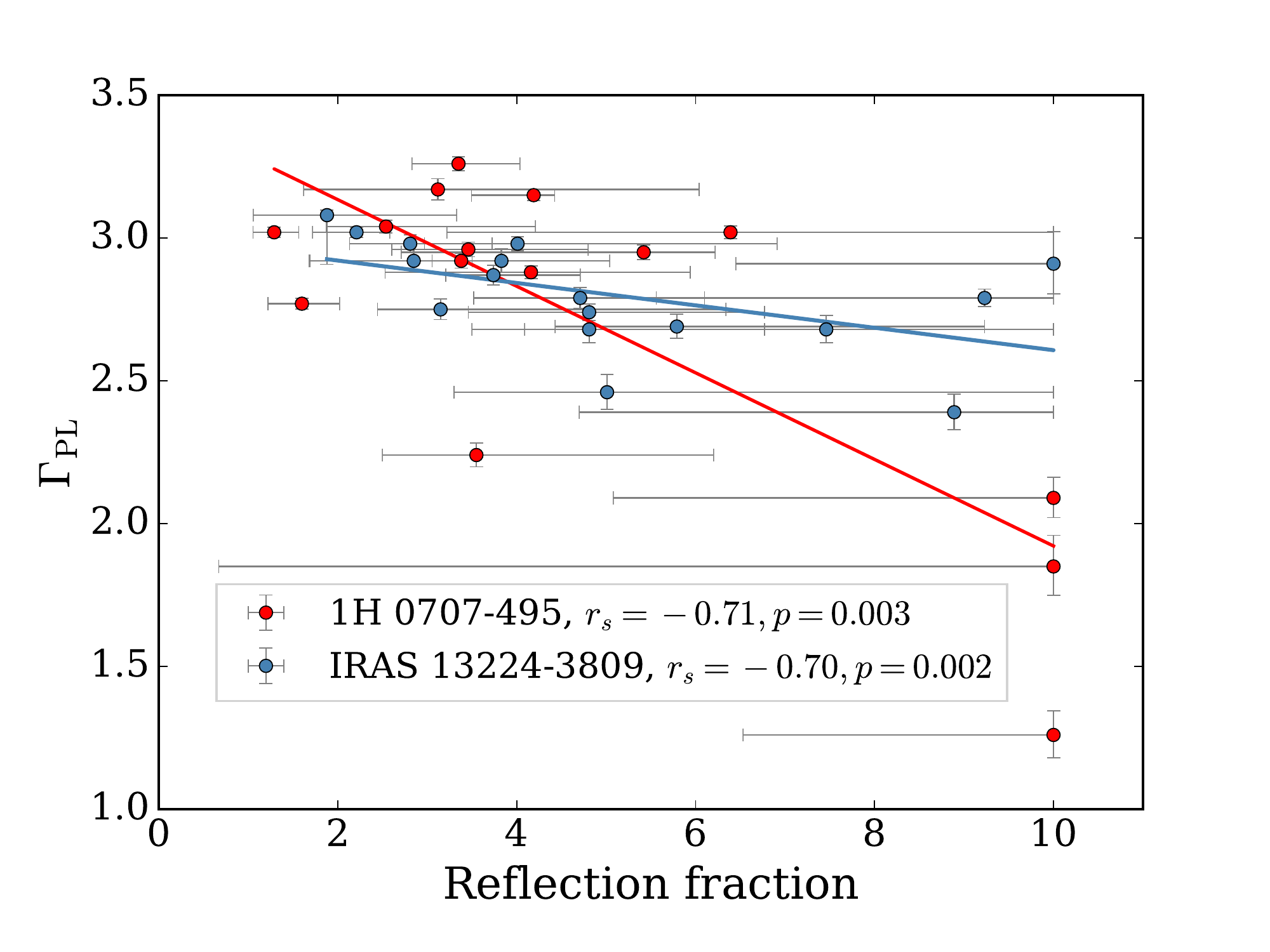}
    \caption{A plot of the RF behaviour against the power law $\Gamma$ for 1H 0707-495 (red) and IRAS 13224-380 (blue) for the individual observations with the best linear fits respectively.}
    \label{fig:1H_IRAS_RF_plawGamma}
\end{figure}

The area of the reflecting region and the location of the X-ray source or emitter can influence the amount of observed reflection. For example, photons emitted from an X-ray source close to the black hole will be gravitationally focused towards the inner regions and unable to escape to infinity and contribute to the strength of the continuum leading to a higher reflection fraction \citep{10.1111/j.1365-2966.2011.19676.x}. \cite{10.1093/mnras/stu2524} found that a low RF may be observed when a patchy or flaring covering fraction is \textless0.85 and Compton scatters the reflection at the ISCO. The highest RF was seen in NGC 1365 and PG 1247+267 where this parameter reached its maximum in all data groups. This dominance was absent in Ark564, Mrk 766, NGC 7314, PG1211+143, and the smallest RF was evident in NGC 7469. The significant correlations reported above are consistent with recent findings reported by \cite{2020MNRAS.495.3373E}, where a significant correlation between RF and the $\Gamma_{\text{PL}}$ in a sample of 14 AGN (including 3 from this study) as observed by \textit{NuSTAR}. They explained the scenario where the rate of cooling of the corona is directly linked to the slope of the power law as a higher input of seed photons from a large area of the accretion disc enters the corona and enables stronger cooling the plasma and hence the steeper power law. This leads to a higher fraction of X-ray illumination on the same region of the disc producing a higher value of RF. The observed reflection fraction depends on the geometry of the accretion disc, therefore correlations between RF and $\Gamma_{\text{PL}}$ may be driven by changes in the disc and corona geometry. Figure \ref{fig:1H_IRAS_RF_plawGamma} shows the correlations between RF and the $\Gamma_{\text{PL}}$ for 1H 0707-495 shown in red with a linear best fit where the slope and intercept was -0.15 and 3.44 respectively. This relationship for IRAS 13224-3809 is shown in blue with a linear fit where the slope and intercept was -0.04 and 2.99 respectively. It is clear that a better understanding can be gained by considering the individual orbits rather than combining spectra into groups. 

Furthermore, previous literature showed that the lag-frequency profiles of 1H 0707-495 and IRAS 13224-3809 exhibited a wavy-like feature on their negative lags \citep[e.g.][]{2018MNRAS.480.2650C}. Although its presence is less noticeable in this work, which is probably due to a large bin size, we can still observe hints of this feature in both AGN towards high frequencies before the phase wrapping occurs (Figure \ref{fig:1H_lo_hi_lags}). The wavy feature on negative lags is likely produced by an extended corona \citep[see, e.g][and discussion therein]{2019MNRAS.487..667C,2020MNRAS.498.3184C}. This, together with RF and $\Gamma_{\text{PL}}$ correlation derived here (Figure \ref{fig:1H_IRAS_RF_plawGamma}), strongly suggests that the variability of 1H 0707-495 and IRAS 13224-3809 could be driven by the dynamics of an extended corona, perhaps simplified by two co-axial X-ray sources varying in height and luminosity \citep{chainakun_investigating_2017,2019ApJ...878...20C}, rather than the evolution of the lamppost scenario.   

\begin{figure}
    \centering
    \includegraphics[trim={0.2cm 0.5cm 1.5cm 0.5cm},clip,scale=0.47]{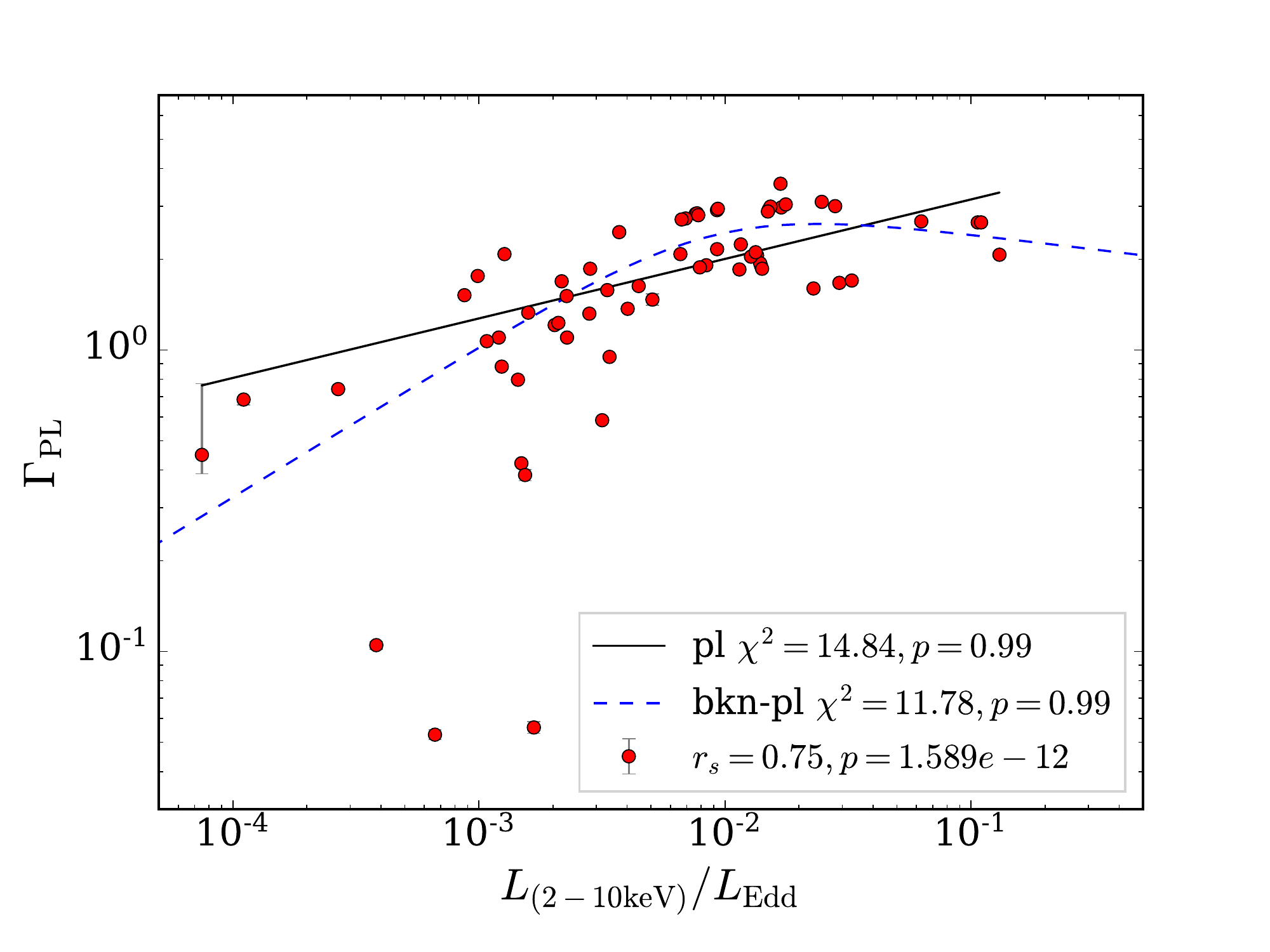}
    \caption{The power law photon index evolution as a function of the Eddington fraction for all AGN groups fitted with a power law and a broken-power law. The photon index decreases at lower Eddington fractions, but a broken power law model is not required by the data.}
    \label{fig:Gamma_v_Eddfrac_plfits}
\end{figure}

Both $\Gamma_{\text{PL}}$ and $\Gamma_{\text{REF}}$ increase as the luminosity increases until the 2 -- 10 keV Eddington fraction is $\sim0.02$, after which the spectral indices decrease as the luminosity continues to increase, suggesting variations of the inner regions and accretion flow. However when $\Gamma_\text{PL}$ and $\Gamma_\text{REF}$ are plotted against the Eddington ratio, the downturn in the photon index is not clearly evident as seen in Figure~\ref{fig:gamma-lum-relation}. We investigate the Eddington fraction turnover further by fitting a power law and a broken-power law, obtaining $\chi^2$ values of 14.84 and 11.78 respectively as shown in Figure~\ref{fig:Gamma_v_Eddfrac_plfits}. Whilst the data are marginally better fit by a broken power law, but this difference is not statistically significant. Nevertheless, this turnover point is still clearly evident in a small fraction of the luminosity for all AGN groups, furthermore the same significantly correlated $\Gamma_{\text{PL}}$ -- Eddington fraction profile is evident in the individual observations of 1H 0707-495 and IRAS 13224-3809. We investigated the JED-SAD framework and two-phase accretion scenario as reported in GBHBs, where the $\Gamma$ turnover at $L_\text{Bol}/L_\text{Edd}\sim0.02$ has been found \citep[e.g.][]{2021arXiv210913592M,2015MNRAS.447.1692Y,2011MNRAS.417..280S}. However, this $\Gamma$  evolution cannot be clearly seen in the reverberating AGN samples. In contrast, we find hints of the $\Gamma$ turnover at $L_{(2-10\text{keV})}/L_\text{Edd}\sim0.02$, the origin of which remains unclear. We suggest that this turnover point could be consequential of accretion rate changes and/or modification of the inner geometry of the accretion disc and corona.

\section{Conclusions}

We have presented the time lag estimates as a function of frequency for 20 AGN in a variety of flux states and recovered the lag amplitude and black hole mass scaling relationship first discovered by DM13. In addition, these lags are positively correlated to the luminosity. We have shown that modelling the same varied flux states as used in the lag analysis can provided a good description of the observed data using both reflection and absorption modelling.
There is a very strong correlation between the disc reflection and power law flux and a moderate correlation between the covering fraction and $\Gamma_{\text{PL}}$ for all AGN. We also find significant correlations with the covering fraction and flux in IRAS 13224-3809 in particular, suggesting that orbiting clouds may play a significant role in the observed spectral variability without changing the spectral profile. Whilst this phenomenon is not strongly evident across the sample, there are hints this could help explain the variability of AGN where the spectral shape remains constant and modulated by non-uniform orbiting clouds.

We also find strong correlations between the reflection fraction RF and $\Gamma_{\text{PL}}$ and also the reflected flux which may be linked to the changing disc-corona geometry. However, we acknowledge that different components may be required to achieve global descriptions of observed AGN variability and the dynamics of AGN are best studied using long duration individual (single-orbit) observations for each source. 

Finally we have confirmed all grouped correlations are evident across the AGN sample via independent modelling and retrieved hints of a $\Gamma$ evolutionary turnover point in the AGN sample. We suggest this turnover point could be due to changes in the accretion rate and/or modification of the inner geometry of the accretion disc and corona.

\section*{Acknowledgements}
This research has made use of software provided
by the High Energy Astrophysics Science Archive Research Center (HEASARC) and has also used the NASA/IPAC Extragalactic Database (NED). This work was carried out using the computational facilities of the Advanced Computing Research Centre, University of Bristol -- \url{http://www.bristol.ac.uk/acrc/}. SH would like to thank the STFC for funding and the Bristol, Cardiff \& Swansea CDT Team for support. PC acknowledges funding support from (i) Suranaree University of Technology (SUT), (ii) Thailand Science Research and Innovation (TSRI), and (iii) National Science Research and Innovation Fund (NSRF), project no. 160355. We wish to thank the anonymous reviewer for comments and suggestions which has improved the quality of this manuscript.

\section*{Data availability}
The data underlying this article were accessed from {\it XMM-Newton} Observatory (\url{http://nxsa.esac.esa.int}). The derived data generated in this research can be downloaded via \url{http://www.star.bris.ac.uk/steff/hancock.html}. This data is also available via Github pages (in development) at \url{https://github.com/Steff075}.

\bibliographystyle{mnras}
\bibliography{ecm}

\bsp
\label{lastpage}

\appendix

\onecolumn
\section{AGN basic data}
\begin{table*}
\centering
\caption[AGN sample basic data]{AGN selected for this study listing the redshift $z$, Eddington luminosity $L_\text{Edd}$, the bolometric luminosity $L_\text{bol}$, Eddington ratio $\lambda_{\text{Edd}}$, the black hole mass $M$ and the Luminosity distance $D_L$. The numbers in brackets indicate the references for each item where: (1) \cite{2010MNRAS.401.2419Z}; (2) \cite{2016MNRAS.462..511K}; (3) \cite{2003MNRAS.343..164B}; (4) \cite{1994MNRAS.271..958B}; (5) \cite{2012A&A...542A..83P}; (6) \cite{2013MNRAS.429.2917F}; (7) \cite{2012A&A...544A..80G}; (8) \cite{2004MNRAS.348.1415V}; (9) \cite{NED2019}; (10) \cite{2015MNRAS.452..333C}; (11) \cite{2015PASP..127...67B}; (12) \cite{2009ApJ...705..199B}; (13) \cite{2011A&A...535A.113C}; (14) \cite{2002ApJ...579..530W}; (15) \cite{1996AJ....111..696K}; (16) \cite{2017MNRAS.466.3259M}; (17) \cite{2012MNRAS.422..129Z}; (18) \cite{2013ApJS..207...19B}; (19) \cite{2005MNRAS.356..524V}; (20) \cite{2013ApJ...767..121Z}; (21) \cite{1994A&A...288..425S}; (22) \cite{2003A&A...403..481B}; (23) \cite{2006MNRAS.372.1275P}; (24) \cite{2008ApJ...678..693M}; (25) \cite{chainakun_investigating_2017}; (26) \cite{2016A&A...590A..77L}; (27) \cite{2012MNRAS.420.1848D}; (28) \cite{2014A&A...562A..44G}; (29) \cite{2020NatAs...4..597A}; (30) \cite{2002ApJ...571..733V}; (31) \cite{Bechtold_2002}.}
\begin{adjustbox}{max width=\textwidth}
\begin{tabular}{lcccccc}\hline 
Source & Redshift  & $L_\text{Edd}$  & $L_\text{bol}$ &  $\lambda_{\text{Edd}}$ & $\log M$ & $D_L$ (9)\\
& $(z)$ & (erg s$^{-1}$) & (erg s$^{-1}$) &  & ($M_{\odot}$) & pc\\
\hline 
1H 0707-495 & 0.0411(1) &  $2.57 \times 10^{44}$ & $2.69 \times 10^{44}$(2) & 1.05 & $6.31\pm0.50$(3) & $1.74 \times 10^{8}$\\  
Ark~564 & 0.024(4)  & $2.35 \times 10^{44}$ & $2.29 \times 10^{44}$(2) & 0.976 & $6.27\pm0.50$(5) & $9.85 \times 10^{7}$ \\ 
IRAS 13224-3809  & 0.0406(6) & $7.95 \times 10^{44}$ & $5.50 \times 10^{45}$(2) & 6.91 & $6.28\pm0.20$(29) & $2.88\times 10^{8}$\\ 
MCG-6-30-15  &  0.007749(8) & $2.51 \times 10^{44}$ & $1.20 \times 10^{44}$(9) & 0.478 & $6.30^{+0.16}_{-0.24}$(11) & $3.58 \times 10^{7}$\\ 
Mrk~335 & 0.0285(10) & $2.14 \times 10^{45}$ & $1.26 \times 10^{45}$(2) & 0.588 & $7.23\pm0.04$(11) & $1.03 \times 10^{8}$ \\ 
Mrk~766 & 0.01293(12) & $8.32 \times 10^{44}$ & $1.95 \times 10^{44}$(28) & 0.233 &  $6.822^{+0.05}_{-0.06}$(11) & $5.76 \times 10^{4}$\\ 
Mrk~841 & 0.0365(13) & $4.17 \times 10^{46}$ & $6.92 \times 10^{45}$(14) & 0.166 & $7.88\pm0.10$(30) & $1.57 \times 10^{8}$\\ 
NGC~1365 & 0.0045(7) & $ 5.01 \times 10^{45}$ & $ 9.77 \times 10^{43}$(2) & 0.0195 & $7.6\pm0.50$(7) & $2.12\times 10^{7}$\\ 
NGC~3516 & 0.00886(15) & $ 3.13 \times 10^{45}$ & $ 1.95 \times 10^{44}$(14) & 0.0623 &  $7.40^{+0.04}_{-0.06}$(11)& $3.57\times 10^{7}$\\ 
NGC~4051  & 0.0023(16) & $1.70 \times 10^{44}$ & $1.82 \times 10^{43}$(2) & 0.0107 & $5.89^{+0.08}_{-0.15}$(11)& $1.27\times 10^{7}$\\ 
NGC~4151  & 0.0033(17) & $5.63 \times 10^{45}$ & $1.02 \times 10^{44}$(2) & 0.0182 & $7.56\pm0.05$(11)& $1.71\times 10^{7}$\\ 
NGC~4395  & 0.0011(18)& $3.54 \times 10^{43}$ & $1.51 \times 10^{41}$(19) & 0.00423 & $5.45^{+0.13}_{-0.15}$(11)& $8.03\times 10^{6}$\\
NGC~5548  & 0.01718(12 & $6.58 \times 10^{45}$ & $6.17 \times 10^{44}$(2) & 0.0937 & $7.72\pm0.02$(11)& $7.45\times 10^{7}$\\ 
NGC~6860  & 0.0149(20) & $5.02 \times 10^{45}$ & $5.13 \times 10^{43}$(2) & 0.0102 & $7.6\pm0.50$(11)& $6.07\times 10^{7}$\\ 
NGC~7314 & 0.0048(21)  &  $6.31 \times 10^{44}$ & $9.55 \times 10^{42}$(2) & 0.0151 & $6.7\pm0.50$(21)& $1.54\times 10^{7}$\\ 
NGC~7469 & 0.0164(22) & $1.14 \times 10^{45}$ & $1.26 \times 10^{45}$(9) & 1.11  & $6.96\pm0.05$(11)& $6.27\times 10^{7}$\\ 
PG 1211+143 & 0.0809(23)  & $5.13 \times 10^{45}$ & $1.48 \times 10^{46}$(9) & 2.88  & $7.61\pm0.50$(5)& $3.58\times 10^{8}$\\ 
PG 1244+026 & 0.0482(25) & $2.29 \times 10^{45}$ & $4.17 \times 10^{44}$(2) & 0.182  & $7.26\pm0.50$(24)& $2.10\times 10^{8}$\\
PG 1247+267 & 2.043 (31) & $1.05 \times 10^{47}$ & $2.19 \times 10^{47}$(9) & 2.09 & $8.919\pm0.50$(26)& $1.57\times 10^{10}$\\ 
REJ 1034+396 & 0.04(27) & $5.02 \times 10^{44}$ & $3.31 \times 10^{44}$(2) & 0.660 & $6.18\pm0.50$(27)& $1.84\times 10^{8}$ \\ 
\hline
\end{tabular}
\end{adjustbox}
\label{lit_info}
\end{table*}

\clearpage
\section{AGN sample selection}

\begin{longtable}{cccccc}
\caption[The AGN sample list]{The full sample, listing the source, observation ID, year, exposure time, photon counts and grouping where lc and hc refer to observations containing a count rate $<5$ and $>5~\text{cts s}^{-1}$ respectively.} \\
\hline\\
{\textbf{Source}} & {\textbf{Obs ID}} & {\textbf{Year}} & {\textbf{Exposure [eff] (ks)}} & {\textbf{Total counts}} & {\textbf{Group}}\\
\hline
\label{table:obs_log}
1H0707-495 & 0110890201 & 2000 & 46[41] & 4.22 $\times~10^4$ & med(lc)\\
& 0148010301 & 2002 & 80[76] & 2.66 $\times~10^5$ & hi(hc)\\
& 0506200201 & 2007 & 41[38] & 2.45 $\times~10^4$ & lo(lc)\\
& 0506200301 & & 41[39] & 7.12 $\times~10^4$ & med(hc)\\ 
& 0506200401 & & 43[41] & 1.63 $\times~10^5$ & hi(hc)\\ 
& 0506200501 & & 47[41] & 2.02 $\times~10^5$ & hi(hc)\\ 
& 0511580101 & 2008 & 124[111] & 4.13 $\times~10^5$ & hi(hc) \\ 
& 0511580201 & & 124[93] & 4.54 $\times~10^5$ & hi(hc)\\ 
& 0511580301 & & 123[84] & 4.12 $\times~10^5$ & hi(hc)\\
& 0511580401 & & 122[81] & 2.78 $\times~10^5$ & hi(hc)\\
& 0653510301 & 2010 & 117[112] & 4.06 $\times~10^5$ & hi(hc)\\
& 0653510401 &  &128[118] & 6.58 $\times~10^5$ & hi(hc)\\
& 0653510501 &  &128[93] & 4.16 $\times~10^5$ & hi(hc)\\
& 0653510601 &  &129[105] & 5.50 $\times~10^5$ & hi(hc)\\
& 0554710801 & 2011 & 98[86] & 2.68 $\times~10^4$  & lo(lc)\\
\hline
{Ark~564} & 0006810101 & 2000 & 35[10] & 3.90 $\times ~10^5$ & hi\\ 
& 0206400101 & 2005 & 102[96] & 2.68 $\times~10^6$ & hi\\
& 0670130201 & 2011 & 60[59] & 2.65 $\times~10^6$ & hi\\
& 0670130301 & & 56[55] & 1.39 $\times~10^6$ & hi\\
& 0670130401 & & 64[55] & 1.40 $\times~10^6$ & hi\\
& 0670130501 & & 67[67]	& 2.42 $\times~10^6$ & hi\\
& 0670130601 & & 61[53] & 1.41 $\times~10^6$ & hi\\
& 0670130701 & & 64[41] & 6.42 $\times~10^5$ & lo\\
& 0670130801 & & 58[57] & 1.83 $\times~10^6$ & hi\\
& 0670130901 & & 56[56]	& 2.30 $\times~10^6$ & hi\\       
\hline
{IRAS13224-3809} & 0110890101 & 2002 & 64[61] & 1.01 $\times~10^5$ & med(lc)\\
& 0673580101 & 2011 & 133[49] & 1.09 $\times~10^5$ & med(lc)\\
& 0673580201 & & 132[99] & 1.74 $\times~10^5$ & med(hc)\\
& 0673580301 & & 129[82] & 8.89 $\times~10^4$ & lo(lc)\\
& 0673580401 & & 135[113] & 8.89 $\times~10^4$ & med(hc)\\
& 0780560101 & 2016 & 141[141] & 6.86 $\times~10^4$ & med(hc)\\
& 0780561301 & & 141[127] & 2.81 $\times~10^5$ & med(hc)\\
& 0780561401 & & 141[126] & 2.11 $\times~10^5$ & med(hc)\\
& 0780561501 & & 141[126] & 1.76 $\times~10^5$ & med(hc)\\
& 0780561601 & & 141[137] & 4.08 $\times~10^5$ & med(hc)\\
& 0780561701 & & 141[123] & 2.23 $\times~10^5$ & med(hc)\\
& 0792180101 & & 141[123] & 1.83 $\times~10^5$ & med(hc)\\
& 0792180201 & & 141[129] & 2.51 $\times~10^5$ & med(hc)\\
& 0792180301 & & 141[129] & 1.08 $\times~10^5$ & lo(lc)\\
& 0792180401 & & 141[120] & 5.22 $\times~10^5$ & hi(hc)\\
& 0792180501 & & 138[122] & 2.19 $\times~10^5$ & med(lc)\\
& 0792180601 & & 136[122] & 5.38 $\times~10^5$  & hi(hc)\\
\hline
{MCG-6-30-15} & 0029740101 & 2001 & 89[80] &  1.37 $\times~10^6$ & hi\\  
& 0029740701 & & 129[122] & 2.26 $\times~10^6$ & hi\\  
& 0029740801 & & 130[124] & 2.10 $\times~10^6$ & hi\\  
& 0111570101 & 2000 & 46[43] & 3.96 $\times~10^5$ & lo\\
& 0111570201 & & 66[41] & 5.06 $\times~10^5$ & lo\\
& 0693781201 & 2013 & 134[121] & 2.70 $\times~10^6$ & hi\\  
& 0693781301 & & 134[130] & 1.71 $\times~10^6$ & lo\\  
& 0693781401 & & 49[49] & 4.72 $\times~10^5$ & lo\\
\hline
{Mrk~335} & 0306870101 & 2006 & 133[120] & 1.80 $\times~10^6$ & hi\\  
& 0600540501 & 2009 & 83[80] & 2.73 $\times~10^5$ & lo\\
& 0600540601 & & 132[107] & 2.44 $\times~10^5$ & lo\\
\hline
{Mrk~766} & 0096020101 & 2000 & 59[27] & 2.45 $\times~10^5$ & med\\
& 0109141301 & 2001 & 130[104] & 1.79 $\times~10^6$ & hi\\  
& 0304030101 & 2005 & 96[78] & 2.22 $\times~10^5$ & lo\\
& 0304030301 & & 99[98] & 5.97 $\times~10^5$ & med\\
& 0304030401 & & 99[92] & 7.36 $\times~10^5$ & med\\
& 0304030501 & & 96[73] & 7.31 $\times~10^5$ & med\\
& 0304030601 & & 98[85] & 7.07 $\times~10^5$ & med\\
& 0304030701 & & 34[29] & 2.08 $\times~10^5$ & med\\
\hline
{Mrk~841} & 0070740101 & 2001 & 123[108] & 1.26 $\times~10^5$ & hi\\
& 0070740301 & & 148[122] & 1.41 $\times~10^5$ & hi\\
& 0205340201 & 2005 & 73[43] & 1.62 $\times~10^5$ & lo\\
& 0205340401 & & 30[18] & 8.91 $\times~10^4$ & lo\\
\hline
{NGC~1365} & 0151370101 & 2003 & 19[13] & $8.68\times~10^3$ & lo \\
& 0151370201 & & 11[2] & $1.08 \times~10^3$ & lo \\
& 0151370701 & & 11[8] & $7.61 \times~10^3$ & lo \\
& 0205590301 & 2004 & 60[48] & $7.63 \times~10^4$ & lo\\
& 0205590401 & & 69[33] & $3.17 \times~10^4$ & lo\\
& 0505140201 & 2007 & 129[38] & $2.08 \times~10^4$ & lo\\
& 0505140401 & 2007 & 131[107] & $6.27 \times~10^4$ & lo\\
& 0505140501(1) & 2007 & 131[88] & $5.45 \times~10^4$ & lo\\
& 0505140501(2) & 2007 & 131[53] & $3.12 \times~10^4$ & lo\\
& 0692840201 & 2012 & 139[101] & $1.02 \times~10^5$ & lo\\
& 0692840301 &  & 126[44] & $1.20 \times~10^4$ & hi\\
& 0692840401  & 2013 & 134[87] & $3.46 \times~10^5$ & hi\\
& 0692840501(1) &  & 135[64] & $1.10 \times~10^5$ & lo\\
& 0692840501(2) & & 135[34] & $4.36 \times~10^4$ & lo\\
\hline
{NGC~3516} & 0107460601 & 2001 & 128[114] & 4.34 $\times~10^5$ & lo \\
& 0107460701 & & 130[121] & 2.83 $\times~10^5$ & lo\\
& 0401210401 & 2006 & 52[51] & 8.95 $\times~10^5$ & hi \\
& 0401210501 & & 69[61] & 9.82 $\times~10^5$ & hi \\
& 0401210601 & & 68[62] & 5.39 $\times~10^5$ & med \\ 
& 0401211001 & & 68[58] & 9.07 $\times~10^5$ & hi \\
\hline
{NGC~4051} & 0109141401 & 2001 & 122[106] & 1.90 $\times~10^6$ & hi \\
& 0157560101 & 2002 & 52[42] & 1.71 $\times~10^5$ & lo\\
& 0606320101 & 2009 & 46[45] & 3.21 $\times~10^5$ & lo\\
& 0606320201 & & 46[42] & 4.81 $\times~10^5$ & hi\\
& 0606320301 & & 46[21] & 2.92 $\times~10^5$ & hi\\
& 0606320401 & & 45[18] & 6.25 $\times~10^4$ & hi\\
& 0606321301 & & 33[30] & 4.86 $\times~10^5$ & hi\\
& 0606321401 & & 42[35] & 3.62 $\times~10^5$ & lo\\
& 0606321501 & & 42[36] & 3.72 $\times~10^5$ & hi\\
& 0606321601 & & 42[39] & 7.76 $\times~10^5$ & hi\\
& 0606321701 & & 45[28] & 1.43 $\times~10^5$ & lo\\
& 0606321801 & & 44[40] & 2.99 $\times~10^5$ & lo\\
& 0606321901 & & 45[36] & 1.34 $\times~10^5$ & lo\\
& 0606322001 & & 40[37] & 2.58 $\times~10^5$ & lo\\
& 0606322101 & & 44[24] & 4.89 $\times~10^4$ & lo\\
& 0606322201 & & 44[36] & 1.34 $\times~10^5$ & lo\\
& 0606322301 & & 43[35] & 2.65 $\times~10^5$ & lo\\
\hline
{NGC~4151} & 0112310101 & 2000 & 33[30] & 1.31 $\times~10^5$ & lo\\
& 0112830201 & & 62[57] & 3.07 $\times~10^5$ & lo \\
& 0112830501 & & 23[20] & 1.06 $\times~10^5$ & lo\\
& 0143500101 & 2003 & 19[19] & 2.90 $\times~10^5$ & hi\\
& 0143500201 & & 19[18] & 2.94 $\times~10^5$ & hi\\
& 0143500301 & & 19[19] & 3.75 $\times~10^5$ & hi\\
& 0402660101 & 2006 & 40[40] & 1.56 $\times~10^5$ & lo\\
& 0402660201 & & 53[34] & 2.06 $\times~10^5$ & lo\\
\hline
{NGC~4395} & 0142830101 & 2003 & 113[90] & 9.22 $\times~10^4$ & hi\\
& 0744010101 & 2014 & 54[52] & 1.57 $\times~10^4$ & lo\\
& 0744010201 & & 53[48] & 3.03 $\times~10^4$ & lo\\
\hline
{NGC~5548} & 0089960301 & 2001 & 96[84] &  1.24 $\times~10^6$ & hi\\
& 0720110801 & 2013 & 58[52] & 1.56 $\times~10^5$ & lo\\
& 0720110901 & & 57[55] & 1.51 $\times~10^5$ & lo\\
& 0720111001 & & 57[53] & 1.47 $\times~10^5$ & lo\\
& 0720111101 & & 57[35] & 1.26 $\times~10^5$ & lo\\
& 0720111201 & & 57[56] & 1.88 $\times~10^5$ & lo\\
& 0720111301 & & 57[50] & 1.59 $\times~10^5$ & lo\\
& 0720111401 & & 57[52] & 1.39 $\times~10^5$ & lo\\
& 0720111501 & & 57[53] & 1.35 $\times~10^5$ & lo\\
& 0720111601 & 2014 & 57[56] & 2.03 $\times~10^5$ & lo\\
\hline
{NGC~6860} & 0552170301 & 2009 & 123[117] & 8.36 $\times~10^5$ & --\\
\hline
{NGC~7314} & 0111790101 & 2001 & 45[43] & 2.72 $\times~10^5$ & hi\\
& 0311190101 & 2006 & 84[74] & 3.46 $\times~10^5$ & lo\\
& 0725200101 & 2013 & 140[122] & 1.30 $\times~10^6$ & lo\\
& 0725200301 & & 132[128] & 1.09 $\times~10^6$ & lo\\
\hline
{NGC~7469} & 0112170101 & 2000 & 19[18] & 2.18 $\times~10^5$ & lo\\
& 0112170301 & & 25[23] & 3.46 $\times~10^5$ & hi\\
& 0207090101 & 2004 & 85[85] & 1.30 $\times~10^6$ & hi\\
& 0207090201 & & 79[78] & 1.09 $\times~10^6$ & lo\\
\hline
{PG1211+143} & 0112610101 & 2001 & 56[53] & 1.96 $\times~10^5$ & lo\\
& 0208020101 & 2004 & 60[46] & 1.91 $\times~10^5$ & lo\\
& 0502050101 & 2007 & 65[45] & 3.60 $\times~10^5$ & hi\\
& 0502050201 & & 51[35] & 2.23 $\times~10^5$ & hi\\
& 0745110101 & 2014 & 87[78] & 3.02 $\times~10^5$ & hi\\
& 0745110201 & & 104[98] & 2.61 $\times~10^5$ & lo\\
& 0745110301 & & 102[54] & 2.19 $\times~10^5$ & lo\\
& 0745110401 & & 100[91] & 4.36 $\times~10^5$ & hi\\
& 0745110501 & & 58[55] & 3.35 $\times~10^5$ & hi\\
& 0745110601 & & 95[92] & 5.41 $\times~10^5$ & hi\\
& 0745110701 & & 99[96] & 4.35 $\times~10^5$ & hi\\
\hline
{PG1244+026} & 0675320101 & 2011 & 124[123] & 7.40 $\times~10^5$ & hi\\
& 0744440101 & 2014 & 119[108] & 4.01 $\times~10^5$ & lo\\
& 0744440201 & & 120[92] & 4.23 $\times~10^5$ & lo\\
& 0744440301 & & 122[121] & 5.87 $\times~10^5$ & lo\\
& 0744440401 & & 129[127] & 5.34 $\times~10^5$ & lo\\
& 0744440501 & 2015 & 120[118] & 4.54 $\times~10^5$ & lo\\
\hline
{PG1247+267} & 0143150201 & 2003 & 34[32] & 8.16 $\times~10^3$ & -- \\
\hline
{REJ1034+396} & 0506440101 & 2007 & 93[84] & 5.78 $\times~10^5$ & lo\\
& 0561580201 & 2009 & 70[54] & 2.41 $\times~10^5$ & hi\\
& 0655310101 & 2010 & 52[45] & 1.50 $\times~10^5$ & lo\\
& 0655310201 & & 54[50] & 1.64 $\times~10^5$ & lo\\
\hline
\end{longtable}

\section{The lag-frequency results}
\begin{longtable}{ccccc}
\caption[The AGN sample]{The lag-frequency results for all AGN groups detailing the spectral flux (erg cm$^{-2}$ s$^{-1}$), the maximum soft reverberation lag found and the frequency at which they occur. All fluxes have been corrected for Galactic absorption.} \\
\hline\\ 
{\textbf{Source}} & {\textbf{Group}} & {\textbf{Flux$(0.3 - 10~\text{keV}$)}} & {\textbf{Lag (s)}} & {\textbf{Lag-frequency (Hz)}} \\
\hline

1H0707-495 & Combined &	$3.05 \times 10^{-12}$ & $29.1 \pm 3.6$ & $1.55 \times 10^{-3}$ \\ 
& hi & $3.54 \times10^{-12}$  & $29.0 \pm 3.8$ & $1.55 \times 10^{-3}$  \\ 
& hi-cts\textgreater5 & $3.47 \times 10^{-12}$ & $26.6 \pm 3.1$ & $ 1.61 \times 10^{-3}$ \\
& Med & $1.09 \times10^{-12}$ & $34.0 \pm 11.8$ & $1.61 \times 10^{-3}$ \\
& lo & $4.19 \times 10^{-13}$ & $28.7 \pm 10.5$ & $1.61 \times 10^{-3}$  \\
& lo-cts\textless5 & $5.16 \times 10^{-13}$ & $49.0 \pm 35.4$ & $5.46 \times 10^{-4}$ \\
\hline 
Ark 564 & Combined & $1.17 \times10^{-10}$ & $36.2 \pm 10.5$ & $6.07 \times 10^{-4}$ \\ 
& hi & $1.19 \times 10^{-10}$ & $61.1 \pm 19.3$ & $6.07 \times 10^{-4}$ \\ 
& lo & $2.50 \times 10^{-11}$ & $15.6 \pm 5.6$ & $2.21 \times 10^{-3}$ \\  
\hline 
IRAS 13224-3809 & Combined & $1.85 \times10^{-12}$ & $ 39.3 \pm 9.6$  & $ 5.06 \times 10^{-4}$ \\
& hi & $ 3.50 \times 10^{-12}$ & $43.7 \pm 24.0$ & $ 5.06 \times 10^{-4}$ \\
& hi-cts\textgreater5 & $ 1.80 \times 10^{-12}$ & $45.3 \pm 11.9$ & $ 4.36 \times 10^{-4}$ \\
& med & $1.41 \times 10^{-12}$ & $37.7 \pm 11.5$ & $ 5.06 \times 10^{-4}$ \\
& lo & $5.68 \times 10^{-13}$ & $67.1 \pm 11.2$ & $9.66 \times 10^{-4}$ \\
& lo-cts\textless5 & $ 1.08 \times 10^{-12}$ & $65.4 \pm 21.9$ & $4.36 \times 10^{-4}$ \\
\hline 
MCG-6-30-15 & Combined & $5.68 \times10^{-11}$ & $15.9 \pm 5.9$  & $9.66 \times 10^{-4}$ \\
& hi & $ 6.45 \times 10^{-11}$ & $15.2 \pm 16.1$ & $9.66 \times 10^{-4}$ \\
& lo & $ 4.34 \times 10^{-11}$ & $16.1 \pm 7.1$ & $9.66 \times 10^{-4}$ \\
\hline 
Mrk 335 & Combined & $1.57 \times 10^{-11}$ & $132.7 \pm 36.4$ & $2.65 \times 10^{-4}$ \\
& hi & $2.99 \times 10^{-11}$ & $141.4 \pm 56.3$ & $2.65 \times 10^{-4}$ \\
& lo & $6.48 \times 10^{-12}$ & $24.0 \pm 47.2$ & $5.06 \times 10^{-4}$ \\
\hline 
Mrk 766 & Combined & $3.44 \times 10^{-11}$ & $23.9 \pm 6.7$ & $9.66 \times 10^{-4}$ \\
& hi & $4.03 \times 10^{-11}$ & $35.2 \pm 13.8$ & $9.66 \times 10^{-4}$ \\
& med & $2.15 \times 10^{-11}$ & $10.2 \pm 8.4$ & $9.66 \times 10^{-4}$ \\
& lo & $9.19 \times 10^{-12}$ & $157.6 \pm 98.2$ & $2.65 \times 10^{-4}$ \\
\hline 
Mrk 841 & Combined & $1.73 \times10^{-11}$ & $265.85 \pm 217.5$  & $1.02 \times 10^{-4}$ \\
& hi & $2.54 \times 10^{-11}$ & $212.0 \pm 122.4$ & $4.77 \times 10^{-4}$ \\
& lo & $1.421 \times 10^{-11}$ & $562.8 \pm 121.0$ & $1.02 \times 10^{-4}$ \\ 
\hline 
NGC 1365 & Combined & $9.61 \times10^{-12}$ & $144.2 \pm 113.4$ & $7.27 \times 10^{-5}$ \\
& hi & $2.48 \times 10^{-11}$ & $108.9 \pm 104.3$ & $7.27 \times 10^{-5}$ \\
& lo & $6.54\times 10^{-12}$ &  $156.7 \pm 95.1$ & $2.65 \times 10^{-4}$ \\
\hline 
NGC 3516 & Combined & $4.05 \times 10^{-10}$ & $256.6 \pm 144.4$ & $7.27 \times 10^{-5}$ \\
& hi & $5.60 \times 10^{-11}$ & $296.3 \pm 229.6$ & $7.27 \times 10^{-5}$ \\
& med & $4.32 \times 10^{-11}$ & $131.5 \pm 213.2$ & $7.27 \times 10^{-5}$ \\
& lo & $2.25 \times 10^{-11}$ & $143.7 \pm 79.6$ & $2.65 \times 10^{-4}$ \\
\hline 
NGC 4051 & Combined & $2.30 \times 10^{-11}$ & $17.2 \pm 7.1$ & $9.66 \times 10^{-4}$ \\
& hi & $3.21 \times 10^{-11}$ & $17.2 \pm 10.0$ & $ 5.06 \times 10^{-4}$ \\
& lo & $1.50 \times 10^{-11}$ & $19.8 \pm 6.7$ & $9.66 \times 10^{-4}$ \\
\hline 
NGC 4151 & Combined & $9.45 \times10^{-11}$ & $488.0 \pm 278.6$ & $1.39 \times 10^{-4}$ \\
& hi & $2.38 \times 10^{-10}$ & $585.5 \pm 380.8$ & $1.39 \times 10^{-4}$ \\
& lo & $1.15 \times 10^{-11}$ & $41.9 \pm 142.2$ & $2.65 \times 10^{-4}$ \\
\hline 
NGC 4395 & Combined & $6.28 \times10^{-12}$ & $ 23.9 \pm 16.2$ & $5.06 \times 10^{-4}$ \\
& hi & $6.37 \times 10^{-12}$ & $22.4 \pm 17.4$ & $5.06 \times 10^{-4}$ \\
& lo & $6.18 \times 10^{-12}$ & $59.2 \pm 26.7$ & $8.60 \times 10^{-4}$ \\
\hline 
NGC 5548 & Combined & $3.47 \times10^{-11}$ & $156.7 \pm 55.9$ & $2.65 \times 10^{-4}$ \\
& hi & $5.42 \times 10^{-11}$ & $197.3 \pm 98.7$ & $2.65 \times 10^{-4}$\\
& lo & $3.06 \times 10^{-11}$ & $300.4 \pm 240.3$ & $1.39 \times 10^{-4}$ \\
\hline 
NGC 6860 & 2009 & $2.92 \times 10^{-11}$ & $186.7 \pm 192.5$ & $1.94 \times 10^{-4}$ \\
\hline 
NGC 7314 & Combined & $2.71 \times10^{-11}$ & $1.6 \pm 5.8$ & $1.84 \times 10^{-3}$ \\
& hi & $4.88 \times 10^{-11}$ & $104.2 \pm 101.7$ & $2.65 \times 10^{-4}$ \\
& lo & $2.41 \times 10^{-11}$ & $1.2 \pm 2.9$ & $2.95 \times 10^{-3}$ \\
\hline 
NGC 7469 & Combined & $4.42 \times10^{-11}$ & $82.2 \pm 51.1$ & $3.71 \times 10^{-4}$ \\
& hi & $4.45 \times 10^{-11}$ & $292.3 \pm 80.3$ & $3.71 \times 10^{-4}$ \\
& lo & $4.40 \times 10^{-11}$ & $18.4 \pm 12.1$ & $1.35 \times 10^{-3}$ \\
\hline 
PG 1211+143 & Combined & $5.88 \times 10^{-12}$ & $215.6 \pm 112.7$ & $8.33 \times 10^{-5}$ \\
& hi & $6.32 \times 10^{-12}$ & $162.1 \pm 140.4$ & $8.33 \times 10^{-5}$ \\
& lo & $4.87 \times 10^{-12}$ & $313.8 \pm 170.7$ & $7.27 \times 10^{-5}$ \\
\hline 
PG 1244+026 & Combined & $6.17 \times10^{-12}$ & $54.5 \pm 20.3$ & $5.06 \times 10^{-4}$ \\
& hi & $7.88 \times 10^{-12}$ & $72.8 \pm 45.2$ & $5.06 \times 10^{-4}$\\
& lo & $5.84 \times 10^{-12}$ & $45.0 \pm 23.0$ & $5.06 \times 10^{-4}$ \\
\hline 
PG 1247+267 & 2003 & $7.62 \times10^{-13}$ & $498.6 \pm 513.2$ & $1.16 \times 10^{-4}$\\
\hline 
REJ 1034+396 & Combined & $2.44 \times 10^{-12}$ & $55.5 \pm 68.5$ & $2.65 \times 10^{-4}$ \\
& hi & $2.11 \times 10^{-12}$ & $3.0 \pm 5.0$ & $3.52 \times 10^{-3}$ \\
& lo & $2.84 \times 10^{-12}$ & $72.9 \pm 72.0$ & $2.65 \times 10^{-4}$ \\
\hline 
\label{table:lag_results}
\end{longtable}

\end{document}


\maketitle

\addcontentsline{toc}{section}{List of Tables}
\thispagestyle{empty}
\listoftables

\addcontentsline{toc}{section}{List of Figures}
\thispagestyle{empty}
\listoffigures

\begin{landscape}
\begin{longtable}{ccccccccc}
\caption[All AGN groups RELXILL spectral fits]{The best spectral fits for AGN groups. Error bars indicate the 90\% confidence interval. The table shows the model flux (2-10 keV erg cm$^{-2}$ s$^{-1}$), photon index $\Gamma$, ionisation parameter $\log\xi$ (erg cm s$^{-1}$), iron abundance $A_\text{Fe}$ (solar), reflection fraction $RF$, disk inclination $i$ (deg) and the covering fraction (if applied).} \\ \hline
\label{spec_results}
\multicolumn{1}{c}{Source} & \multicolumn{1}{c}{Group} & \multicolumn{1}{c}{$F_{2-10}$ kev} & \multicolumn{1}{c}{$\Gamma_\texttt{Relxill}$} & \multicolumn{1}{c}{$\log \xi$} & \multicolumn{1}{c}{$AF_e$} & \multicolumn{1}{c}{RF} & \multicolumn{1}{c}{$\textit{i}$} & \multicolumn{1}{c}{Cvr Frac} \\ \hline 
\endfirsthead

\multicolumn{9}{c}%
{{\tablename\ \thetable{} -- continued from previous page}} \\
\hline \multicolumn{1}{c}{Source} & \multicolumn{1}{c}{Group} & \multicolumn{1}{c}{$F_{2-10}$ kev} & \multicolumn{1}{c}{$\Gamma_\texttt{Relxill}$} & \multicolumn{1}{c}{$\log \xi$} & \multicolumn{1}{c}{$AF_e$} & \multicolumn{1}{c}{RF} & \multicolumn{1}{c}{$\textit{i}$} & \multicolumn{1}{c}{Cvr Frac} \\ \hline 
\endhead

\hline \multicolumn{9}{r}{{Continued on next page}} \\ 
\endfoot

\hline \hline
\endlastfoot

\hline
1H0707-495 & Combined & 9.27 $\times 10^{-13}$ & $3.38^{+0.02}_{-0.02}$ &  $2.43^{+0.03}_{-0.03}$ & $0.50^{+0.03}_{-0.00}$ & $2.14^{+0.15}_{-0.15}$ & $74.90^{+0.99}_{-1.52}$ & $0.65^{+0.02}_{-0.02}$\\
& hi & 1.04 $\times 10^{-12}$ & $3.40^{+0.00}_{-0.02}$ & $2.44^{+0.08}_{-0.03}$ & $0.50^{+0.04}_{-0.00}$ & $2.13^{+0.15}_{-0.14}$ & $76.68^{+0.69}_{-1.04}$ & $0.63^{+0.01}_{-0.01}$\\
& hi cts s$^{-1}$ & 1.02 $\times 10^{-12}$ & $3.40^{+0.00}_{-0.02}$ & $2.56^{+0.08}_{-0.08}$ & $0.50^{+0.10}_{-0.00}$ & $2.16^{+0.15}_{-0.14}$ & $80.00^{+0.00}_{-0.70}$ & $0.65^{+0.02}_{-0.02}$\\
& med & 5.58 $\times 10^{-13}$ & $2.80^{+0.14}_{-0.34}$  & $3.58^{+0.16}_{-0.32}$ & $10.00^{+0.00}_{-1.64}$ & $4.98^{+3.30}_{-2.70}$ & $76.10^{+1.09}_{-0.57}$ & $0.83^{+0.03}_{-0.16}$\\
& lo & 2.99 $\times 10^{-13}$ & $2.73^{+0.12}_{-0.16}$ & $2.85^{+0.18}_{-0.15}$ & $9.59^{+0.41}_{-0.49}$ & $9.95^{+0.05}_{-4.98}$ & $80.00^{+0.00}_{-2.99}$ & $0.95^{+0.00}_{-0.01}$\\ 
& lo cts s$^{-1}$ & 3.28 $\times 10^{-13}$ & $2.95^{+0.35}_{-0.29}$ & $0.87^{+0.57}_{-0.29}$ & $10.00^{+0.00}_{-6.01}$ & $7.31^{+2.69}_{-3.85}$ & $80.00^{+0.00}_{-3.18}$ & $0.63^{+0.16}_{-0.00}$\\ \hline

Ark 564 & Combined & 1.87 $\times 10^{-11}$ & $2.36^{+0.06}_{-0.03}$ & $1.30^{+0.40}_{-0.25}$ & $4.99^{+5.01}_{-2.02}$ & $0.64^{+0.44}_{-0.26}$ & $75.60^{+1.81}_{-1.76}$ & -- \\ 
& hi & 1.92 $\times 10^{-11}$ & $2.31^{+0.01}_{-0.04}$  & $3.65^{+0.12}_{-0.29}$ & $10.00^{+0.00}_{-2.61}$ & $0.05^{+0.24}_{-0.01}$ & $22.10^{+22.40}_{-21.40}$ & -- \\
& lo & 1.09$\times 10^{-11}$ & $2.25^{+0.09}_{-0.09}$  & $3.48^{+0.20}_{-1.37}$ & $0.50^{+0.41}_{-0.00}$ & $8.69^{+1.31}_{-7.09}$ & $80.00^{+0.00}_{-12.40}$ & -- \\ \hline 

IRAS 13224-3809 & Combined & 6.36 $\times 10^{-13}$ & $3.25^{+0.04}_{-0.02}$ &  $2.13^{+0.06}_{-0.06}$ & $0.50^{+0.11}_{-0.00}$ & $3.20^{+0.36}_{-0.37}$ & $77.51^{+1.39}_{-1.18}$ & $0.60^{+0.02}_{-0.02}$ \\ 
& hi & 1.23 $\times 10^{-12}$ &  $3.17^{+0.09}_{-0.08}$ &  $1.41^{+0.39}_{-0.01}$ & $2.79^{+1.40}_{-2.02}$ & $2.16^{+0.41}_{-0.24}$ & $77.82^{+2.18}_{-2.19}$ & $0.29^{+0.15}_{-0.09}$ \\ 
& hi cts s$^{-1}$ & 6.45 $\times 10^{-13}$ &  $3.13^{+0.09}_{-0.08}$ &  $0.92^{+0.17}_{-0.04}$ & $0.50^{+0.24}_{-0.00}$ & $4.02^{+0.99}_{-1.02}$ & $77.29^{+1.48}_{-1.64}$ & $0.22^{+0.07}_{-0.05}$ \\ 
& med & 5.30 $\times 10^{-13}$ &  $3.27^{+0.00}_{-0.00}$ & $2.12^{+0.01}_{-0.01}$ & $0.86^{+0.01}_{-0.01}$ & $3.09^{+0.22}_{-0.21}$ & $72.30^{+0.35}_{-0.40}$ & $0.63^{+0.00}_{-0.00}$ \\ 
& lo & 2.57 $\times 10^{-13}$ &  $3.14^{+0.06}_{-0.21}$ & $1.87^{+0.16}_{-0.13}$ & $0.50^{+0.71}_{-0.00}$ & $9.24^{+0.76}_{-4.06}$ & $65.46^{+3.35}_{-2.32}$ & $0.74^{+0.03}_{-0.11}$ \\
& lo cts s$^{-1}$ & 4.21 $\times 10^{-13}$ &  $3.20^{+0.00}_{-0.00}$ &  $1.83^{+0.06}_{-0.08}$ & $0.50^{+0.01}_{-0.00}$ & $2.92^{+0.37}_{-0.38}$ & $74.68^{+1.09}_{-1.41}$ & $0.64^{+0.00}_{-0.00}$ \\ \hline

MCG-6-30-15 & Combined & 4.12 $\times 10^{-11}$  & $2.00^{+0.02}_{-0.12}$ & $3.01^{+0.02}_{-0.12}$ & $10.00^{+0.00}_{-0.36}$ & $10.00^{+0.00}_{-4.99}$ & $32.16^{+2.79}_{-2.42}$ & --\\ 
& hi & 4.64 $\times 10^{-13}$ & $2.00^{+0.13}_{-0.10}$ & $3.01^{+0.03}_{-0.11}$ & $10.00^{+0.00}_{-0.52}$ & $10.00^{+0.00}_{-0.54}$ & $32.06^{+2.47}_{-2.21}$ & --\\
& lo & 3.24 $\times 10^{-11}$ & $1.78^{+0.06}_{-0.08}$ & $2.82^{+0.18}_{-0.29}$ & $10.00^{+0.00}_{-5.54}$ & $0.77^{+0.52}_{-0.31}$ & $41.98^{+5.97}_{-5.66}$ & --\\  \hline 

Mrk 335 & Combined  & 9.58 $\times 10^{-12}$ & $2.82^{+0.28}_{-0.20}$ & $0.82^{+0.25}_{-0.39}$ & $0.78^{+1.47}_{-0.23}$ & $10.00^{+0.00}_{-4.38}$ & $69.62^{+8.97}_{-26.89}$	& --\\
& hi/2006  & 1.67 $\times 10^{-11}$ & $1.00^{+0.05}_{-0.05}$ & $3.10^{+0.04}_{-0.01}$ & $5.00^{+0.99}_{-0.70}$ & $10.00^{+0.00}_{-3.83}$ & $13.02^{+4.47}_{-8.02}$ & --\\ 
& lo/2009  & 4.83 $\times 10^{-12}$ & $1.75^{+0.23}_{-0.12}$ &  $0.10^{+3.13}_{-0.10}$ & $3.70^{+6.30}_{-2.50}$ & $1.13^{+4.64}_{-0.74}$ & $32.38^{+8.79}_{-27.38}$ & --\\  \hline

Mrk 766 & Combined & 1.53 $\times 10^{-11}$ & $1.89^{+0.01}_{-0.01}$ & $3.40^{+0.56}_{-0.04}$ & $0.50^{+0.02}_{-0.00}$ & $4.33^{+0.41}_{-0.08}$ & $29.92^{+6.55}_{-1.25}$ &  $0.95^{+0.00}_{-0.13}$\\ 
& hi & 2.46 $\times 10^{-11}$ & $2.04^{+0.04}_{-0.03}$ & $3.23^{+0.09}_{-0.14}$ & $0.56^{+0.12}_{-0.06}$ & $2.04^{+2.11}_{-0.76}$ & $41.97^{+4.00}_{-3.98}$ & -- \\ 
& med & 1.43 $\times 10^{-11}$ & $1.97^{+0.00}_{-0.00}$ & $3.00^{+0.01}_{-0.06}$ & $10.00^{+0.00}_{-0.35}$ & $0.34^{+0.04}_{-0.09}$ & $38.27^{+1.81}_{-2.44}$ & -- \\ 
& lo & 7.32 $\times 10^{-12}$ & $1.53^{+0.12}_{-0.09}$ & $3.00^{+0.05}_{-0.15}$ & $2.46^{+2.47}_{-0.71}$ & $2.55^{+7.45}_{-1.18}$ & $19.42^{+10.99}_{-14.42}$ &  $0.39^{+0.56}_{-0.08}$ \\ \hline 

Mrk 841 & Combined & 1.22 $\times 10^{-11}$ & $2.00^{+0.49}_{-0.35}$ & $3.00^{+0.29}_{-0.15}$ & $9.61^{+0.39}_{-5.53}$ & $10.00^{+0.00}_{-5.00}$ & $69.04^{+2.26}_{-17.38}$ & --\\
& hi/2001 & 1.55 $\times 10^{-11}$ & $2.49^{+0.36}_{-0.08}$ & $3.70^{+0.62}_{-0.73}$ & $1.92^{+2.18}_{-1.14}$ & $2.94^{+7.06}_{-2.40}$ & $43.26^{+26.34}_{-7.97}$ & --\\ 
& lo/2005  & 1.08 $\times 10^{-11}$ & $1.80^{+0.41}_{-0.31}$ & $3.00^{+0.31}_{-0.16}$ & $10.00^{+0.00}_{-0.00}$ & $9.79^{+0.21}_{-0.00}$ & $68.84^{+3.06}_{-10.90}$ & --\\ \hline   
NGC 1365 & Combined & 8.91 $\times 10^{-12}$ & $1.59^{+0.04}_{-0.14}$ & $3.00^{+0.02}_{-0.00}$ & $0.70^{+0.24}_{-0.16}$ & $10.00^{+0.00}_{-6.50}$ & $5.00^{+1.98}_{-0.00}$ & $0.95^{+0.00}_{-0.44}$\\ 
& hi & 2.16 $\times~10^{-11}$ & $3.14^{+0.00}_{-0.01}$ & $0.41^{+0.04}_{-0.03}$ & $0.50^{+0.53}_{-0.00}$ & $10.00^{+0.00}_{-0.21}$ & $6.24^{+14.37}_{-1.24}$ & $0.93^{+0.00}_{-0.02}$ \\
& lo & 6.03 $\times~10^{-12}$ & $2.28^{+0.12}_{-0.07}$ & $2.27^{+0.04}_{-0.07}$ & $4.48^{+0.37}_{-0.48}$ & $10.00^{+0.00}_{-0.07}$ & $6.76^{+2.83}_{-1.76}$ & $0.94^{+0.00}_{-0.00}$ \\ \hline 

NGC 3516 & Combined & 3.27 $\times 10^{-11}$ & $1.96^{+0.05}_{-0.07}$ & $3.00^{+0.00}_{-0.01}$ & $5.00^{+0.22}_{-0.15}$ & $8.88^{+1.12}_{-0.54}$ & $50.49^{+1.09}_{-0.83}$ & $0.81^{+0.01}_{-0.01}$\\ 
& hi/2006 & 4.53 $\times 10^{-11}$ & $1.00^{+0.04}_{-0.00}$ & $2.75^{+0.04}_{-0.04}$ & $0.75^{+0.08}_{-0.07}$ & $1.75^{+0.25}_{-0.29}$ & $54.27^{+1.30}_{-1.55}$ & $0.95^{+0.00}_{-0.00}$\\ 
& med/2006 & 3.68 $\times 10^{-11}$ & $1.00^{+0.22}_{-0.00}$ & $1.57^{+0.19}_{-0.19}$ & $0.76^{+0.21}_{-0.22}$ & $2.91^{+2.91}_{-0.80}$ & $12.46^{+2.77}_{-7.46}$ & $0.57^{+0.05}_{-0.02}$\\ 
& lo/2001 & 1.99 $\times 10^{-11}$ & $2.00^{+0.07}_{-0.36}$ & $0.00^{+1.15}_{-0.00}$ & $10.00^{+0.00}_{-6.19}$ & $9.90^{+0.10}_{-1.11}$ & $57.79^{+2.14}_{-3.44}$ & $0.69^{+0.02}_{-0.05}$\\  \hline 

NGC 4051 & Combined & 1.64 $\times 10^{-11}$ & $1.84^{+0.05}_{-0.04}$ & $3.05^{+0.01}_{-0.04}$ & $0.50^{+0.06}_{-0.00}$ & $10.00^{+0.00}_{-4.78}$ & $5.05^{+0.16}_{-0.05}$ & $0.39^{+0.01}_{-0.05}$ \\ 
& hi & 2.14 $\times 10^{-11}$ & $2.29^{+0.10}_{-0.09}$ & $3.34^{+0.32}_{-0.17}$ & $0.50^{+0.10}_{-0.00}$ & $10.00^{+0.00}_{-5.39}$ & $15.45^{+3.93}_{-10.44}$ & $0.39^{+0.06}_{-0.07}$ \\ 
& lo & 1.20 $\times 10^{-11}$ & $1.49^{+0.12}_{-0.00}$ & $3.04^{+0.00}_{-0.19}$ & $3.91^{+0.92}_{-0.71}$ & $6.75^{+0.31}_{-2.62}$ & $15.56^{+3.73}_{-6.84}$ & $0.39^{+0.00}_{-0.00}$ \\ \hline

NGC 4151 & Combined & 9.22 $\times 10^{-11}$ & $1.61^{+0.10}_{-0.12}$ & $2.88^{+0.10}_{-0.12}$ & $2.67^{+0.95}_{-1.08}$ & $10.00^{+0.00}_{-4.53}$ & $5.00^{+4.07}_{-0.00}$ & $0.95^{+0.00}_{-0.00}$ \\
& hi & 2.33 $\times 10^{-10}$ & $1.10^{+0.00}_{-0.09}$ & $0.00^{+2.71}_{-0.00}$ & $10.00^{+0.00}_{-6.64}$ & $0.65^{+1.02}_{-0.24}$ & $26.57^{+2.28}_{-4.66}$ & $0.95^{+0.00}_{-0.02}$ \\
& lo & 5.32 $\times 10^{-11}$ & $2.74^{+0.06}_{-0.01}$ & $1.37^{+0.05}_{-0.04}$ & $0.50^{+0.15}_{-0.00}$ & $7.37^{+0.00}_{-0.44}$ & $5.37^{+2.66}_{-0.37}$ & $0.95^{+0.00}_{-0.00}$ \\ \hline

NGC 4395 & Combined & 5.89 $\times 10^{-12}$ & $1.06^{+0.00}_{-0.04}$ & $0.34^{+2.66}_{-0.34}$ & $10.00^{+0.00}_{-0.89}$ & $0.40^{+0.00}_{-0.18}$ & $39.23^{+2.71}_{-3.46}$ & $0.64^{+0.07}_{-0.03}$\\
& hi/2003  & 5.73 $\times 10^{-12}$ & $1.00^{+0.00}_{-0.00}$  & $0.00^{+2.35}_{-0.00}$ & $10.00^{+0.00}_{-7.78}$ & $0.87^{+1.03}_{-0.65}$ & $9.54^{+15.46}_{-4.54}$ & $0.60^{+0.04}_{-0.04}$\\ 
& lo/2014 & 6.12 $\times 10^{-12}$ & $1.26^{+0.16}_{-0.26}$ & $2.39^{+0.59}_{-2.09}$ & $10.00^{+0.00}_{-7.05}$ & $10.00^{+0.00}_{-7.20}$ & $26.66^{+4.63}_{-3.29}$ & $0.95^{+0.00}_{-0.01}$\\ \hline

NGC 5548 & Combined & 3.05 $\times~10^{-11}$ & $1.77^{+0.36}_{-0.36}$ & $0.07^{+1.28}_{-0.07}$ & $10.00^{+0.00}_{-6.19}$ & $6.09^{+3.91}_{-1.29}$ & $67.34^{+2.65}_{-1.60}$ & $0.54^{+0.04}_{-0.06}$ \\
& hi/2001 & 4.00 $\times~10^{-11}$ & $2.80^{+0.30}_{-0.62}$ & $2.15^{+0.70}_{-0.42}$ & $0.50^{+0.62}_{-0.00}$ & $1.31^{+0.94}_{-0.68}$ & $5.00^{+3.89}_{-0.00}$ & $0.48^{+0.19}_{-0.15}$ \\
& lo & 2.84 $\times~10^{-11}$ & $1.60^{+0.20}_{-0.43}$ & $0.70^{+1.16}_{-0.70}$ & $10.00^{+0.00}_{-2.95}$ & $10.00^{+0.00}_{-3.21}$ & $68.81^{+1.47}_{-3.36}$ & $0.84^{+0.00}_{-0.00}$ \\ \hline

NGC 6860 & 2009 & $2.23 \times~10^{-11}$ & $3.20^{+0.20}_{-0.31}$ & $3.99^{+0.35}_{-0.44}$ & $9.72^{+0.28}_{-0.21}$ & $2.19^{+1.68}_{-0.89}$ & $44.47^{+6.12}_{-5.26}$ & $0.89^{+0.03}_{-0.03}$ \\  \hline 

NGC 7314 & Combined & 2.32 $\times~10^{-11}$ & $2.09^{+0.04}_{-0.06}$ & $1.78^{+0.07}_{-0.04}$ & $10.00^{+0.00}_{-1.08}$ & $0.68^{+0.14}_{-0.13}$ & $46.68^{+1.57}_{-1.22}$  & $0.95^{+0.00}_{-0.00}$\\
& hi/2001 & 4.03 $\times~10^{-11}$ & $2.15^{+0.18}_{-0.09}$ & $2.75^{+0.32}_{-0.27}$ & $10.00^{+0.00}_{-7.48}$ & $0.57^{+0.25}_{-0.14}$ & $42.35^{+3.52}_{-1.50}$  & $0.95^{+0.00}_{-0.00}$\\
& lo & 2.07 $\times~10^{-11}$ & $2.06^{+0.00}_{-0.66}$ & $1.78^{+0.09}_{-0.05}$ & $10.00^{+0.00}_{-1.49}$ & $0.64^{+0.16}_{-0.15}$ & $46.88^{+1.67}_{-1.47}$  & $0.95^{+0.00}_{-0.00}$\\ \hline

NGC 7469 & Combined &  $2.95 \times~10^{-11}$ & $2.39^{+0.14}_{-0.29}$ & $3.48^{+0.75}_{-0.40}$ & $9.24^{+0.76}_{-4.16}$ & $0.30^{+0.14}_{-0.08}$ & $45.05^{+6.14}_{-6.29}$ &  $0.47^{+0.14}_{-0.09}$ \\
& hi & $2.91 \times~10^{-11}$ & $2.62^{+0.16}_{-0.27}$ & $3.70^{+0.40}_{-0.65}$ & $10.00^{+0.00}_{-2.27}$ & $0.76^{+0.34}_{-0.27}$ & $74.88^{+2.67}_{-3.16}$ &  $0.55^{+0.04}_{-0.13}$ \\
& lo & $2.97 \times~10^{-11}$ & $2.29^{+0.45}_{-0.34}$ & $3.30^{+0.88}_{-0.38}$ & $8.50^{+1.53}_{-4.51}$ & $0.38^{+0.29}_{-0.17}$ & $44.35^{+5.24}_{-9.80}$ &  $0.55^{+0.35}_{-0.17}$ \\ \hline

PG 1211+143 & Combined &  $3.68 \times~10^{-12}$ & $2.06^{+0.07}_{-0.07}$ &  $2.40^{+0.30}_{-0.21}$ & $2.27^{+0.99}_{-1.04}$ & $2.61^{+2.15}_{-1.18}$ & $15.73^{+5.13}_{-6.58}$ & --  \\ 
& hi &  $3.85 \times~10^{-12}$ & $2.03^{+0.09}_{-0.16}$ &  $2.70^{+0.15}_{-0.35}$ & $2.02^{+1.06}_{-0.87}$ & $2.28^{+2.79}_{-0.00}$ & $6.24^{+0.13}_{-1.24}$ & --  \\ 
& lo &  $3.31 \times~10^{-12}$ & $1.80^{+0.00}_{-0.00}$ & $2.30^{+0.14}_{-0.14}$ & $0.60^{+0.20}_{-0.10}$ & $1.62^{+1.13}_{-0.00}$ & $24.27^{+1.58}_{-1.84}$ & -- \\ \hline 

PG 1244+026 & Combined & $2.62 \times~10^{-12}$ & $1.94^{+0.33}_{-0.27}$ & $1.45^{+0.56}_{-0.40}$ & $0.79^{+0.46}_{-0.24}$ & $7.61^{+2.39}_{-3.29}$ & $79.97^{+0.03}_{-19.29}$ & -- \\
& hi & $2.93 \times~10^{-12}$ & $2.36^{+0.48}_{-0.22}$ & $1.69^{+2.28}_{-1.70}$ & $10.00^{+0.00}_{-9.50}$ & $0.88^{+0.93}_{-0.54}$ & $45.28^{+0.17}_{-2.89}$ & -- \\
& lo & $2.54 \times~10^{-12}$ & $2.05^{+0.19}_{-0.33}$ & $1.43^{+0.58}_{-0.13}$ & $0.69^{+0.31}_{-0.19}$ & $9.54^{+0.46}_{-5.43}$ & $43.99^{+2.83}_{-3.51}$ & -- \\ \hline 

PG 1247+267 & 2003 & $4.22 \times~10^{-13}$ & $2.53^{+0.57}_{-0.28}$ & $0.05^{+0.75}_{-0.05}$ & $4.63^{+5.37}_{-4.13}$ & $10.00^{+0.00}_{-6.41}$ & $5.00^{+35.51}_{-0.00}$ & $0.73^{+0.22}_{-0.33}$ \\ \hline 

REJ 1034+396 & Combined & $1.03 \times~10^{-12}$ & $1.54^{+0.24}_{-0.27}$ & $3.30^{+0.13}_{-0.29}$ & $10.00^{+0.00}_{-3.70}$ & $10.00^{+0.00}_{-4.19}$ & $5.00^{+37.72}_{-0.00}$ & -- \\ 
& hi & $7.20 \times~10^{-13}$ & $2.27^{+0.71}_{-0.41}$ & $2.90^{+0.54}_{-2.90}$ & $6.19^{+3.81}_{-5.69}$ & $1.89^{+8.11}_{-1.30}$ & $80.00^{+0.00}_{-5.99}$ & -- \\ 
& lo & $1.12 \times~10^{-12}$ & $1.74^{+0.17}_{-0.14}$ & $0.46^{+3.08}_{-0.46}$ & $10.00^{+0.00}_{-4.30}$ & $10.00^{+0.00}_{-4.37}$ & $5.00^{+35.88}_{-0.00}$ & -- \\ \hline 
\end{longtable}

\begin{longtable}{cccccccccl}
\caption[1H0707-495 individual orbits RELXILL spectral fits]{The best spectral fits for 1H0707-495 individual orbits. Error bars indicate the 90\% confidence interval. The table shows the model flux (2-10 keV), photon index $\Gamma$, ionisation parameter $\log\xi$, iron abundance $A_\text{Fe}$, reflection fraction $RF$, disk inclination $i$ (deg), covering fraction and $nH$.}\\

\hline \multicolumn{1}{c}{Obs ID} & \multicolumn{1}{c}{$F_{2-10}$ kev} & \multicolumn{1}{c}{$\Gamma_\texttt{Relxill}$} & \multicolumn{1}{c}{$\log \xi$} & \multicolumn{1}{c}{$AF_e$} & \multicolumn{1}{c}{RF} & \multicolumn{1}{c}{$\textit{i}$} & \multicolumn{1}{c}{Cvr Frac} & \multicolumn{1}{c}{$nH$}\\  
\multicolumn{1}{c}{} & \multicolumn{1}{c}{(erg cm$^{-2}$ s$^{-1}$)} & \multicolumn{1}{c}{} & \multicolumn{1}{c}{(erg cm s$^{-1}$)} & \multicolumn{1}{c}{(solar)} & \multicolumn{1}{c}{} & \multicolumn{1}{c}{(Deg)} & \multicolumn{1}{c}{} & \multicolumn{1}{l}{($10^{22}$ cm$^{-2}$)}\\ \hline 
0110890201 & $4.28 \times~10^{-13}$ & $2.64^{+0.01}_{-0.02}$ & $0.76^{+0.31}_{-0.34}$ & $0.50^{+1.94}_{-0.00}$ & $10.00^{+0.00}_{-4.92}$ & $77.95^{+2.04}_{-0.30}$ &  $0.74^{+0.01}_{-0.11}$ & $12.80^{+4.02}_{-5.20}$ \\
0148010301 & $1.12 \times~10^{-12}$ & $3.01^{+0.01}_{-0.01}$ & $0.38^{+2.54}_{-0.14}$ & $0.50^{+1.88}_{-0.00}$ & $4.16^{+1.78}_{-1.63}$ & $80.00^{+0.00}_{-2.41}$ &  $0.80^{+0.15}_{-0.19}$ & $259.70^{+143.00}_{-94.40}$  \\
0506200201 & $2.46 \times~10^{-13}$ & $2.40^{+0.28}_{-0.31}$ & $3.29^{+0.35}_{-0.49}$ & $5.04^{+4.96}_{-2.37}$ & $10.00^{+0.00}_{-8.33}$ & $78.18^{+1.81}_{-64.00}$ &  $0.80^{+0.01}_{-0.38}$ & $4.77^{+1.55}_{-1.56}$\\
0506200301 & $6.87 \times~10^{-13}$ & $2.49^{+0.18}_{-0.16}$ & $1.11^{+1.22}_{-0.01}$ & $1.08^{+4.06}_{-0.58}$ & $3.55^{+2.65}_{-1.05}$ & $65.77^{+6.08}_{-2.55}$ &  $0.12^{+0.64}_{-0.01}$ & $24.51^{+61.00}_{-11.72}$\\ 
0506200401 & $1.07 \times~10^{-12}$ & $3.31^{+0.01}_{-0.01}$ & $1.19^{+0.29}_{-0.01}$ & $0.50^{+2.13}_{-0.00}$ & $3.12^{+0.29}_{-1.50}$ & $67.44^{+11.48}_{-4.13}$ &  $0.37^{+0.12}_{-0.10}$ & $2.01^{+1.03}_{-1.58}$ \\ 
0506200501 & $1.48 \times~10^{-12}$ & $3.16^{+0.13}_{-0.11}$ & $2.06^{+0.27}_{-0.23}$ & $0.50^{+0.91}_{-0.00}$ & $3.38^{+0.00}_{-0.17}$ & $76.74^{+3.25}_{-3.87}$ &  $0.46^{+0.12}_{-0.14}$ & $5.81^{+1.15}_{-1.19}$ \\ 
0511580101 & $1.02 \times~10^{-12}$ & $3.14^{+0.12}_{-0.01}$ & $2.31^{+0.13}_{-0.25}$ & $0.59^{+0.62}_{-0.01}$ & $1.60^{+0.42}_{-0.38}$ & $77.10^{+2.53}_{-3.73}$ &  $0.54^{+0.01}_{-0.01}$ & $5.32^{+0.69}_{-0.91}$   \\ 
0511580201 & $1.46 \times~10^{-12}$ & $3.36^{+0.00}_{-0.01}$ & $2.37^{+0.26}_{-0.21}$ & $0.50^{+0.38}_{-0.00}$ & $1.29^{+0.28}_{-0.24}$ & $79.01^{+0.98}_{-3.31}$ &  $0.53^{+0.01}_{-0.01}$ & $3.74^{+1.20}_{-2.98}$\\ 
0511580301 & $1.06 \times~10^{-12}$ & $3.36^{+0.01}_{-0.01}$ & $2.21^{+0.15}_{-0.21}$ & $0.50^{+0.62}_{-0.00}$ & $2.54^{+1.67}_{-0.69}$ & $72.82^{+3.15}_{-3.27}$ &  $0.52^{+0.01}_{-0.13}$ & $0.29^{+3.91}_{-0.29}$\\
0511580401 & $8.51 \times~10^{-13}$ & $3.40^{+0.00}_{-0.15}$ & $1.92^{+0.17}_{-0.51}$ & $0.50^{+1.85}_{-0.00}$ & $5.42^{+0.80}_{-2.71}$ & $69.29^{+6.91}_{-10.75}$ &  $0.51^{+0.00}_{-0.19}$ & $6.99^{+1.42}_{-1.11}$ \\
0554710801 & $2.79 \times~10^{-13}$ & $2.82^{+0.15}_{-0.22}$ & $1.82^{+0.26}_{-0.45}$ & $0.50^{+0.75}_{-0.00}$ & $10.00^{+0.00}_{-3.47}$ & $80.00^{+0.00}_{-3.91}$ &  $0.93^{+0.01}_{-0.01}$ & $5.47^{+1.79}_{-2.77}$\\
0653510301 & $7.84 \times~10^{-13}$ & $3.40^{+0.00}_{-0.01}$ & $2.17^{+0.13}_{-0.12}$ & $0.50^{+1.01}_{-0.00}$ & $6.39^{+0.36}_{-0.32}$ & $71.59^{+3.03}_{-4.19}$ &  $0.55^{+0.01}_{-0.01}$ & $346.00^{+146.97}_{-31.50}$\\
0653510401 & $1.02 \times~10^{-12}$ & $3.25^{+0.00}_{-0.00}$ & $0.76^{+0.01}_{-0.00}$ & $2.28^{+1.22}_{-1.78}$ & $4.19^{+0.24}_{-0.70}$ & $80.00^{+0.00}_{-0.60}$ &  $0.95^{+0.00}_{-0.07}$ & $6.48^{+0.66}_{-0.49}$\\
0653510501 & $7.58 \times~10^{-13}$ & $3.36^{+0.01}_{-0.01}$ & $2.14^{+0.17}_{-0.34}$ & $0.50^{+0.44}_{-0.00}$ & $3.46^{+1.34}_{-0.86}$ & $73.35^{+2.86}_{-4.62}$ &  $0.95^{+0.01}_{-0.14}$ & $235.22^{+59.95}_{-97.90}$\\
0653510601 & $7.88 \times~10^{-13}$ & $3.30^{+0.00}_{-0.00}$ & $0.74^{+0.00}_{-0.00}$ & $0.50^{+0.49}_{-0.00}$ & $3.35^{+0.69}_{-0.52}$ & $80.00^{+0.00}_{-0.21}$ &  $0.95^{+0.17}_{-0.00}$ & $0.13^{+1.36}_{-1.67}$\\
\end{longtable}

\clearpage

\begin{longtable}{cccccccccl}
\caption[IRAS 13224-3809 individual orbits RELXILL spectral fits]{The best spectral fits for IRAS 13224-3809 individual orbits. Error bars indicate the 90\% confidence interval. The table shows the model flux (2-10 keV), photon index $\Gamma$, ionisation $\log\xi$, iron abundance $A_\text{Fe}$, reflection fraction $RF$, disk inclination $i$ (deg), covering fraction and $n_H$.}\\

\hline \multicolumn{1}{c}{Obs ID} & \multicolumn{1}{c}{$F_{2-10}$ kev} & \multicolumn{1}{c}{$\Gamma_\texttt{Relxill}$} & \multicolumn{1}{c}{$\log \xi$} & \multicolumn{1}{c}{$AF_e$} & \multicolumn{1}{c}{RF} & \multicolumn{1}{c}{$\textit{i}$} & \multicolumn{1}{c}{Cvr Frac} & \multicolumn{1}{c}{$n_H$}\\  
\multicolumn{1}{c}{} & \multicolumn{1}{c}{(erg cm$^{-2}$ s$^{-1}$)} & \multicolumn{1}{c}{} & \multicolumn{1}{c}{(erg cm s$^{-1}$)} & \multicolumn{1}{c}{(solar)} & \multicolumn{1}{c}{} & \multicolumn{1}{c}{(Deg)} & \multicolumn{1}{c}{} & \multicolumn{1}{l}{($10^{22}$ cm$^{-2}$)}\\ \hline 
0110890101 & $4.65 \times~10^{-13}$ & $3.21^{+0.17}_{-0.22}$ & $2.26^{+0.24}_{-0.27}$ & $0.50^{+0.39}_{-0.00}$ & $7.46^{+2.54}_{-3.37}$ & $67.55^{+6.60}_{-3.12}$ &  $0.70^{+0.07}_{-0.18}$ &  $0.97^{+17.50}_{-0.59}$\\
0673580101 & $6.24 \times~10^{-13}$ & $3.23^{+0.12}_{-0.09}$ & $1.93^{+0.17}_{-0.40}$ & $6.93^{+3.07}_{-2.55}$ & $2.85^{+0.65}_{-1.17}$ & $64.92^{+9.72}_{-5.23}$ &  $0.40^{+0.15}_{-0.14}$ &  $0.13^{+9.90}_{-0.13}$\\
0673580201 & $5.07 \times~10^{-13}$ & $3.11^{+0.18}_{-0.16}$ & $1.77^{+0.42}_{-0.38}$ & $0.50^{+2.97}_{-0.00}$ & $3.83^{+1.21}_{-0.76}$ & $80.00^{+0.00}_{-12.69}$ &  $0.60^{+0.15}_{-0.14}$  &  $4.25^{+2.26}_{-1.23}$\\
0673580301 & $2.67 \times~10^{-13}$ & $3.14^{+0.19}_{-0.12}$ & $1.46^{+0.72}_{-0.34}$ & $0.50^{+0.41}_{-0.00}$ & $5.01^{+4.99}_{-1.71}$ & $63.96^{+12.39}_{-4.38}$ &  $0.64^{+0.18}_{-0.11}$  &  $0.32^{+23.23}_{-0.01}$ \\
0673580401 & $5.34 \times~10^{-13}$ & $3.14^{+0.16}_{-0.15}$ & $1.45^{+0.39}_{-0.15}$ & $0.50^{+2.91}_{-0.00}$ & $9.23^{+0.71}_{-3.67}$ & $79.07^{+0.93}_{-9.90}$ &  $0.57^{+0.09}_{-0.01}$ &  $2.74^{+1.35}_{-2.04}$ \\
0780560101 & $3.75 \times~10^{-13}$ & $3.40^{+0.00}_{-0.06}$ & $2.14^{+0.22}_{-0.14}$ & $0.50^{+1.50}_{-0.00}$ & $10.00^{+0.00}_{-3.55}$ & $67.41^{+3.70}_{-2.26}$ &  $0.83^{+0.03}_{-0.03}$ &  $0.36^{+0.05}_{-0.05}$  \\
0780561301 & $5.37 \times~10^{-13}$ & $3.32^{+0.07}_{-0.07}$ & $2.17^{+0.20}_{-0.13}$ & $0.50^{+0.28}_{-0.00}$ & $2.81^{+0.92}_{-0.68}$ & $75.91^{+3.31}_{-2.98}$ &  $0.62^{+0.07}_{-0.04}$ &  $4.60^{+2.02}_{-3.03}$  \\
0780561401 & $6.33 \times~10^{-13}$ & $3.26^{+0.14}_{-0.15}$ & $2.03^{+0.29}_{-0.26}$ & $0.50^{+1.37}_{-0.00}$ & $3.74^{+0.98}_{-0.53}$ & $73.70^{+3.61}_{-3.12}$ &  $0.65^{+0.05}_{-0.13}$ &  $6.77^{+0.05}_{-0.06}$  \\
0780561501 & $3.25 \times~10^{-13}$ & $3.36^{+0.04}_{-0.12}$ & $1.91^{+0.16}_{-0.19}$ & $0.96^{+1.71}_{-0.46}$ & $4.81^{+1.96}_{-1.31}$ & $66.95^{+6.15}_{-4.11}$ &  $0.78^{+0.04}_{-0.05}$ &  $288.20^{+64.3}_{-18.18}$  \\
0780561601 & $7.61 \times~10^{-13}$ & $3.30^{+0.07}_{-0.09}$ & $2.18^{+0.21}_{-0.14}$ & $0.50^{+0.26}_{-0.00}$ & $4.01^{+2.90}_{-1.04}$ & $77.41^{+2.58}_{-2.54}$ &  $0.60^{+0.07}_{-0.07}$ &  $6.38^{+1.44}_{-1.57}$  \\
0780561701 & $3.99 \times~10^{-13}$ & $3.32^{+0.08}_{-0.15}$ & $2.10^{+0.22}_{-0.31}$ & $0.50^{+2.79}_{-0.00}$ & $3.15^{+3.19}_{-0.75}$ & $70.19^{+4.91}_{-3.86}$ &  $0.66^{+0.07}_{-0.15}$ &  $0.03^{+3.19}_{-0.56}$  \\
0792180101 & $3.73 \times~10^{-13}$ & $3.19^{+0.09}_{-0.13}$ & $1.92^{+0.13}_{-0.21}$ & $0.99^{+2.11}_{-0.49}$ & $5.79^{+3.44}_{-1.36}$ & $69.02^{+4.20}_{-2.33}$ &  $0.76^{+0.04}_{-0.07}$ &  $4.31^{+1.40}_{-2.74}$  \\
0792180201 & $4.17 \times~10^{-13}$ & $3.30^{+0.07}_{-0.17}$ & $2.11^{+0.19}_{-0.10}$ & $0.50^{+0.22}_{-0.00}$ & $4.71^{+1.39}_{-1.19}$ & $69.15^{+2.82}_{-1.97}$ &  $0.67^{+0.04}_{-0.12}$ &  $4.46^{+0.68}_{-2.61}$  \\
0792180301 & $2.48 \times~10^{-13}$ & $3.33^{+0.07}_{-0.17}$ & $1.84^{+0.54}_{-0.71}$ & $2.80^{+3.91}_{-2.30}$ & $8.89^{+1.11}_{-4.19}$ & $58.47^{+12.91}_{-9.32}$ &  $0.76^{+0.08}_{-0.07}$ &  $4.86^{+1.99}_{-1.97}$  \\
0792180401 & $1.22 \times~10^{-12}$ & $3.01^{+0.04}_{-0.03}$ & $1.11^{+0.16}_{-0.08}$ & $2.35^{+0.98}_{-1.13}$ & $2.21^{+0.37}_{-0.49}$ & $79.84^{+0.16}_{-2.72}$ &  $0.15^{+0.08}_{-0.09}$ &  $1.50^{+6.63}_{-0.68}$  \\ 
0792180501 & $3.25 \times~10^{-13}$ & $3.16^{+0.15}_{-0.16}$ & $1.43^{+0.39}_{-0.11}$ & $0.50^{+1.04}_{-0.00}$ & $4.81^{+1.96}_{-1.35}$ & $70.06^{+9.06}_{-3.77}$ &  $0.49^{+0.15}_{-0.11}$ &  $6.38^{+1.44}_{-1.62}$  \\
0792180601 & $1.14 \times~10^{-12}$ & $3.30^{+0.08}_{-0.06}$ & $1.48^{+0.49}_{-0.11}$ & $2.92^{+1.63}_{-1.85}$ & $1.88^{+1.45}_{-0.84}$ & $74.39^{+2.52}_{-2.21}$ &  $0.29^{+0.08}_{-0.08}$ &  $2.28^{+2.76}_{-1.57}$  \\
\end{longtable}

\end{landscape}

\begin{figure}
\centering
\includegraphics[scale=0.4]{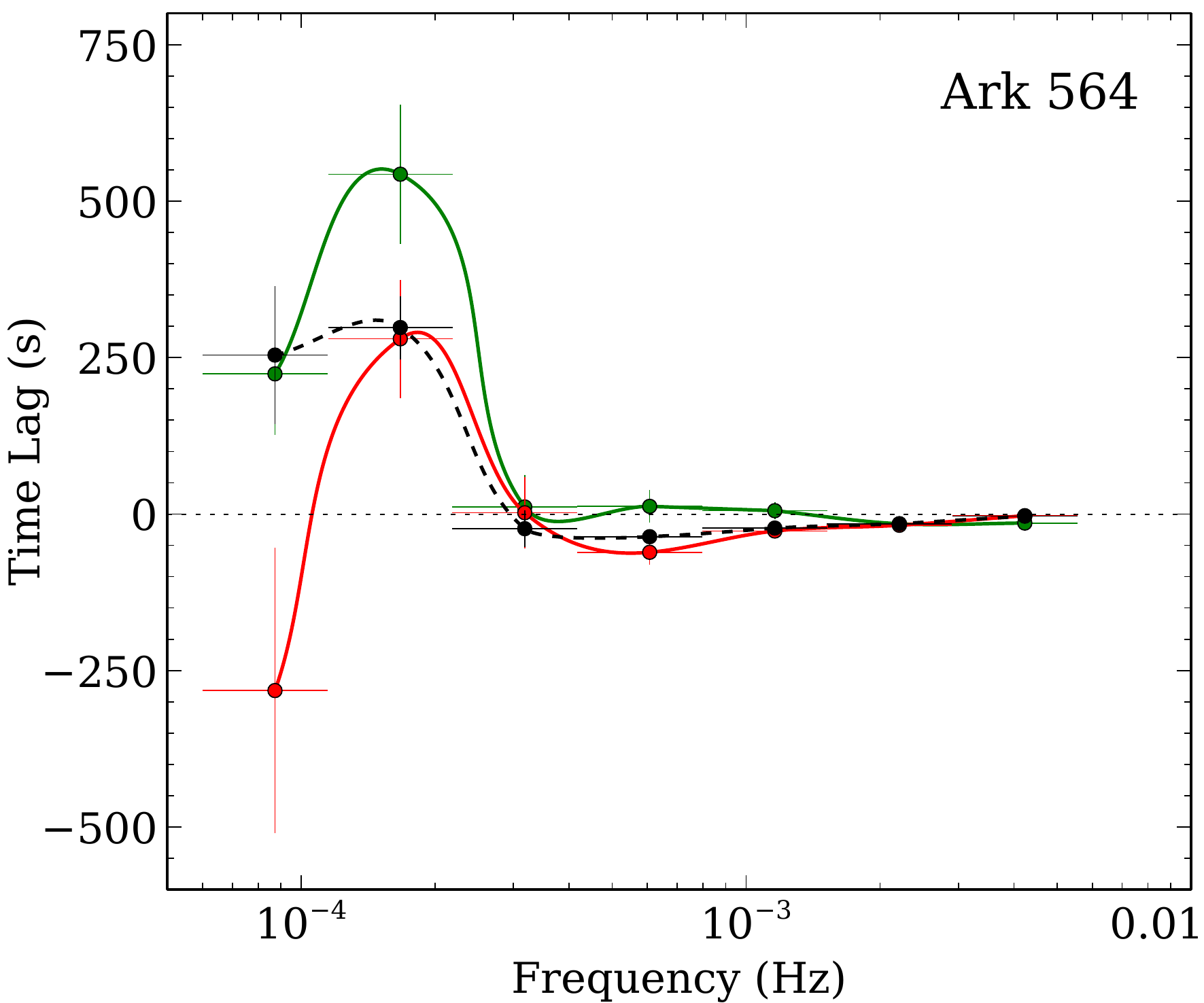}
\includegraphics[scale=0.4]{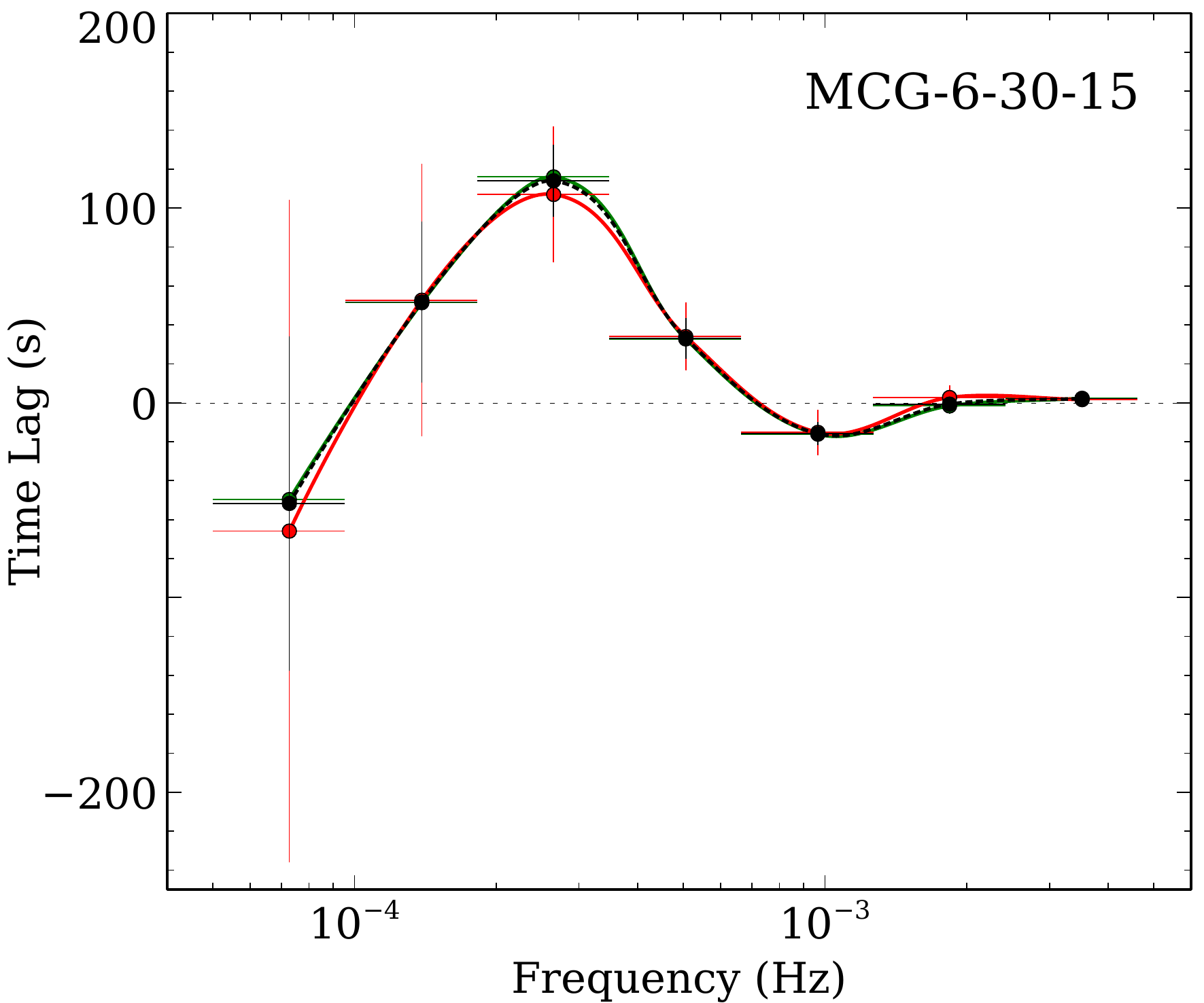}\\
\vspace{0.5cm}
\includegraphics[scale=0.4]{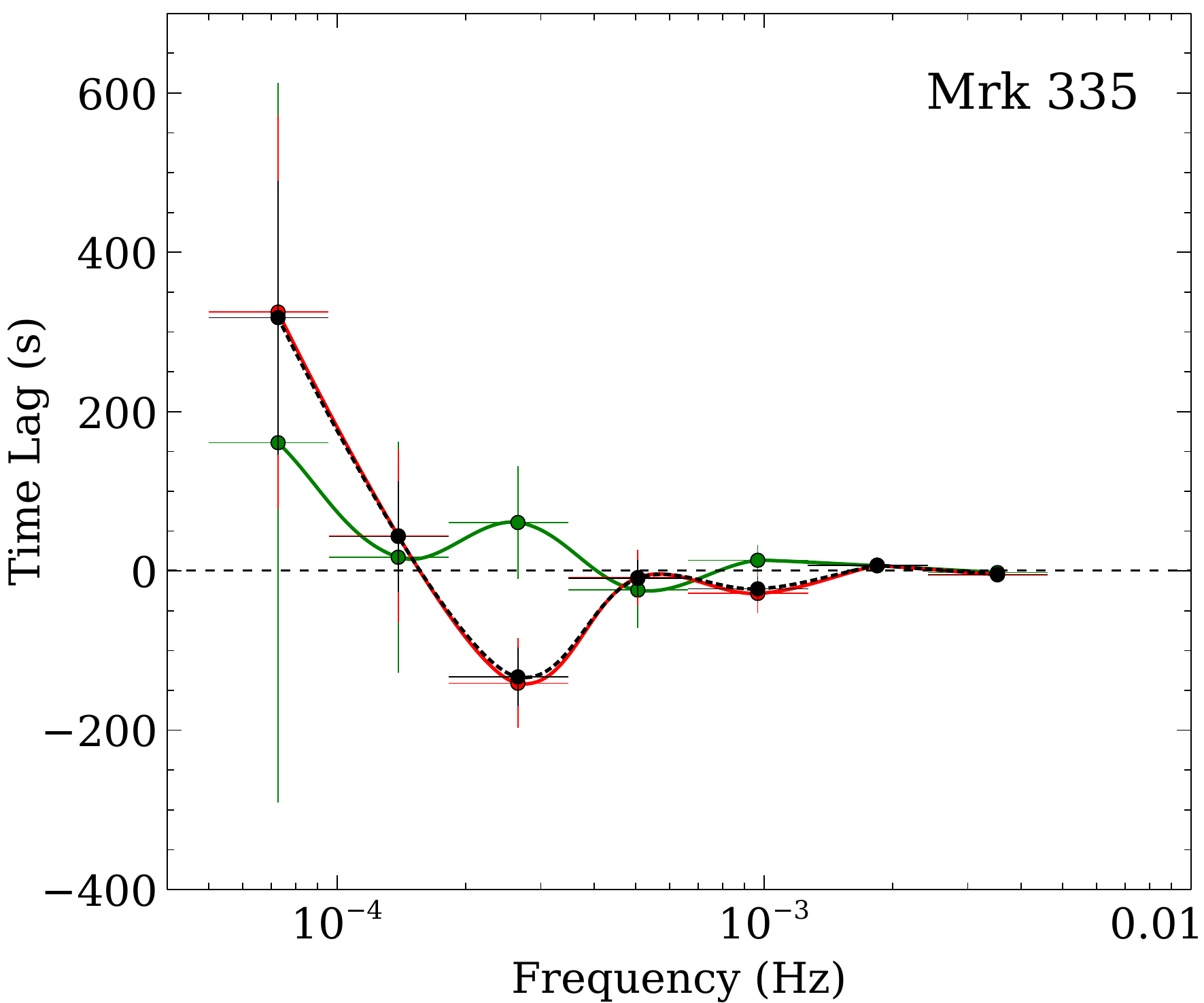}
\includegraphics[scale=0.4]{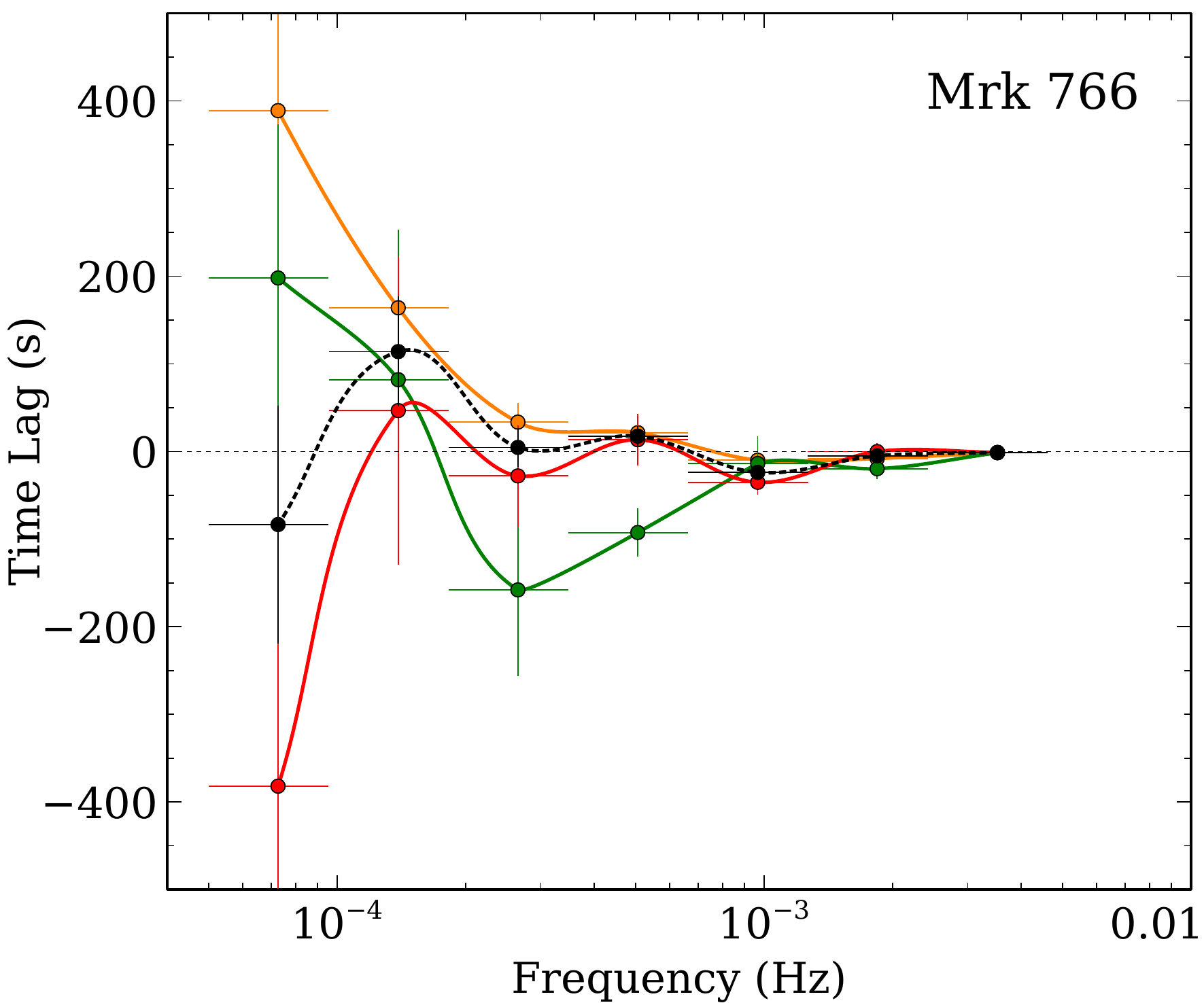}\\
\vspace{0.5cm}
\includegraphics[scale=0.4]{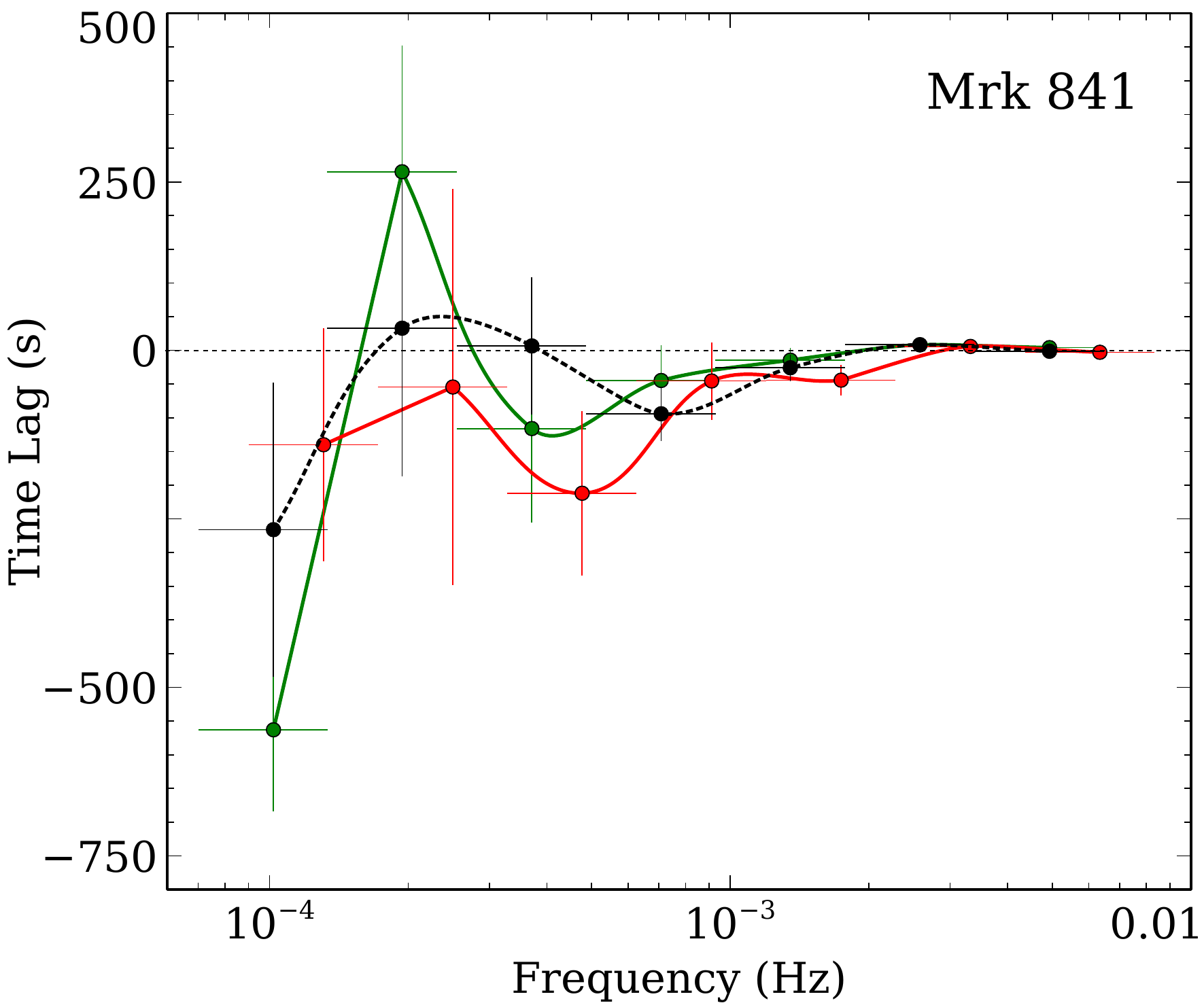}
\includegraphics[scale=0.4]{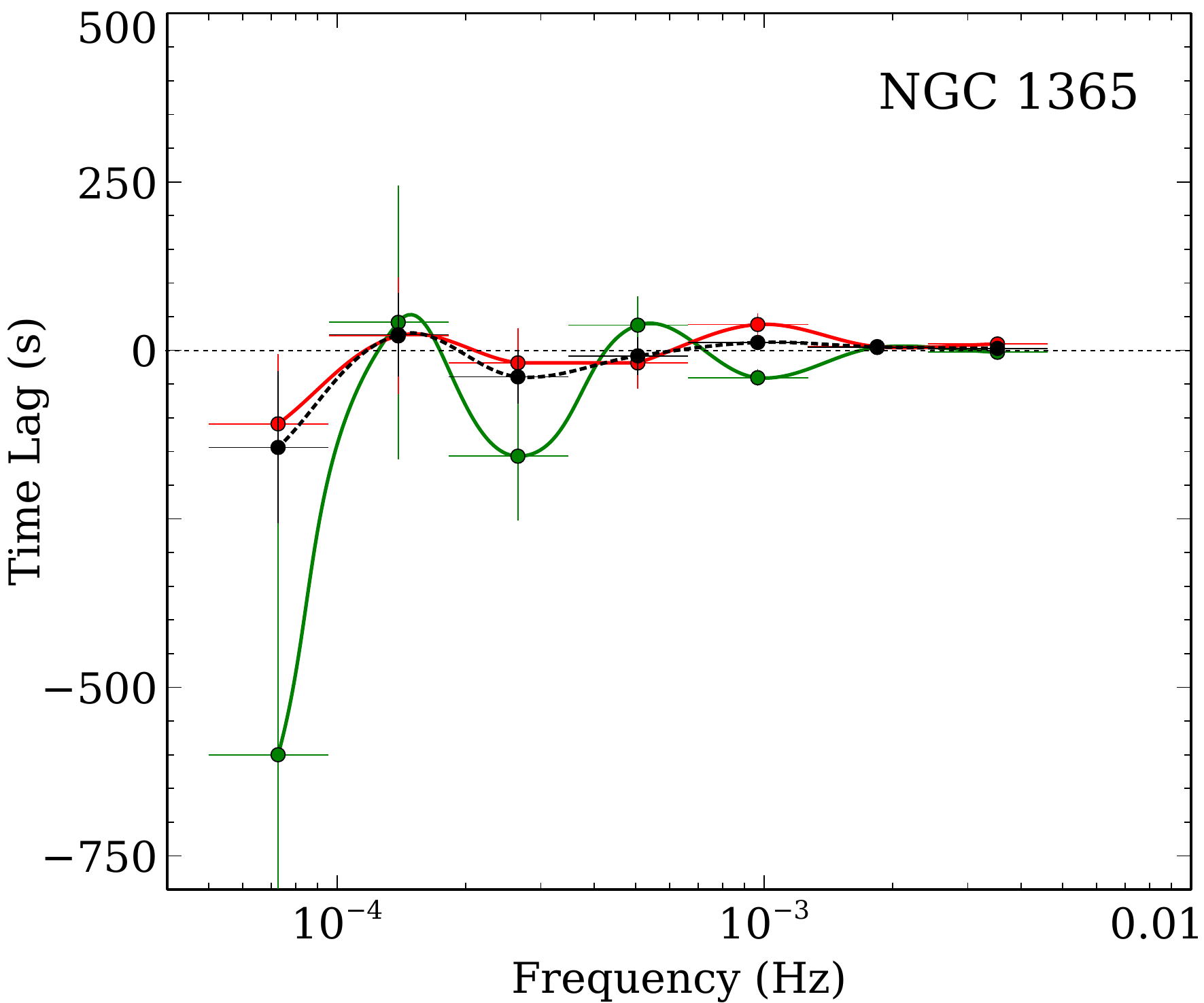}
\caption[The lag-frequency results]{The lag-frequency results for all other AGN in the sample list, showing the combined lags (black dashed lines), high flux (red), medium flux (amber) and low flux lags (green).  }
\label{fig:lagfreq-results}
\end{figure}

\begin{figure}
    \ContinuedFloat
\centering
\includegraphics[scale=0.4]{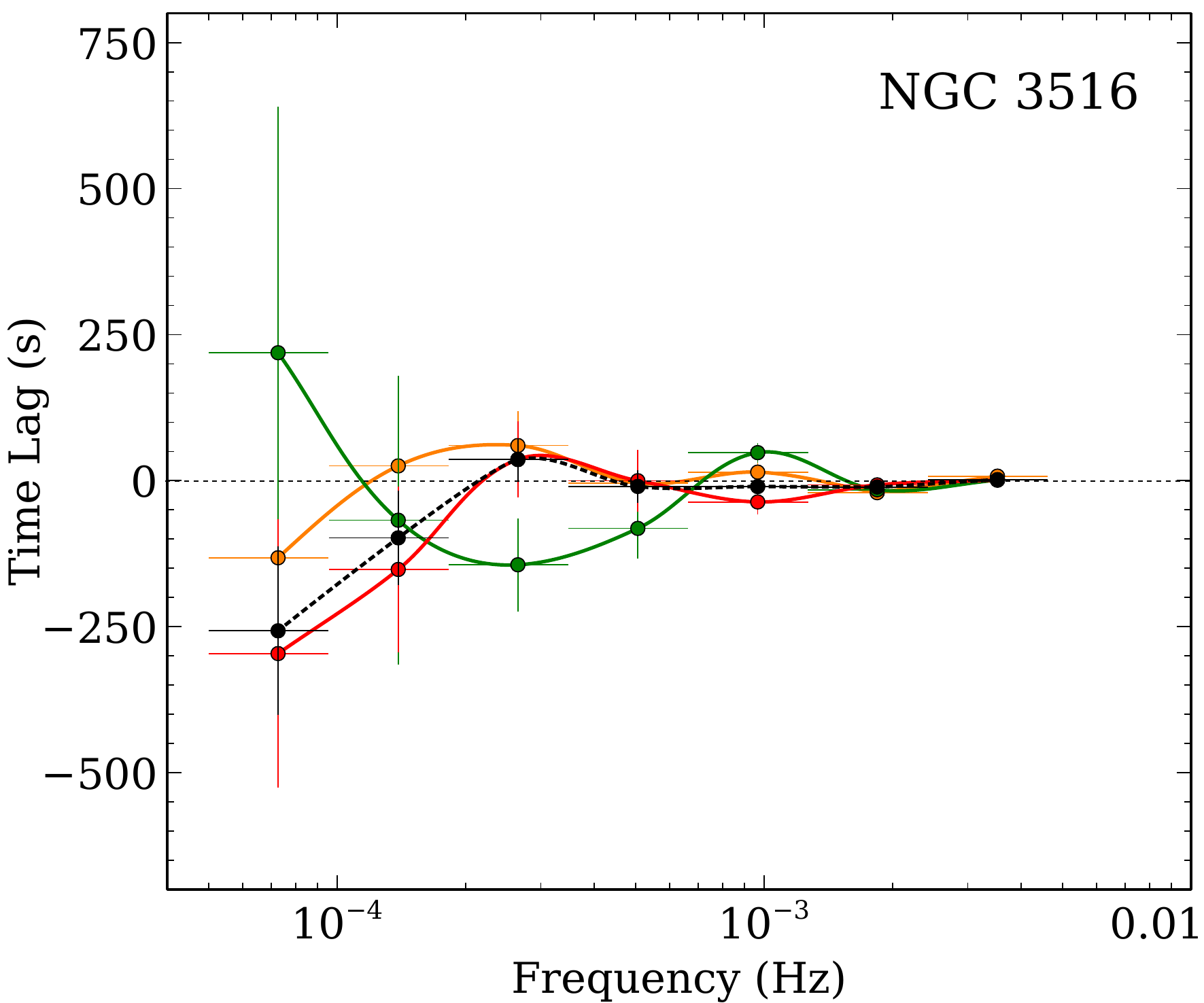}
\includegraphics[scale=0.4]{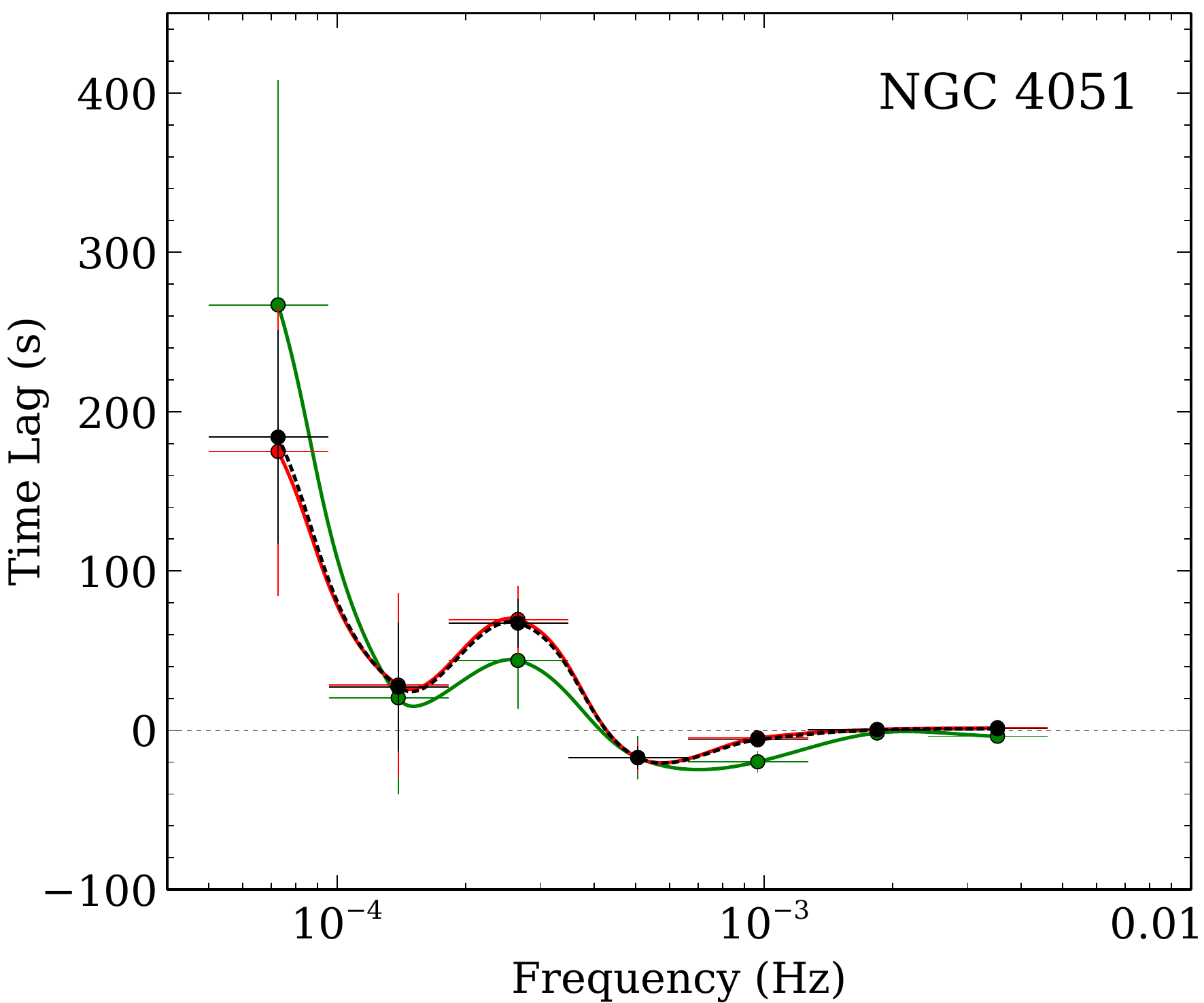}\\
\vspace{0.5cm}
\includegraphics[scale=0.4]{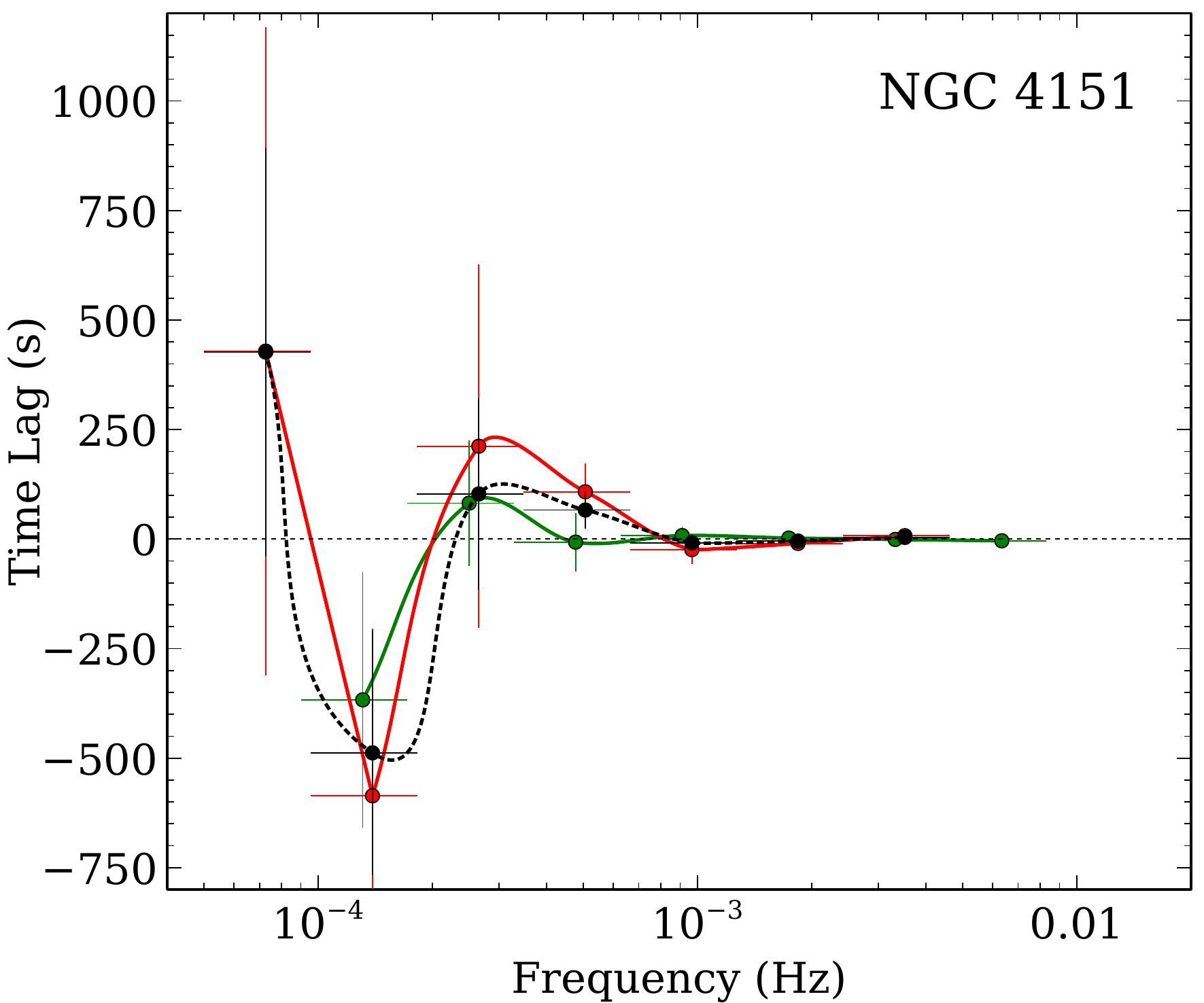}
\includegraphics[scale=0.4]{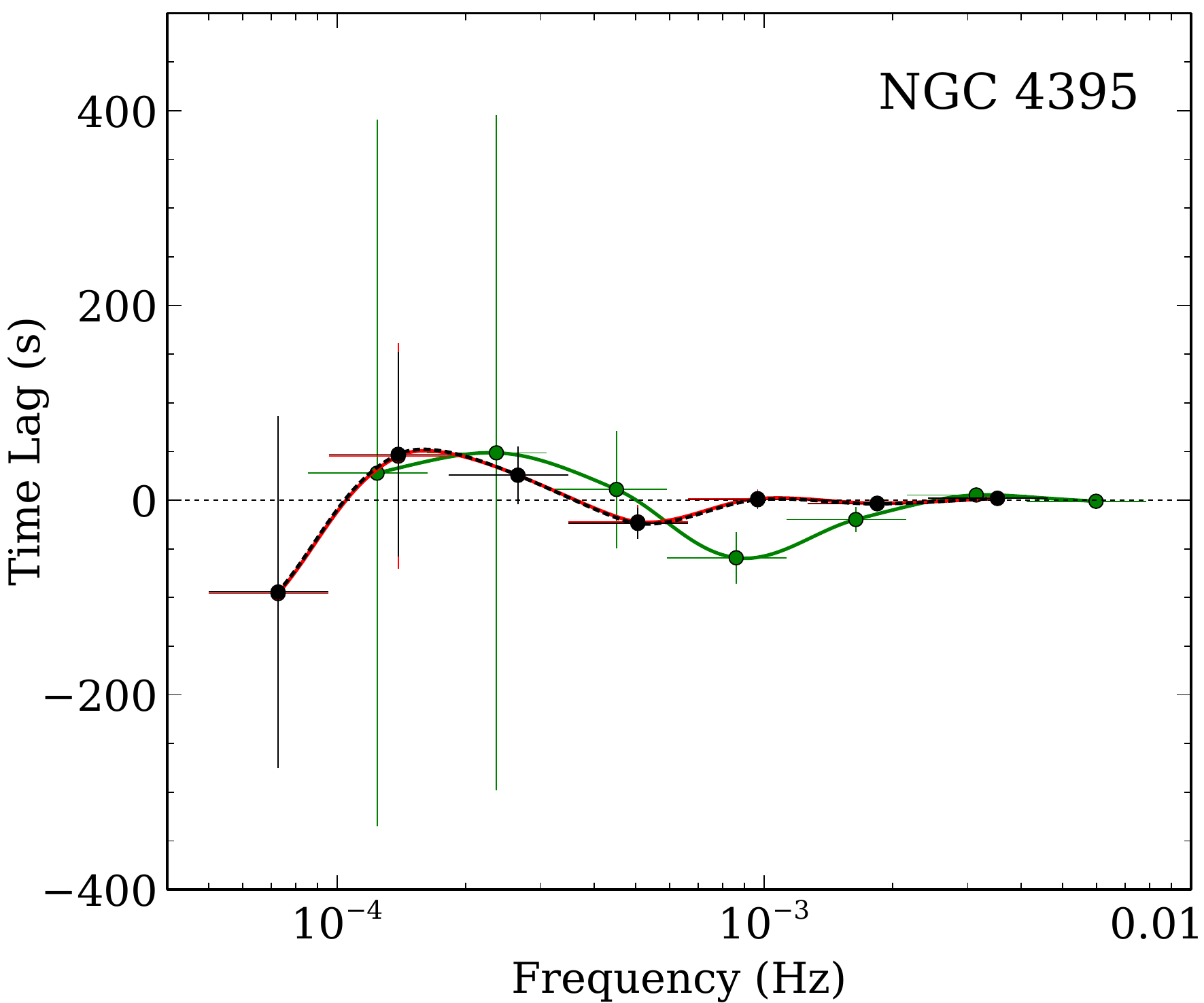}\\
\vspace{0.5cm}
\includegraphics[scale=0.4]{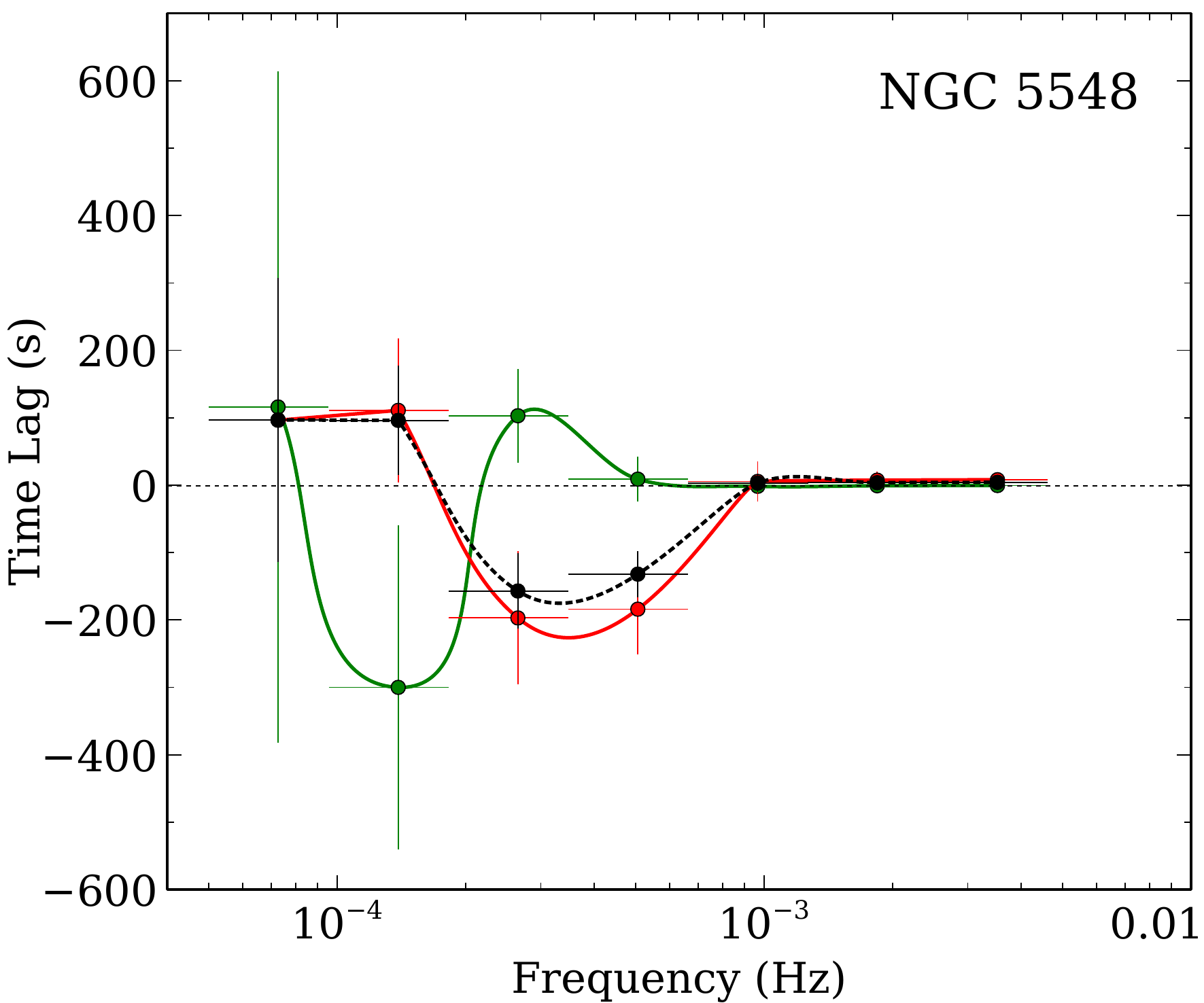}
\includegraphics[scale=0.4]{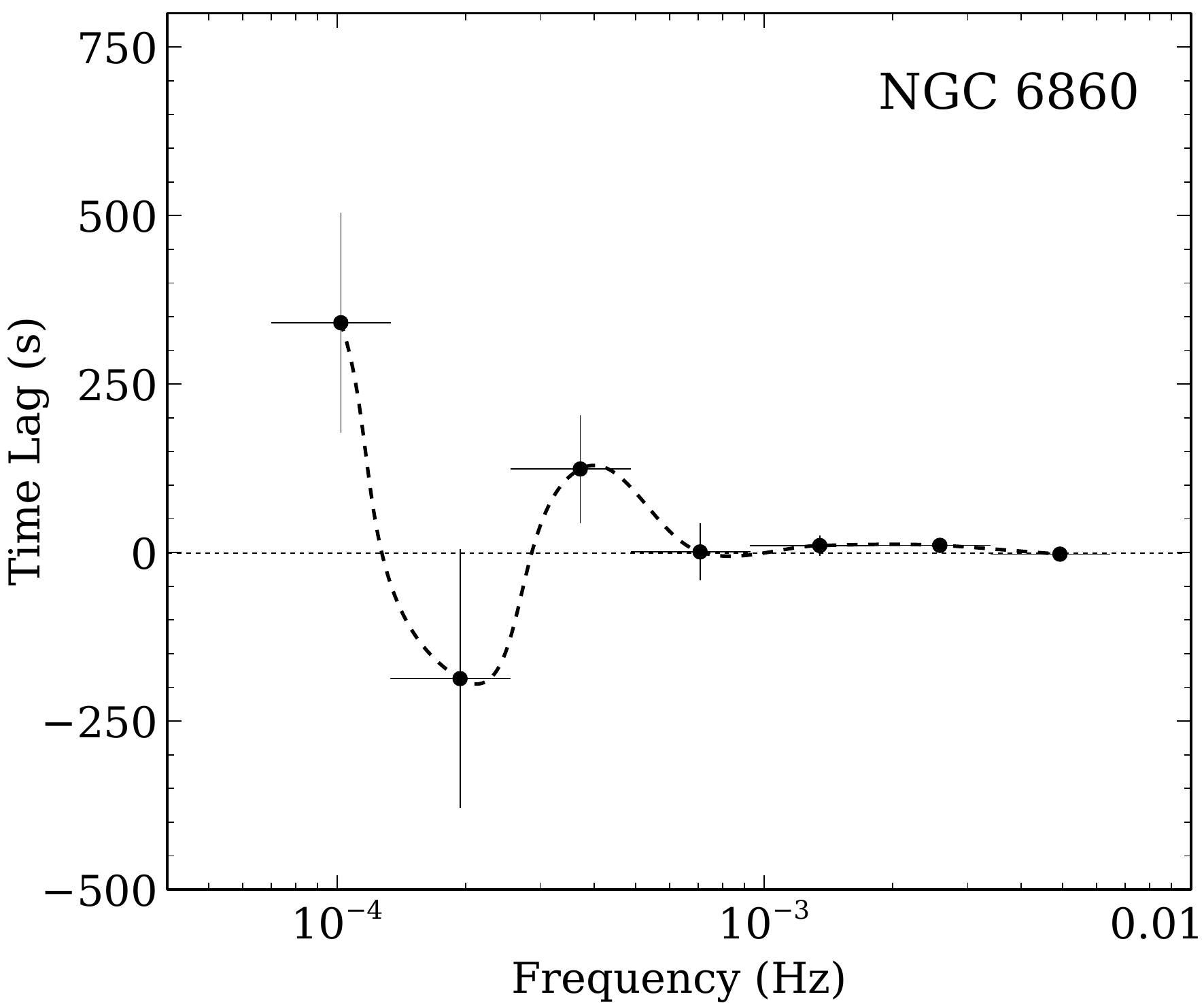}
\caption[The lag-frequency results (continued)]{The lag-frequency results (continued).}
\end{figure}

\begin{figure}
    \ContinuedFloat
\centering
\includegraphics[scale=0.4]{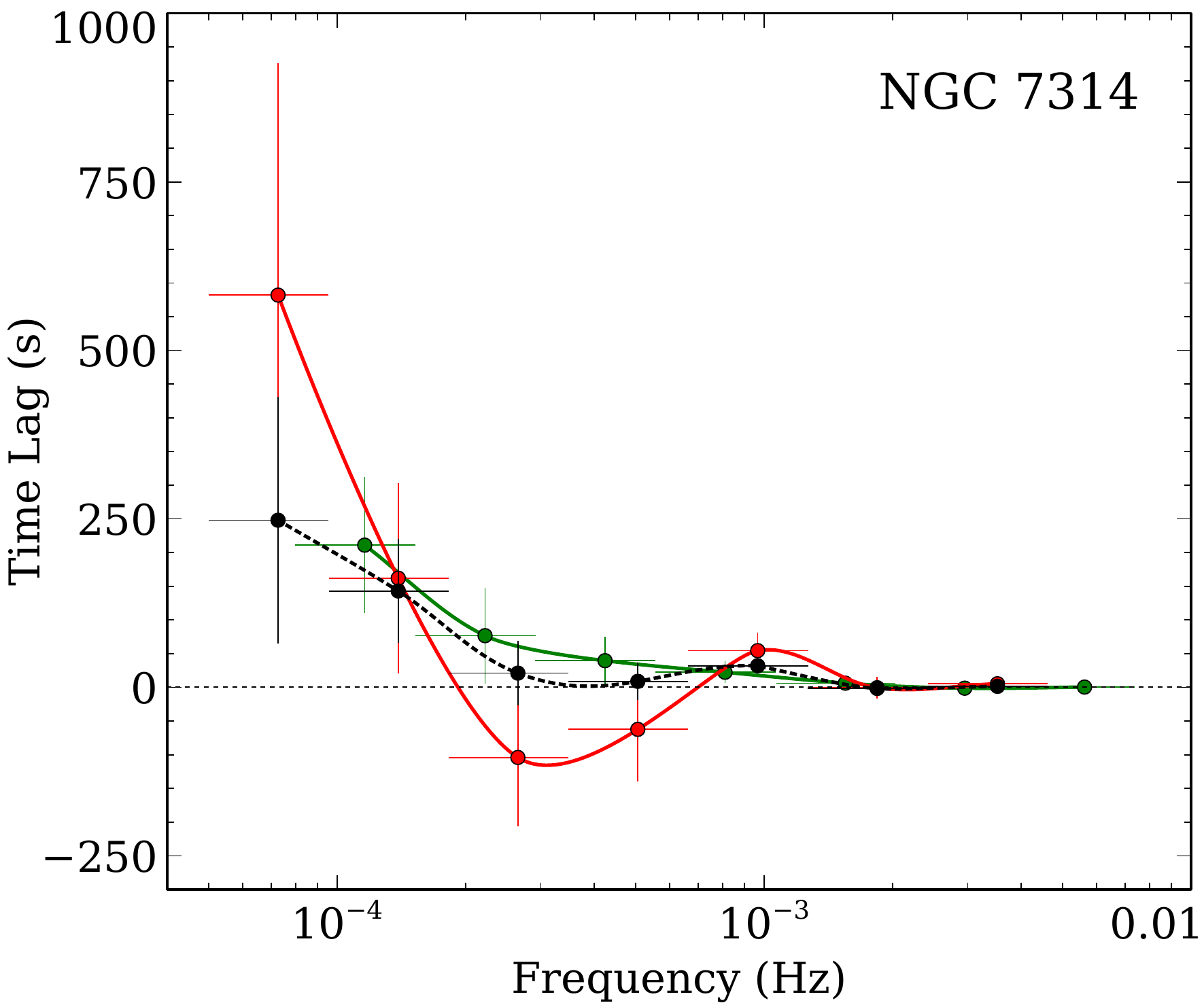}
\includegraphics[scale=0.4]{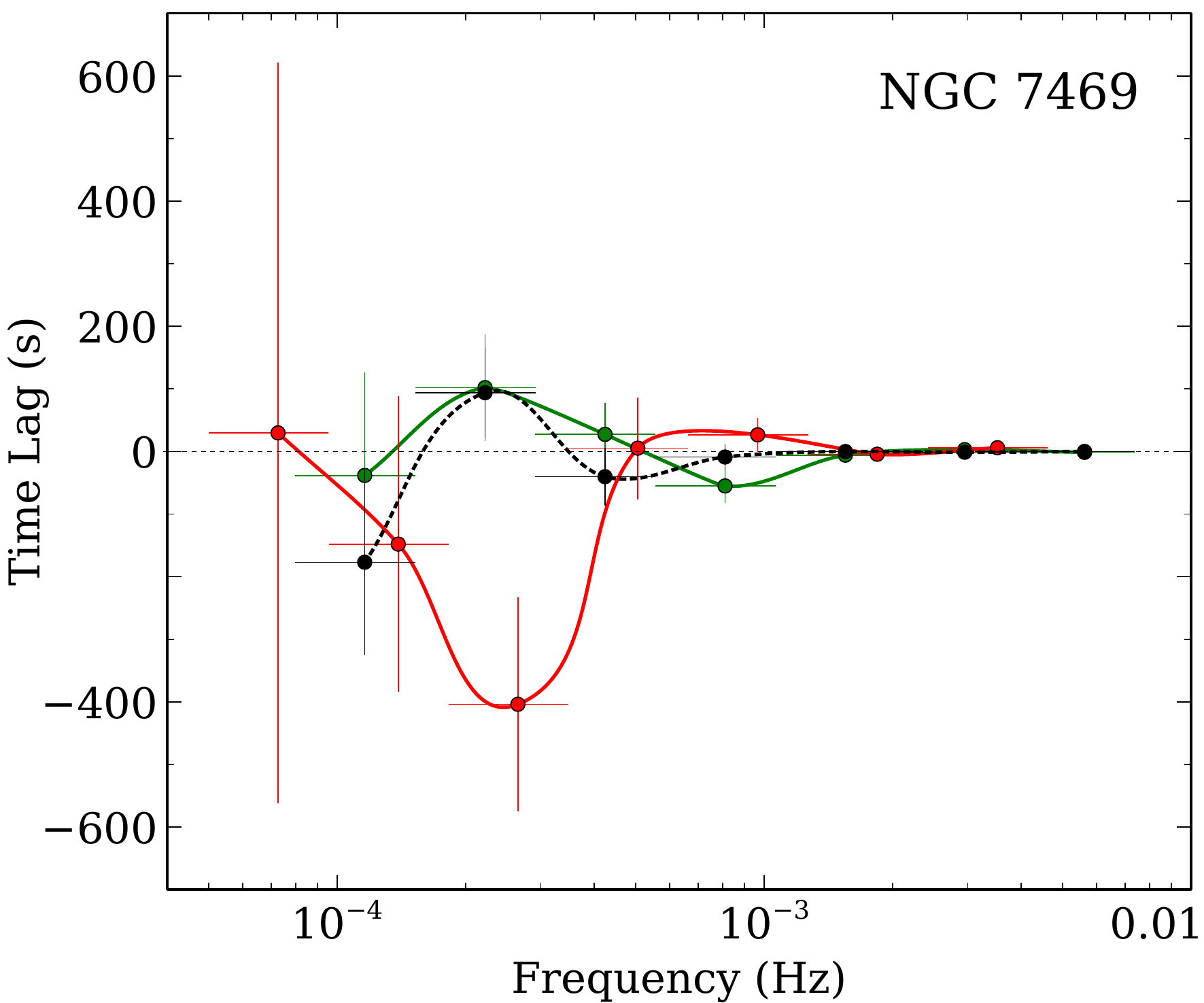}\\
\vspace{0.5cm}
\includegraphics[scale=0.4]{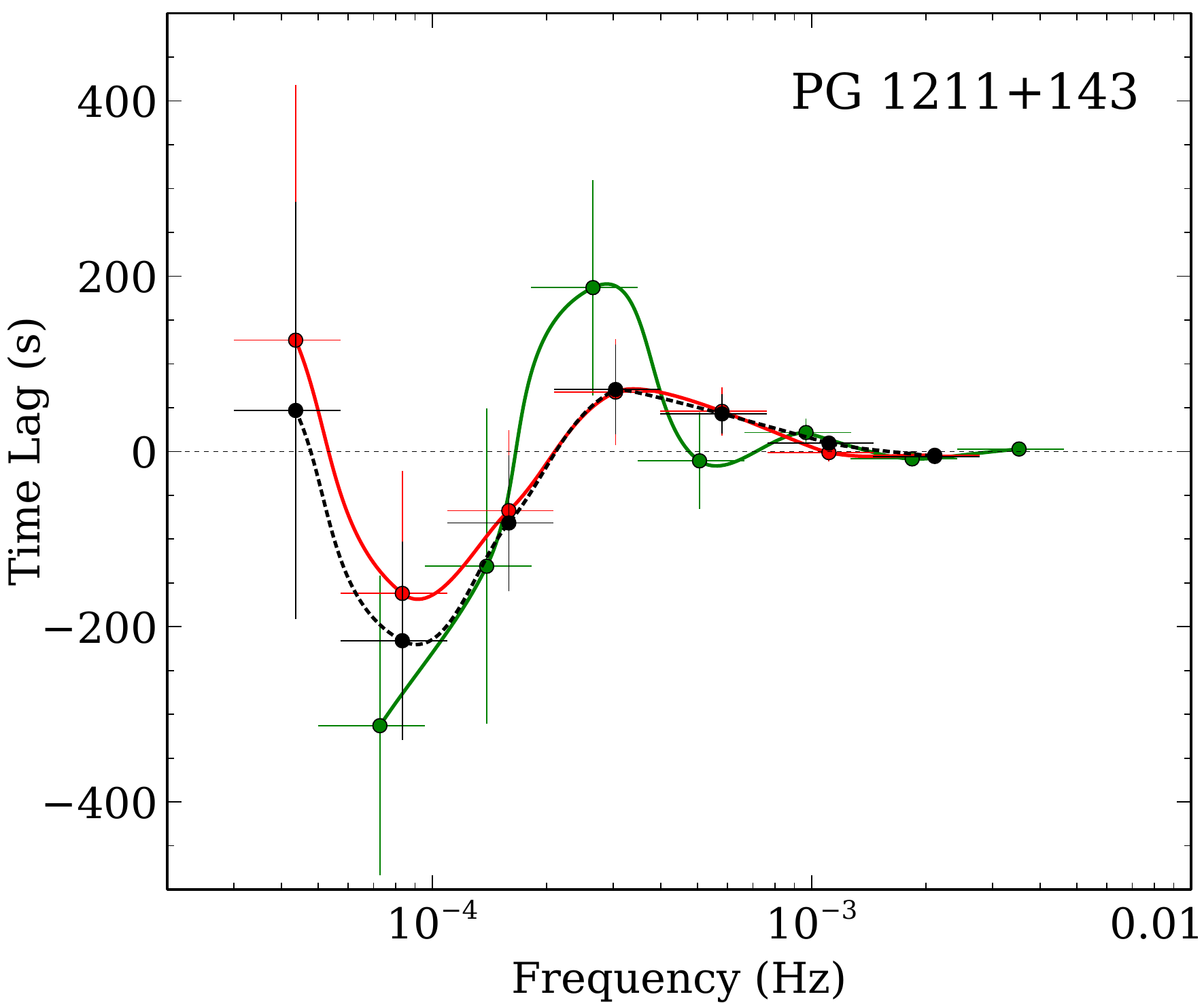} 
\includegraphics[scale=0.4]{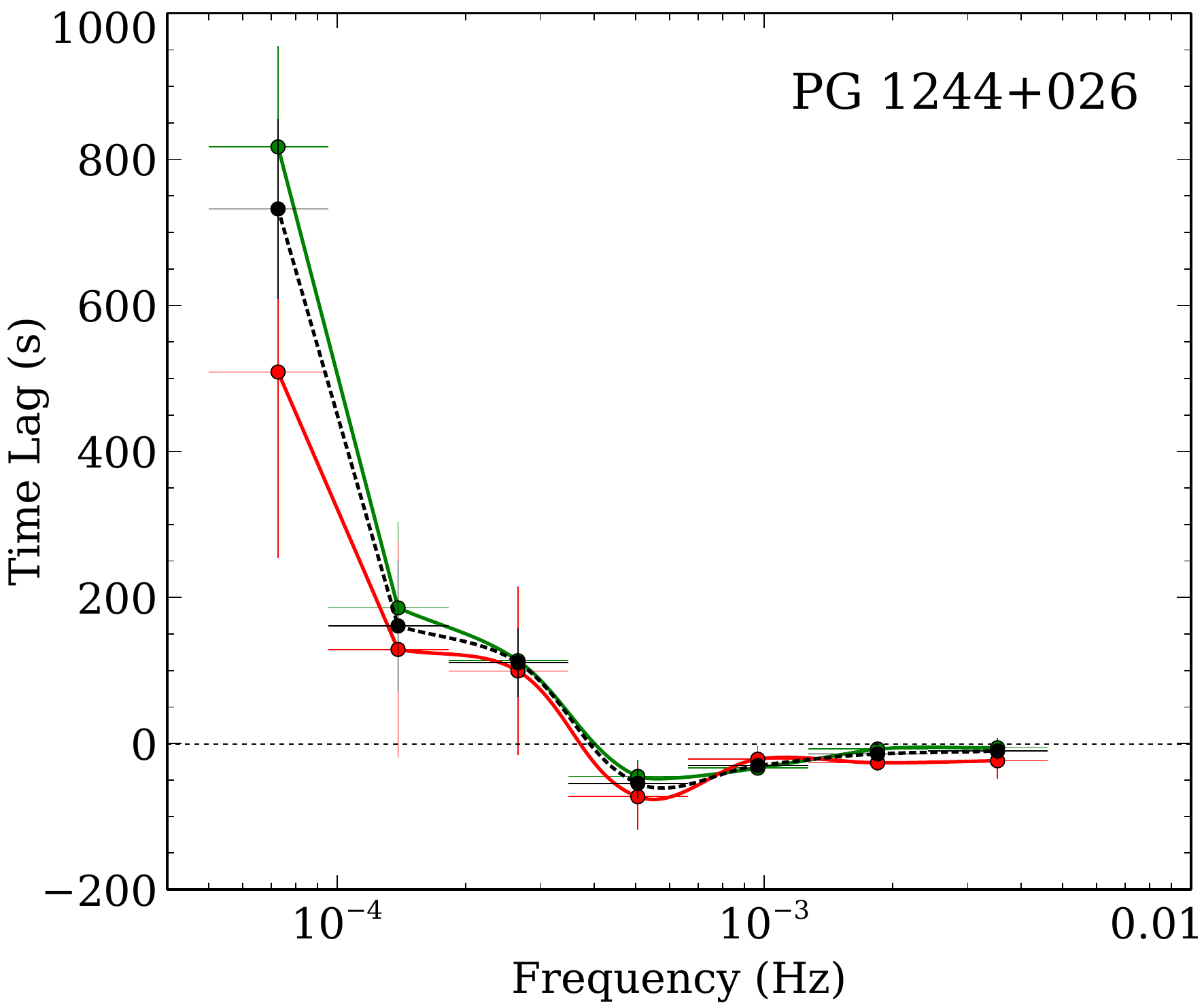}\\
\vspace{0.5cm}
\includegraphics[scale=0.4]{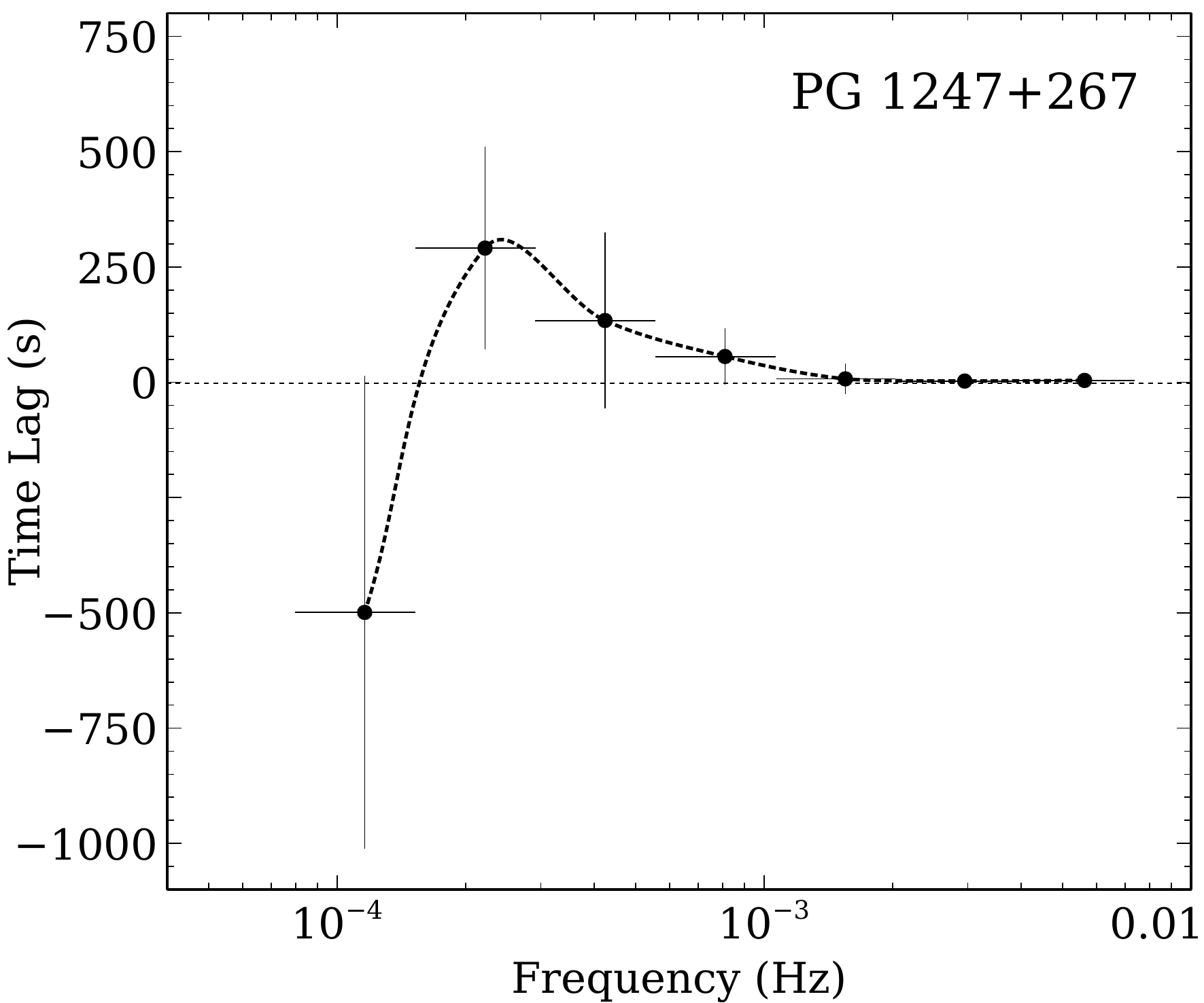}
\includegraphics[scale=0.4]{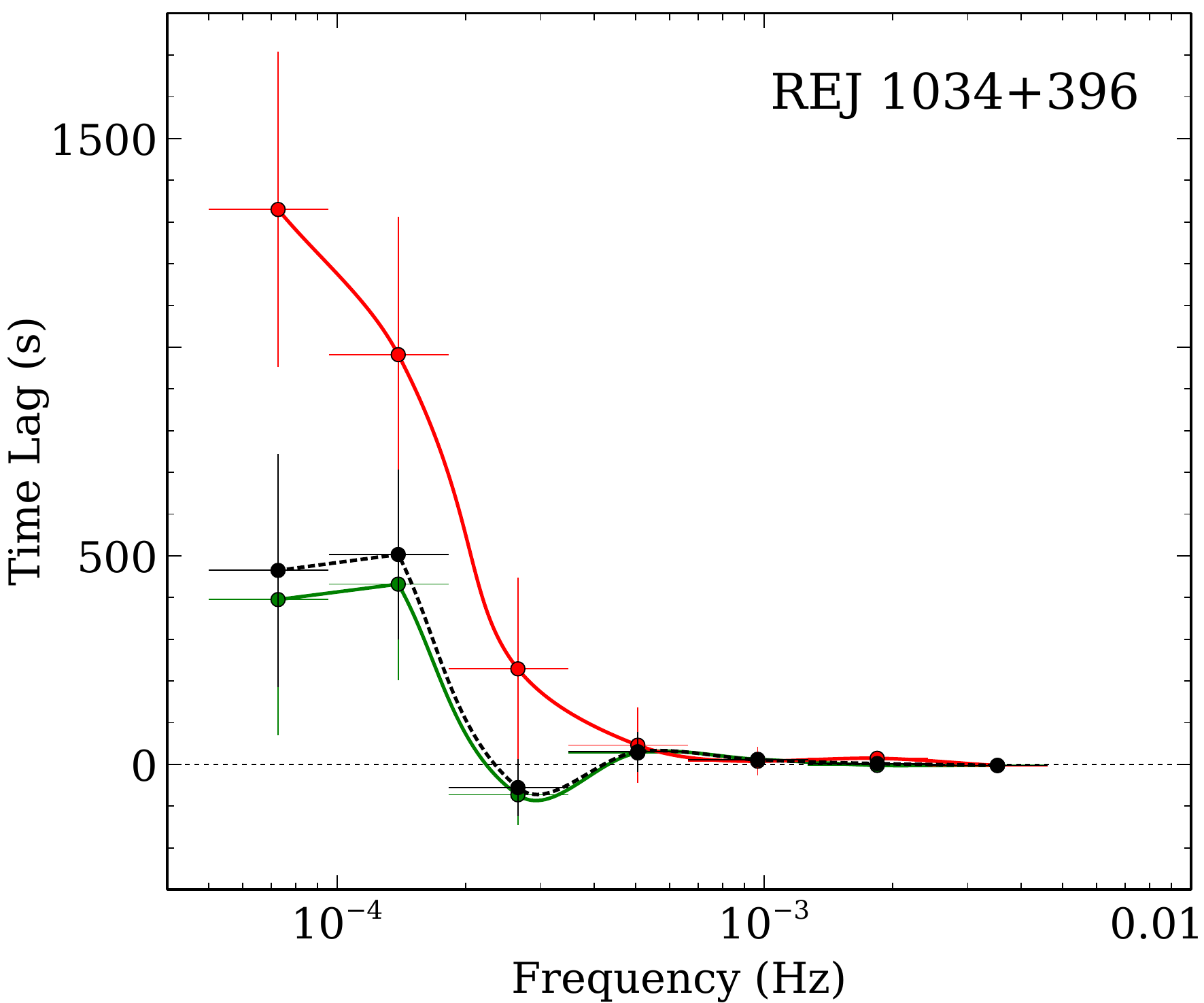}
\caption[The lag-frequency results (continued)]{The lag-frequency results (continued).}
\end{figure}